\documentclass[11pt,a4paper]{article}

\usepackage{jheppub}
\usepackage{physics}
\usepackage{mathtools}
\usepackage{amssymb}
\usepackage{comment}
\usepackage{subcaption}

\DeclareMathOperator{\erfc}{erfc}
\DeclareMathOperator{\sgn}{sgn}

\title{
Falling through the horizon of a quantum black hole
} 

\author{Victor Franken,}
\author{Thomas G. Mertens,}
\author{Bruno de S. L. Torres}

\affiliation{Department of Physics and Astronomy, Ghent University, \\ Krijgslaan 299, 9000 Gent, Belgium}

\emailAdd{victor.franken@ugent.be}
\emailAdd{thomas.mertens@ugent.be}
\emailAdd{bruno.desouzaleaotorres@ugent.be}

\abstract{We study quantum-gravitational effects on the response of an infalling detector as it crosses the horizon of a near-extremal black hole in the framework of quantum JT gravity. These effects are incorporated via the gravitational dressing needed to define both the infalling trajectory and the local observables probed by the detector in a diffeomorphism-invariant way. In the black hole exterior, a preferred choice of dressing to the Schwarzian mode of JT gravity can be motivated in connection to geometric modular flow. We show how to extend this dressing to the black hole interior, defining local observables that are gravitationally dressed to both boundaries in the thermofield double state. The gravitational dressing connects the near-horizon region to the IR sector of the Schwarzian theory, leading to measurable effects as the horizon is approached. We find that the infalling detector is able to locally determine the location of the horizon and its temperature, violating the equivalence principle, but without encountering a firewall. 
}

\begin{document}
\maketitle
\section{Introduction}

One of the challenges one faces when trying to describe local physics in quantum gravity is that the very notion of locality in dynamical theories of gravity is hard to define. Diffeomorphism invariance generally precludes an absolute definition of spacetime events, and the concept of local observables can only be understood relative to other dynamical fields or features of the system; in short, local observables in quantum gravity must be defined \emph{relationally}~\cite{Rovelli:1990ph, Giddings:2005id}. This can be achieved in practice through some form of gravitational dressing~\cite{Donnelly:2015hta, Donnelly:2016rvo, Goeller:2022rsx}.

The problem of reconstructing local physics in quantum gravity becomes somewhat more tractable in the context of AdS/CFT, arguably our best understood framework for quantum gravity to date. In AdS/CFT, the asymptotic AdS boundary provides a background structure that can act as a reference frame relative to which bulk quantities are defined. Because the metric at the boundary is not dynamical, this strategy does not face the same problems as in the bulk, where the metric can in general fluctuate. This approach to describing bulk observables in a diffeomorphism-invariant way works particularly well for bulk regions that are in the causal wedge of a single asymptotic boundary. This is the case, for instance, for small perturbations around global AdS or for observables in the exterior region of black holes~\cite{Hamilton:2005ju, Hamilton:2006az}.

The description of local observables in the interior of black holes, however, seems more mysterious. The black hole interior is out of causal contact with the asymptotic boundary, and defining local observables  there while still retaining diffeomorphism invariance explicitly becomes substantially more subtle. Indeed, even in AdS/CFT, a boundary-intrinsic construction of local observables behind black hole horizons remains a hot topic of debate~\cite{Kraus:2002iv,
Almheiri:2013hfa,
Marolf:2013dba,Papadodimas:2012aq,
Papadodimas:2013wnh,
Papadodimas:2013jku,Grinberg:2020fdj,Leutheusser:2021frk}, with many constructions~\cite{Maldacena:2001kr, Hamilton:2006fh,Lewkowycz:2016ukf,Almheiri:2017fbd,Jafferis:2020ora,Gao:2021tzr,deBoer:2022zps,Leutheusser:2022bgi,Leutheusser:2021qhd}, especially in the relatively well-understood case of eternal two-sided black holes, making use of degrees of freedom associated to both asymptotic boundaries to reconstruct the black hole interior. This is deeply tied to the entanglement structure needed for the emergence of the interior from the boundary degrees of freedom~\cite{Maldacena:2001kr,VanRaamsdonk:2010pw}.

The fact that the black hole interior seems to require two disconnected theories, defined on two distinct asymptotic boundaries, makes the horizon appear rather special: it marks the transition between a description of local physics in terms of a single asymptotic boundary and one that involves two disconnected boundaries, crucially relying on their entanglement structure. It is thus natural to ask whether these considerations lead to nontrivial structure at black hole horizons in quantum gravity. 

Relatedly, from the perspective of gravitational dressing, one singles out a feature of the semi-classical state as an anchor (e.g. a holographic boundary, the timelike worldline of a star or spaceship, etc.), with respect to which observables are defined and localized. When an observable crosses a large black hole horizon, however, even though this is semi-classically uneventful for the infaller, the gravitational dressing requires a wild readjustment due to the possible causal disconnect that occurs between probe and anchor. This suggests quantum gravitational effects related to the dressing could have important consequences, including potential violations of the equivalence principle and possibly a firewall at the horizon.

In this paper, we seek to answer this question by studying quantum-gravitational effects on the experience of an infalling observer as they cross the horizon. Defining the observer properly requires a diffeomorphism-invariant definition of its trajectory, which in turn depends on an appropriate form of gravitational dressing that smoothly connects the interior and exterior regions. This gives room for some structure near the horizon, which we probe quantitatively by studying the response of a detector as a function of its affine time. We will find that, indeed, quantum-gravitational effects can lead to nontrivial structure near the horizon, as witnessed by a local particle detector falling into the black hole. These effects allow for a localized observer to detect the location of the horizon and to extract the black hole temperature using only \emph{local} measurements, in contrast to what one would find in semiclassical gravity.

\subsection{Gravitational dressing in JT gravity}\label{subsec:JTdressing}

Our work will make use of a particular form of gravitational dressing of an infalling observer in quantum Jackiw-Teitelboim (JT) gravity~\cite{Jackiw:1984je,Teitelboim:1983ux,Almheiri:2014cka, Mertens:2022irh}. This is a well-known two-dimensional theory of gravity which describes the throat region of higher-dimensional near-extremal black holes~\cite{Fabbri:2005mw,Nayak:2018qej,Iliesiu:2020qvm}, and has led to several insights in recent developments in quantum gravity due to its high degree of solvability. For recent reviews on the fundamentals and applications of JT gravity, see~\cite{Mertens:2022irh, Turiaci:2024cad}. 

The theory is defined on a two-dimensional manifold $M$, whose bulk geometry is non-dynamical and fixed to be a patch of AdS$_2$. The metric can be written in Poincar\'e coordinates as
\begin{equation}
    ds^2 = -\frac{4\,dUdV}{(U-V)^2}.
\end{equation}
The only dynamical degree of freedom of the theory is the trajectory of the cutoff boundary curve $\partial M$. This trajectory is determined by a single function $F(t)$, which defines the Poincar\'e time $T=(U+V)/2$ of the cutoff curve $\partial M$ in terms of the boundary proper time $t$ via $T=F(t)$. The action of JT gravity in the reparametrization mode $F(t)$ is then given by~\cite{Almheiri:2014cka,Jensen:2016pah,Maldacena:2016upp,Engelsoy:2016xyb}
\begin{equation}\label{eq:JTactionSchwarzian}
    S[F] = -C\int dt\,\{F, t\} + S_0\chi(M),
\end{equation}
where $\{F, t\} \equiv \frac{F'''}{F'} - \frac{3}{2}\left(\frac{F''}{F'}\right)^2$ is the Schwarzian derivative, $C \sim 1/G_N$ the Schwarzian coupling constant, and $\chi(M)$ is the Euler characteristics of the manifold $M$. The first term on the right-hand side of Equation~\eqref{eq:JTactionSchwarzian} is usually referred to as the Schwarzian action, and it governs the dynamics of the reparametrization function  $F(t)$, most commonly called the Schwarzian mode. The topological term $S_0\chi(M)$ plays no role classically, but it can become important at the quantum level due to the possibility of topology fluctuations in the gravitational path integral. Our analysis in the present paper will be restricted to JT gravity with disk topology, which contains the leading contributions in the limit $S_0\to\infty$ where topology fluctuations are suppressed. We make some comments on higher-topology effects in Section~\ref{sec:conclusion}, but defer their more systematic analysis to future work~\cite{toap}.

\begin{figure}[ht]
    \centering
        \includegraphics[width=0.3\linewidth]{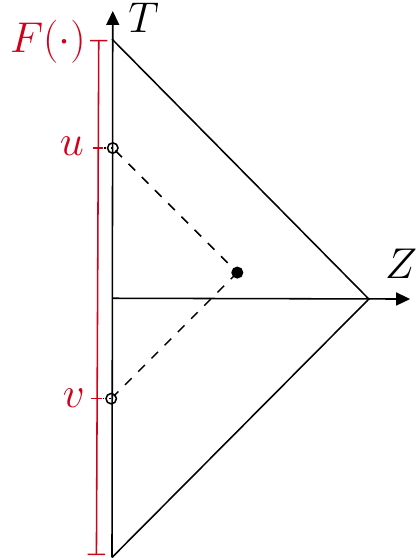}
        \caption{For a given off-shell reparametrization $F(\cdot)$, the gravitational dressing of a point to the boundary is $U=F(u), \,V=F(v)$.
        \label{fig:dressout}}
\end{figure}

We consider a simple matter content made of a minimally coupled massless scalar field. Gravitational dressing of the scalar field to the boundary effectively couples the field with fluctuations in the Schwarzian mode $F(t)$. More concretely, we locate a bulk point $(U,V)$ in the black hole exterior in terms of boundary data via the reparameterization 
\begin{equation}
\label{eq:dressing}
    U=F(u), \quad V=F(v),
\end{equation}
where $u$ and $v$ are the boundary proper times at which future and past-directed lightrays sent from $(U,V)$ meet the boundary. See Figure~\ref{fig:dressout} for an illustration. 
In other words, bulk points are labeled by two boundary times $u$ and $v$, which are mapped to the bulk through the Schwarzian $F(t)$. At the classical level, this is merely a change of coordinates. However, when promoting $F$ to a quantum operator of the Schwarzian theory, this leads to quantum gravitational effects on matter-coupled JT gravity~\cite{Blommaert2019,Mertens:2019bvy, Blommaert:2020yeo, DeVuyst:2022bua, Mertens:2025rpa}. 

It was argued in~\cite{Mertens:2025rpa} that the dressing~\eqref{eq:dressing} is the only viable gravitational dressing in JT gravity that is consistent with the construction of diffeomorphism-invariant observables in quantum gravity via the modular crossed product~\cite{Witten:2021unn}. This provides a strong motivation for this choice of dressing, as the modular crossed product is able to promote the local algebra of operators of a QFT (which, strictly speaking, does not have well-defined density matrices or entropies for subregions~\cite{Witten:2018zxz, Araki:1964lyc, Driessler:1976ky}) to an algebra where local density matrices and renormalized entanglement entropies can be defined~\cite{Chandrasekaran:2022cip, Chandrasekaran:2022eqq, ShadiCrossedProduct, Jensen:2023yxy, Fewster:2024pur}.

\begin{figure}[ht]
        \centering
\includegraphics[width=0.5\linewidth]{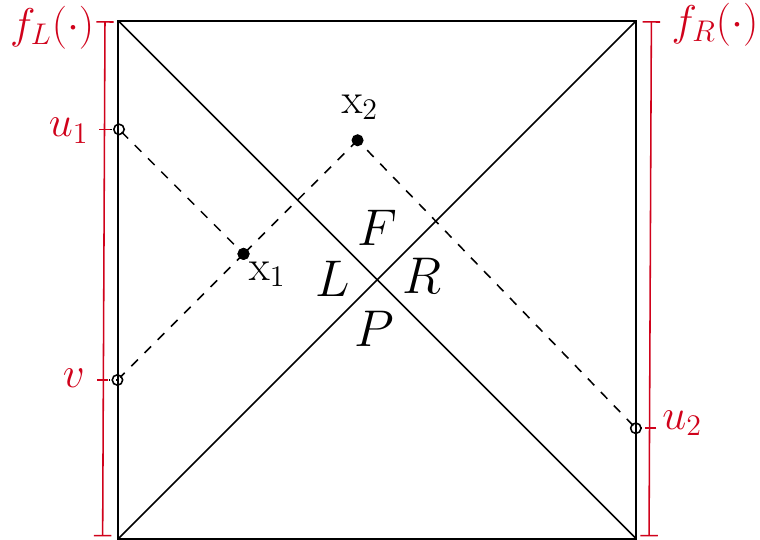}
        \caption{Gravitational dressing of a point $\mathrm{x}_1$ in the exterior and a point $\mathrm{x}_2$ in the interior. The thermal reparametrizations $f_L(\cdot)$ and $f_R(\cdot)$ are related by analytical continuation~\eqref{eq:fcontinuation}. {The notation $L$, $F$, $R$, $P$ for the four wedges (Left, Future, Right, and Past, respectively) will be used throughout the paper.}
        \label{fig:dress}}
\end{figure}

The dressing provided above has a limitation, however, in that it involves anchoring points to the boundary via null rays. As such, it is restricted to the region of the bulk that is causally accessible to the asymptotic boundary. In this work, we propose to generalize the construction above in order to extend the dressing past the black hole horizon. In order to do that, we work in the thermofield double state in JT gravity, which is a pure state of two asymptotic boundaries that appears thermal when restricted to the regions that are causally accessible to just one of the boundaries. The structure of the thermofield double allows one to describe the interior by taking advantage of the fact that the Euclidean continuation
\begin{equation}
\label{eq:continuation}
    u \rightarrow - u \pm i\frac{\beta}{2}
\end{equation}
yields correlation functions of operators behind the horizon just from knowledge of the correlation functions of operators in the exterior. To do this explicitly in the case of JT gravity, we write the Schwarzian mode in its thermal parametrization $F(t) = \tanh\left(\frac{\pi}{\beta}f(t)\right)$ where $\beta$ is the inverse temperature of the thermofield double state in question. The dressing~\eqref{eq:dressing} is then extended by defining
\begin{align}
\begin{split}
\label{eq:fcontinuation}
    f_L(u)&=f(u),\\
    f_R(u) &= -f\left(-u+ i\frac{\beta}{2}\right) + i\frac{\beta}{2},
\end{split}
\end{align}
where $f_L$ and $f_R$ refer to the Schwarzian dressing of points in the left exterior (whose dressing involves exclusively the left boundary) and future interior of the eternal black hole geometry (whose dressing also includes the right boundary), respectively. Geometrically, this can be seen as a modification of the anchoring of $(U,V)$ to the boundary, as shown in Figure~\ref{fig:dress}. When a point is in the interior of the black hole, $v$ remains the boundary time at which a past lightray shot from the point meets the (say left) boundary, while $u$ becomes the boundary time at which a past lightray shot from the point meets the right boundary. We will see that this procedure allows for an efficient computation of gravitationally dressed bulk observables across all regions of the two-sided black hole geometry.

The bulk quantum dressed field is determined in terms of the boundary CFT$_1$ operator through the relation
\begin{equation}
\phi(\mathrm{x}_1) = \int_v^{u_1} dt \,\mathcal{O}_1(t),
\end{equation}
which holds for any choice of time reparametrization $f_L(t)$. This construction is just the HKLL bulk reconstruction formula in AdS$_2$ \cite{Hamilton:2006az}, but now extended to hold off-shell as an insertion in the Schwarzian path integral. The bulk field behind the horizon is defined through the continuation \eqref{eq:continuation} and becomes
\begin{equation}
\phi(\mathrm{x}_2) = \int_{\mathcal{C}_{v,{u_2}}} dt \,\mathcal{O}_1(t)
\end{equation}
along a complexified integration contour $\mathcal{C}_{v,{u_2}}$, which can be visualized as running up from $v$ on the left boundary, and then running down towards $u_2$ on the right boundary, covering the entire boundary region that is spatially separated from $\mathrm{x}_2$ \cite{Hamilton:2006fh}.

\subsection{Unruh-DeWitt detector}\label{sub:UDWintro}

To quantitatively test whether gravitational dressing affects the experience of an infalling observer, we will model the ``observer'' as a local probe based on the Unruh-DeWitt (UDW) detector~\cite{Unruh1976, DeWitt:1980hx}. This consists of a localized quantum system with discrete energy levels that is coupled linearly to a bulk matter field. This type of detector has been extensively used in quantum field theory in curved spacetimes, and is a rather useful tool for the study of several effects in quantum information and QFT from an operational perspective, see \emph{e.g.}~\cite{Unruh1976, Sciama1977, DeWitt:1980hx, Reznik2003, Reznik:2003mnx, jose, Pozas-Kerstjens:2015, kelly}. In the context of JT gravity, a version of the Unruh-DeWitt detector with the gravitational dressing~\eqref{eq:dressing} was used in~\cite{Blommaert:2020yeo} to study the response an accelerated detector along a ``static'' trajectory at fixed radial coordinate $z=(u-v)/2$. This allowed for a study of quantum-gravitational corrections to the thermal bath experienced by static observers in this model, and also provided an operational way to probe the chaotic level statistics of quantum black holes.

The general strategy to probe quantum fields with UDW detectors involves setting up a coupling between detector and field of the form
\begin{equation}\label{eq:generalUDW}
    S_I = q\int\,d\lambda\,\chi(\lambda)\mu(\lambda) O(\mathrm{x}(\lambda)).
\end{equation}
Here, $S_I$ denotes the interaction action (roughly the time integral of an interaction Hamiltonian), $q$ is a coupling constant, $\lambda$ is a time parameter defined along the detector's trajectory, $\mu$ is an observable of the detector, and $O$ is some field observable. $\mathrm{x}(\lambda)$ is the trajectory of the detector in spacetime. The simplest quantity we can compute in this setting is the so-called vacuum excitation probability, which is the probability that a detector will transition from its ground state to an excited state by coupling to a quantum field for some finite amount of time. From the interaction action above, the probability that the detector will transition from a state $\ket{E_i}$ to a state $\ket{E_f}$, where $E$ denotes the internal energy of the detector in its proper frame, is then given at leading order in perturbation theory by
\begin{equation}\label{eq:vacuumexcitationprobintro}
    P_{\text{exc}} \propto \int d\lambda d\lambda ' e^{-i\omega(\lambda-\lambda')}\chi(\lambda)\chi(\lambda')\langle O(\lambda)O(\lambda')\rangle,
\end{equation}
where $\omega=E_f-E_i$, and $\langle O(\lambda)O(\lambda')\rangle$ is the two-point function of the bulk observable $O$ in the initial state of the field. We will review other basic aspects of Unruh-DeWitt model in Section~\ref{sub:UDW}.

In this paper, we place the detector on an infalling null trajectory that starts in the black hole exterior and then crosses into the interior. Quantum-gravitational effects then impact the response of the detector through the gravitational dressing on both the detector itself and the field observables to which the detector couples. The dressing of the detector's infalling trajectory promotes the affine time $\lambda$ to a quantum operator in the Schwarzian theory, given by
\begin{equation}
\label{eq:affineop}
    \lambda - \lambda' = \int_{u'}^u dt\, \mathcal{O}_1(it,iv),
\end{equation}
where $\mathcal{O}_1$ is the Schwarzian bilocal operator~\cite{Mertens:2017mtv}, given explicitly in \eqref{eq:metricSch} later on. One can extend $\lambda$ past the horizon by virtue of the analytic continuation~\eqref{eq:continuation}. Then, the dressing of the bulk observables will also make the matter two-point function $\langle O(\lambda)O(\lambda')\rangle$ an operator of the Schwarzian theory, whose expectation value can also be computed explicitly. For the case of a minimally coupled massless scalar, for instance, the explicit two-point function with gravitational dressing in the black hole exterior reads
\begin{equation}
    W(\mathrm{x},\mathrm{x}') = \frac{1}{4\pi}\int_v^u dt \int_{v'}^{u'}dt'\, \mathcal{O}_1(it,it'),
\end{equation}
with points in the interior again being accessed by the continuation $u\rightarrow -u\pm i\beta/2$.\footnote{The replacement $v\rightarrow -v\pm i\beta/2$ sends the point to the past interior. Finally, replacing both $u$ and $v$ with their analytically continued values takes the point to the opposite exterior region.} The thermal expectation value of the Schwarzian bilocal has an exact integral form, from which we can then compute the response of the infalling detector as it crosses the horizon.

\subsection{Detector-based criterion for a firewall}\label{sec:firewallintro}

If an infalling observer were to detect non-trivial features at the horizon, it is natural to wonder how ``dramatic'' those effects are. In the context of the physics of quantum black holes, this is often embodied in the question of whether quantum gravity leads to the emergence of \emph{firewalls} at or near black hole horizons~\cite{Almheiri:2012rt, Bousso:2012as, Nomura:2012sw, Almheiri:2013hfa, Harlow:2013tf, Stanford:2022fdt, Blommaert:2024ftn,Chandrasekaran:2026gvk}. We would then like to test the presence of a firewall in our model that would prevent some observers from entering the black hole interior. We thus put forth a proposal for an operational definition of a firewall based on the physics of detectors:\\

\textbf{Definition:} \emph{A particle detector is said to encounter a firewall if its excitation probability when coupling smoothly to a quantum field decays at most as a power law in the detector's energy gap $\omega$; i.e., if there is $\alpha\in \mathbb{R}^+$ such that
\begin{equation}
\label{def:firewall}
    \lim_{\omega\to\infty} \omega^{\alpha}P_{\text{exc}}(\omega)>0. 
\end{equation}
Alternatively, the detector crosses the horizon ``safely'' (i.e., without meeting a firewall) if its excitation probability goes to $0$ as $\omega\rightarrow \infty$ faster than any polynomial in the detector's energy gap $\omega$.
\\}

The motivation for this criterion, detailed in Section~\ref{sec:firewall}, ultimately relies on the fact that finite-energy states in semiclassical gravity must share the same short-distance singularity structure as the vacuum in Minkowski space, and it is well-known that the vacuum excitation probability of a smoothly coupled particle detector decays exponentially with the detector's energy gap. Since the high-energy limit of the detector should probe the short-distance structure of the corresponding field correlators, the falloff properties of the excitation probability at very high energies should thus be similar to that in Minkowski. As such, witnessing power law decay draws a sharp distinction from what would be expected in a non-singular state in semiclassical gravity. It is thus with this criterion in mind that we will examine later if an Unruh-DeWitt detector coupled to a suitably gravitationally dressed bulk field experiences a firewall as it crosses the black hole horizon in JT gravity.

\subsection{Outline and summary of main results}

Sections~\ref{sec:background} and~\ref{sec:semiclassical} cover the background material needed to set up our main problem of interest. Section~\ref{sec:background} introduces various patches of the AdS$_2$ geometry that will be useful in later sections, and also reviews the two-point function of a massless scalar field and the Unruh-DeWitt particle detector model. With that setup, in Section~\ref{sec:semiclassical} we present the calculation of the vacuum excitation probability of an Unruh-DeWitt in a classical trajectory in a fixed AdS$_2$ background, which will serve as the semiclassical benchmark against which the quantum-gravitational effects will be compared.

\begin{figure}[ht]
\begin{subfigure}[t] {0.48\linewidth}\centering   
\includegraphics[width=1\linewidth]{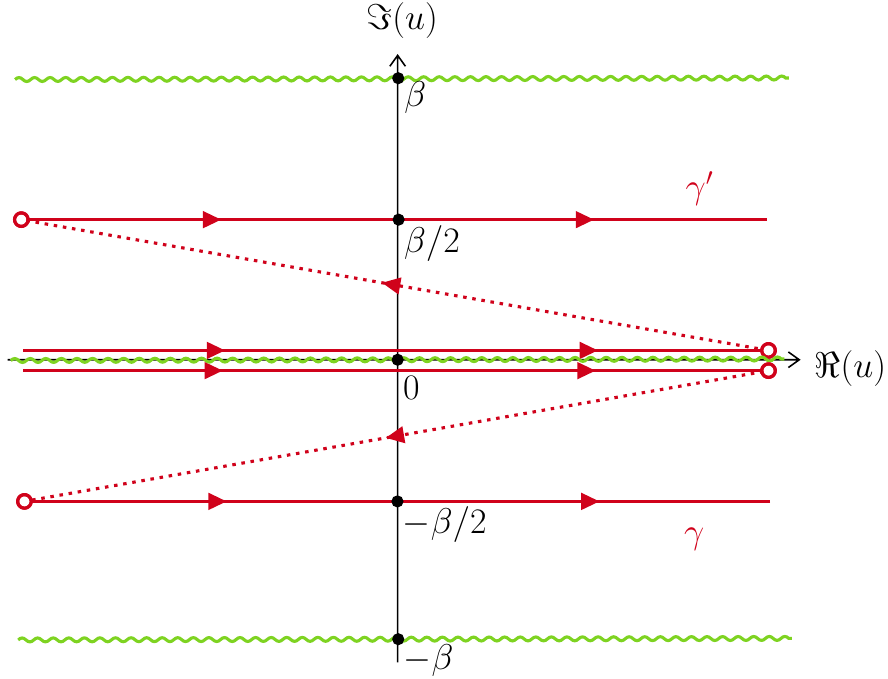}
    \caption{Integration contours $\gamma$ and $\gamma'$ used in \eqref{eq:introP}.}
    \label{fig:contour}
\end{subfigure}
\begin{subfigure}[t] {0.48\linewidth}\centering   
\includegraphics[width=0.8\linewidth]{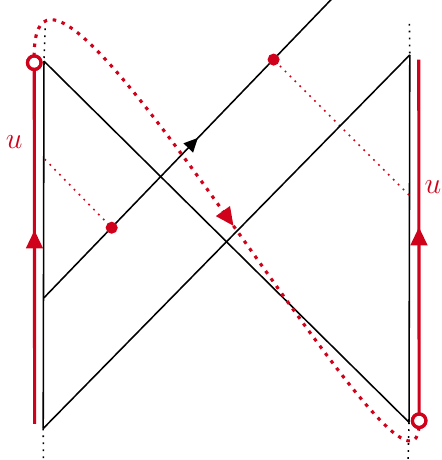}
\caption{Bulk trajectory dressed to the contours $\gamma$ and $\gamma'$. }
\label{fig:contour2}
\end{subfigure}
\caption{The contours $\gamma$ and $\gamma'$ correspond to an infalling trajectory. The first part of the contour on $\mathbb{R}\pm i\epsilon$ correspond to the part of the trajectory in the exterior region, while the part of the contour $\mathbb{R} \pm i\beta/2$ corresponds to the part of the trajectory in the interior region. The two contours are continuous in the sense that observables are continuous at the continuation points, depicted as white dots. Branch cuts of the correlator are depicted by green wiggly lines.}
\end{figure}

In Section~\ref{eq:dressedW}, we present the gravitationally dressed two-point function of a massless scalar field after coupling to the Schwarzian mode, including an extension of the one-sided dressing to also cover points behind the black hole horizon. To do so, we set the Schwarzian mode in the thermofield double and obtain a correlator defined continuously across the horizon, as a function of a unique boundary variable. This variable can be seen as a contour in the complex plane that covers the axes $\mathbb{R}\pm i\epsilon$ and $\mathbb{R}+i\beta/2$,\footnote{The $i\epsilon$ shift is a regularization of the integral as the Schwarzian bilocal has branch cute on $\mathbb{R}\pm i n\beta$ for $n\in\mathbb{N}$. The limit $\epsilon\rightarrow 0$ is to be taken at the end of the computation.} as shown in Figure~\ref{fig:contour}. The choice of sign corresponds to two different contours $\gamma$ and $\gamma'$, and is fixed by imposing that the two-point function is consistent with hermiticity of operators. Practically, $\mathrm{x}$ follows $\gamma$ and $\mathrm{x}'$ follows $\gamma'$. 

In Section~\ref{sec:Schw}, we use these results to compute the excitation probability of the detector in the Schwarzian theory. In the case of the boundary-based computation, we find the expected Schwarzian integral, which gives a thermal response function in the semiclassical limit. For the infalling observer experiment, the excitation probability of the infalling observer can be expressed as a double contour integral
\begin{equation}
\label{eq:introP}
    P_{\rm exc} \propto \int_{\gamma} du \int_{\gamma'} du' \chi(u)\chi(u')e^{-i\omega\int_{u'}^u dt\, \mathcal{O}_1(it,iv)}\mathcal{O}_1(iu,iu'),
\end{equation}
where the contours $\gamma$ and $\gamma'$ are shown in Figure~\ref{fig:contour}, and $\chi(u)$ is a switching function that regulates the strength of the interaction in time. The associated trajectory of the detector and its dressing are shown in Figure~\ref{fig:contour2}.

The double contour integral in Equation~\eqref{eq:introP} is hard to compute in general. To make analytical progress, we consider the near-horizon limit for the affine time:
\begin{equation}
\label{eq:lH}
    \lambda \sim \frac{C}{8\pi Z_0(\beta)}\left\{\begin{array}{cc}
       -(u-v)^{-2} & \text{(exterior)} \\
       (u'+v)^{-2} & \text{(interior)}
    \end{array} \right.,
\end{equation}
where $Z_0(\beta) = \frac{1}{4\pi^2}\left(\frac{2\pi C}{\beta}\right)^{3/2}e^{2\pi^2 C/\beta}$. 
In this regime, we find integral expressions for the excitation probability when the detector is turned on for a small and fixed amount of time, centered around $\lambda_0$. Numerical results then lead us to the following conclusions:
\begin{enumerate}
    \item In the Schwarzian theory, the detector's excitation probability exhibits a smooth and finite peak at the horizon. This means that the infalling observer can detect the horizon's location purely by local measurements.
    \item An infalling observer can foresee the horizon before crossing it. This is because the detector's excitation probability starts to grow (relative to its semiclassical expectation) long before the horizon-crossing moment. This also shows that the detection of the horizon is due to the near-horizon region probing the IR sector of the Schwarzian theory, not to the apparent discontinuity of the dressing, as expected from Equation~\eqref{eq:lH}.
    \item Schwarzian effects are large. The height of the peak in excitation probability is greater than the semiclassical value (which is constant in $\lambda_0$) by multiple orders of magnitude.
    \item The magnitude of the deviation from the semiclassical expectation in the detector response depends on $\beta$, and therefore the detector can measure the black hole temperature locally around the horizon. In particular, the vacuum excitation probability $P_{\text{exc}}$ grows with $\beta$, signaling that quantum fluctuations become stronger at low temperatures (as is generally the case in JT gravity).
    \item The detector does not meet a firewall. Indeed, assuming that the detector can withstand the peak in excitation probability at the horizon, nothing else prevents it from crossing into the interior. In particular, the excitation probability decays exponentially in the excitation energy $\omega$ for sufficiently large energy gaps.
\end{enumerate}
Several directions for future research based on these results are presented in Section~\ref{sec:conclusion}.

\section{Setup}
\label{sec:background}
In this Section, we review some coordinate systems that are convenient to describe patches of AdS$_2$, and also briefly recall the two-point function of a massless scalar on background. We end with a review of some basic aspects of the Unruh-DeWitt detector model that will be useful for later sections of the paper.
\subsection{Patches of AdS\texorpdfstring{$_2$}{2}}

The metric of AdS$_2$ described in global coordinates $(\tau, x)$, is given by
\begin{equation}\label{eq:globalads}
    ds^2 = \dfrac{-d\tau^2 + dx^2}{\sin^2 x} = -\dfrac{4\,dx_+\,dx_-}{\sin^2(x_+-x_-)}.
\end{equation}
The range of the coordinates is $\tau\in(-\infty, +\infty)$ and $x\in (0, \pi)$. In the second equation, we have written the metric in terms of lightcone coordinates $x_{\pm} = (\tau\pm x)/2$. This patch covers the entire AdS space; hence the name \emph{global}. This way of describing AdS$_2$ makes it evident that the spacetime is composed of two disconnected timelike boundaries, sitting at $x=0$ and $x=\pi$. 

The \emph{Poincar\'e patch} of AdS$_2$ is obtained by restricting to the causal hull of a segment with $\tau\in [-\pi, \pi]$ on one of the boundaries. This is just enough time for a lightray to complete a full cycle bouncing between one boundary and the other. The Poincar\'e patch can be parametrized by changing to the coordinates $T, Z$ for which
\begin{align}
    x_{\pm} = \arctan(T\pm Z).
\end{align}
The coordinate transformation is only well-defined in the region between $x_- = -\pi/2$ and $x_+ = \pi/2$. The range in $(T, Z)$ under these definitions is $T\in (-\infty, \infty)$ and $Z\in (0, \infty)$, with $Z=0$ being the asymptotic boundary at $x_- = x_+$. The metric in these coordinates then becomes
\begin{equation}
    ds^2 = \dfrac{-dT^2 + dZ^2}{Z^2}=-\frac{4\,dUdV}{(U-V)^2},
\end{equation}
where $U=T+Z$ and $V=T-Z$. The so-called \emph{black hole patch} of AdS$_2$ is obtained by taking the reparametrization
\begin{equation}\label{eq:poincarebh}
    T\pm Z = \dfrac{\beta}{\pi}\tanh\left[\dfrac{\pi}{\beta}(t \pm \rho)\right],
\end{equation}
which leads to the metric
\begin{equation}\label{eq:blackholepatch}
    ds^2 = \dfrac{4\pi^2}{\beta^2}\dfrac{(-dt^2+d\rho^2)}{\sinh^2\left(\frac{2\pi}{\beta}\rho\right)}.
\end{equation}
The range of the coordinates $t, \rho$ is once again $t \in (-\infty, \infty)$ and $\rho \in (0, \infty)$, with the asymptotic boundary at $\rho=0$. 
Setting $r=r_h\coth \frac{2\pi}{\beta}\rho$, the geometry takes the standard static black hole form
\begin{equation}\label{eq:AdSblackholecoord}
ds^2 = -(r-r_h)^2dt^2 + \frac{dr^2}{r^2-r_h^2}, \qquad r_h = 2\pi/\beta.
\end{equation}
In terms of the double-null coordinates $(u, v)$ defined by $u=t + \rho$ and $v = t - \rho$, if the bulk point $(u,v)$ is in the black hole exterior ($u>v$), the analytically continued bulk point $(u\pm i \beta/2,v)$ corresponds to the time $t\pm i \beta/4$, a quarter of the thermal circle away in Euclidean time. The same analytic continuation then takes the radial coordinate $r$ to the range $0<r<r_h$, which is the black hole interior.

The form~\eqref{eq:blackholepatch} or~\eqref{eq:AdSblackholecoord} of the metric highlights the existence of a Killing horizon at $\rho=\infty$ (or equivalently $r=r_h$), with generator $\chi=\partial_\tau$ and surface gravity $\kappa = 2\pi/\beta$. 
This corresponds to the two-dimensional black hole in AdS$_2$. In terms of $u=t + \rho$ and $v = t - \rho$, the horizon-generating Killing vector is
\begin{equation}
    \chi = \partial_u + \partial_v.
\end{equation}
Lastly, we define Kruskal coordinates adapted to the black hole horizon, given by
\begin{align}
    \tilde{u} = -e^{-\kappa(t+\rho)} = -e^{-\kappa u}, \label{eq:bhexterior1}\qquad\quad
    \tilde{v} =  e^{\kappa (t-\rho)} = e^{\kappa v}, 
\end{align}
in terms of which the horizon-generating Killing vector is
\begin{equation}
    \chi = \kappa\left(-\tilde{u}\,\partial_{\tilde{u}}+\tilde{v}\,\partial_{\tilde{v}}\right).
\end{equation}
and the metric becomes
\begin{equation}
    ds^2 = -\dfrac{4\,d\tilde{u}\,d\tilde{v}}{(1+\tilde{u}\tilde{v})^2}.
\end{equation}
The future horizon corresponds to $\tilde{u}=0$, whereas the past horizon is at $\tilde{v}=0$. The bifurcation surface is at the intersection of the two horizons, at $\tilde{u}=\tilde{v}=0$. The exterior of the black hole in the $(\tilde{u}, \tilde{v})$ coordinates is simply $\tilde{u}<0, \tilde{v}>0$, just like the right Rindler wedge in usual lightcone coordinates in Minkowski space.

\begin{figure}[ht]
    \centering   \includegraphics[width=0.6\textwidth]{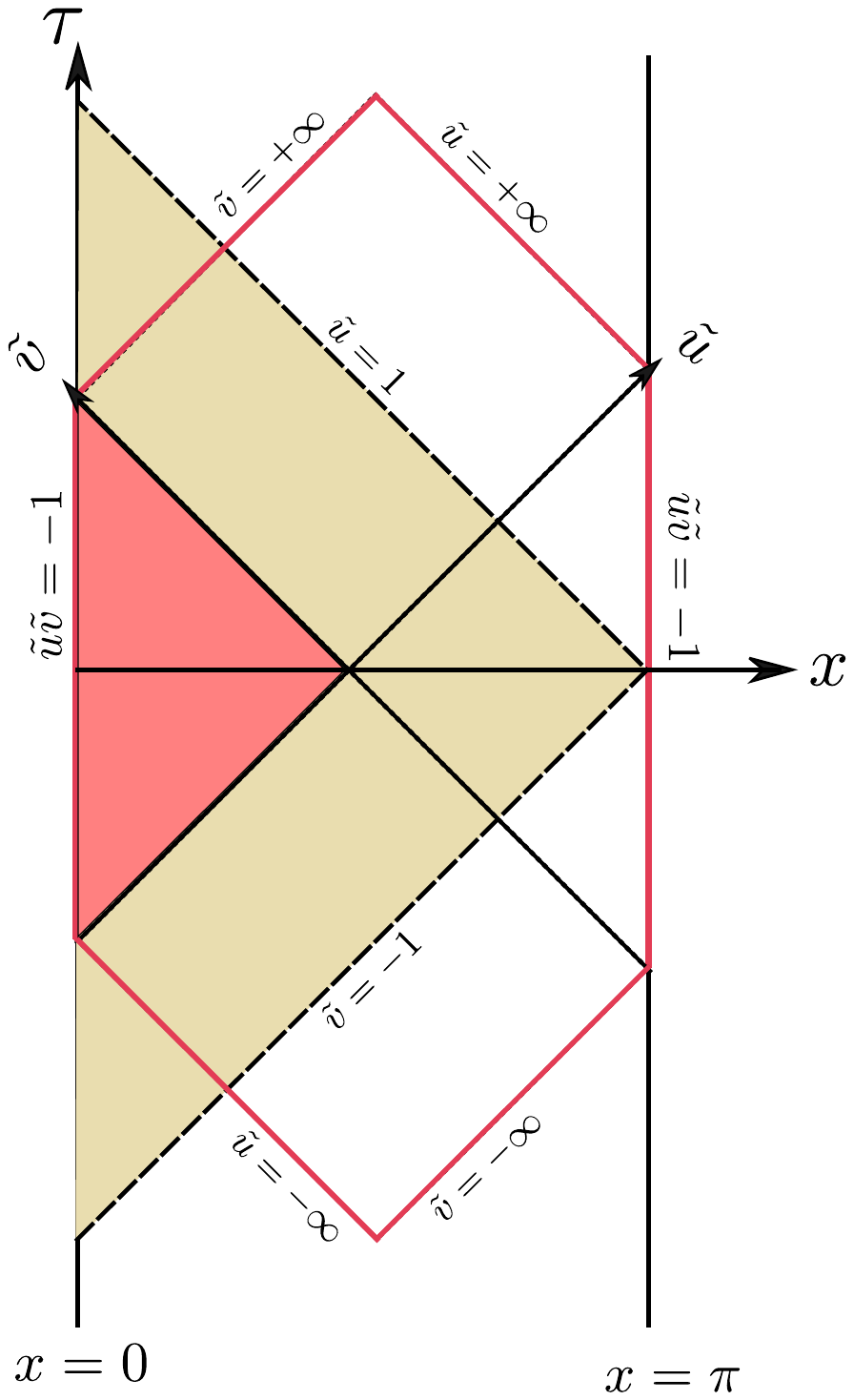}
    \caption{Diagram with a visualization of all the relevant patches of AdS$_2$ being considered. The Poincar\'e patch is in beige, whereas the black hole patch is in red. The Kruskal coordinates $\tilde{u}, \tilde{v}$ cover the region bounded by the six red line segments. The Poincar\'e horizon in Kruskal coordinates is given by the surfaces $\tilde{u}=1$ and $\tilde v = -1$.}
    \label{fig:penrosediagramAdS2}
\end{figure}

The coordinates $\tilde{u}$ and $\tilde{v}$ can of course be extended beyond the horizon, however. In fact, one can show that the relation between $(\tilde{u}, \tilde{v})$ and the global null coordinates $x_\pm$ can be written as 
\begin{align}
    \tan x_+ = \dfrac{\beta}{\pi}\left(\dfrac{1+\tilde u}{1-\tilde u}\right), \qquad
    \tan x_- = \dfrac{\beta}{\pi}\left(\dfrac{\tilde v -1}{\tilde v + 1}\right),
\end{align}
or conversely,
\begin{align}
    \tilde{u} = \dfrac{\pi \tan x_+ - \beta}{\pi \tan x_+ + \beta} \label{eq:KruskalGlobal1}, \qquad     \tilde v = \dfrac{\pi \tan x_- +\beta}{\beta - \pi \tan x_-}. 
\end{align}
This makes it clear that the maximal extension of the $\tilde u, \tilde v$ coordinates in the global patch is the portion of AdS$_2$ between the straight lines $\tan x_{\pm} = - \frac{\beta}{\pi}$, which correspond to $\tilde{u}=\pm\infty$ and $\tilde{v}=\pm\infty$. The pieces of the AdS boundary lying within this region are given by the hyperboloid $\tilde{u}\tilde{v}=-1$. For an easy visualization of the setup above, see Figure~\ref{fig:penrosediagramAdS2}. 

The region covered by the maximal extension of $(\tilde u, \tilde v)$ coincides with the region of AdS$_2$ where the dynamical problem of JT gravity with two causally disconnected boundaries is well-posed~\cite{Harlow:2018tqv}. It is therefore natural to have that as the reference for our treatment of gravitationally dressed observables in the thermofield double state in JT gravity, and we will make use of that later in the paper.

\subsection{Two-point function of massless scalar in AdS\texorpdfstring{$_2$}{2}}
\label{sec:scalarAdS}

For concreteness, the bulk quantum field that we will consider is a minimally coupled massless scalar field, with action given by
\begin{equation}
    S[\phi] = -\dfrac{1}{2}\int d^2 x\sqrt{-g}\nabla^\mu \phi \nabla_\mu \phi.
\end{equation}
The two-point function of a massless scalar field in AdS$_2$ with Dirichlet boundary conditions in the Poincar\'e or global vacuum\footnote{These vacua match, essentially because a Cauchy slice in the Poincar\'e patch is also a Cauchy slice in the global coordinate frame~\cite{Danielsson:1998wt, Spradlin:1999bn}. This is to be contrasted with the black hole patch, which has a different vacuum state.} is~\cite{Spradlin:1999bn}
\begin{equation}\label{eq:wightmanglobalvacuum}
    W(\mathrm{x}, \mathrm{x}') = -\dfrac{1}{4\pi}\log\left(\dfrac{\sin(x_+-x_+')\sin(x_--x_-')}{\sin(x_+-x_-')\sin(x_--x_+')}\right).
\end{equation}
In the Kruskal coordinates $\tilde u, \tilde{v}$ defined by Equation~\eqref{eq:KruskalGlobal1}, this becomes
\begin{equation}\label{eq:WightmanKruskalCoord}
    W(\mathrm x, \mathrm x')= -\dfrac{1}{4\pi}\log\left(-\dfrac{(\tilde u - \tilde u')(\tilde v - \tilde v')}{(1+\tilde u \tilde v')(1 + \tilde v \tilde u')}\right).
\end{equation}
In the left exterior (which is the black hole patch where we originally started), we had $\tilde u = -e^{-2 \pi u/\beta}$ and $\tilde v = e^{2\pi v/\beta}$. Replacing this in the expression above, we get
\begin{equation}\label{eq:wightmanleftexterior}
    W_{\rm LL}(\mathrm x, \mathrm x') =  - \dfrac{1}{4\pi}\log\left(\dfrac{\sinh\left(\dfrac{\pi}{\beta}(u-u')\right)\sinh\left(\dfrac{\pi}{\beta}(v-v')\right)}{\sinh\left(\dfrac{\pi}{\beta}(u-v')\right)\sinh\left(\dfrac{\pi}{\beta}(v-u')\right)}\right),
\end{equation}
where we introduced the subscript $\rm LL$ to indicate that $\mathrm{x}$ and $\mathrm{x}'$ are in the left exterior region of Figure~\ref{fig:penrosediagramAdS2}, in accordance with the convention of Figure~\ref{fig:dress}. The future interior of the two-sided black hole can be accessed if we now parametrize $\tilde u = e^{2\pi u/\beta}$, with the same parametrization of $\tilde v$. If we now keep the point $\mathrm x$ in the left exterior and move the point $\mathrm x'$ to the future interior using the parametrization above, the two-point function becomes
\begin{equation}\label{eq:wightmanonexterioroneinterior}
    W_{\rm LF}(\mathrm x, \mathrm x') =  - \dfrac{1}{4\pi}\log\left(\dfrac{\cosh\left(\dfrac{\pi}{\beta}(u+u')\right)\sinh\left(\dfrac{\pi}{\beta}(v-v')\right)}{\sinh\left(\dfrac{\pi}{\beta}(u-v')\right)\cosh\left(\dfrac{\pi}{\beta}(v+u')\right)}\right),
\end{equation}
where the subscript $F$ indicates that $\mathrm{x}'$ is in the future interior region. By comparing Equations~\eqref{eq:wightmanleftexterior} and~\eqref{eq:wightmanonexterioroneinterior}, we see that the latter can be directly obtained from the former by the substitution
\begin{equation}
    u ' \mapsto -u' \pm i\beta/2. 
\end{equation}
This is clear from the parametrizations of the two coordinate systems as written above. The fact that the analytic continuation of the two-point function on the left exterior region indeed matches the actual two-point function computed in the global vacuum under the parametrization above is yet another check of the thermal nature of the state, as generally expected for the thermofield double. The minus sign above is subtle but important: it signals the fact that the natural notions of time translation on either side of a thermofield double state have opposite orientations. If we want to keep evolving forwards in time in the orientation defined by one of the boundaries, then the time associated to the second boundary must be flipped relative to the direct analytic continuation that gives the two-sided correlation functions.

The point $\mathrm{x}$ can also be moved to the future interior to get the correlation between two operators in the future interior region. As noted above, this can obtained by taking $u\rightarrow -u\pm i\beta/2$ and $u'\rightarrow -u' \pm i\beta/2$ in \eqref{eq:wightmanleftexterior}. The two-point function then reads
\begin{equation}
\label{eq:FFuv}
   W_{\rm FF}(\mathrm{x},\mathrm{x}') = -\frac{1}{4\pi}\log\left(-\frac{\sinh\left(\frac{\pi}{\beta}(u-u')\right)\sinh\left(\frac{\pi}{\beta}(v-v')\right)}{\cosh\left(\frac{\pi}{\beta}(u+v')\right)\cosh\left(\frac{\pi}{\beta}(v+u')\right)}\right).
\end{equation}

Similarly, to get points in the right exterior, we use the same parametrization for $u'$ in the future interior, and also replace $v'$ by $-v' \pm i\beta/2$. If we move both $\mathrm{x}$ and $\mathrm{x}'$ to the right exterior this way, the two-point function becomes
\begin{equation}\label{eq:wightmantwoexteriors}
    W_{\rm RR}(\mathrm x, \mathrm x') =  - \dfrac{1}{4\pi}\log\left(\dfrac{\sinh\left(\dfrac{\pi}{\beta}(u-u')\right)\sinh\left(\dfrac{\pi}{\beta}(v-v')\right)}{\sinh\left(\dfrac{\pi}{\beta}(u-v')\right)\sinh\left(\dfrac{\pi}{\beta}(v-u')\right)}\right).
\end{equation}
This is identical to Equation~\eqref{eq:wightmanleftexterior} -- as expected, since the Hartle-Hawking state (which coincides with the global/Poincar\'e vacuum for AdS$_2$ black holes~\cite{Spradlin:1999bn}) looks the same when restricted to either one of the exterior regions in the two-sided black hole geometry.

\subsection{Infalling Unruh-DeWitt detector}\label{sub:UDW}

To model how the infalling observer will probe physics in the near-horizon region, we make use of an Unruh-DeWitt detector~\cite{Unruh1976, DeWitt:1980hx}, a discrete-level quantum system that couples linearly to the bulk scalar field. The interaction Hamiltonian between detector and field takes the form
\begin{equation}\label{eq:UDWhamiltonian}
    H_{\rm I}(\lambda) = -q\,\chi(\lambda)\mu(\lambda)\,k^{\mu}\nabla_{\mu}\phi(\mathrm{x}(\lambda)),
\end{equation}
where $q$ is a coupling constant, $\lambda$ is a time parameter along the detector's trajectory, $\mathrm{x}(\lambda)$ is the worldline of the detector, $\mu(\lambda)$ is a monopole operator acting on the Hilbert space of the detector, and $k^{\mu}$ is the tangent vector to the detector's trajectory,
\begin{equation}
    k^\mu = \dfrac{d x^\mu}{d\lambda},
\end{equation}
with $x^\mu(\lambda)$ describing the trajectory of the detector in a given coordinate system $\{x^\mu\}$.  This interaction can also be rephrased in terms of the interaction action
\begin{equation}\label{eq:UDWaction}
    S_I = q\int d\lambda\,\chi(\lambda)\mu(\lambda)k^{\mu}\nabla_{\mu}\phi(\mathrm{x}(\lambda)),
\end{equation}
which is just minus the integral of the interaction Hamiltonian~\eqref{eq:UDWhamiltonian} over the detector's time parameter. To account for the fact that the detector only probes the field in a local region of spacetime, we also include a switching function $\chi(\lambda)$, which regulates the strength of the interaction over time and is assumed to be most strongly supported in a finite domain in $\lambda$. The action~\eqref{eq:UDWaction} is precisely of the form briefly described in Equation~\eqref{eq:generalUDW} in Section~\ref{sub:UDWintro}, with the choice \mbox{$O(\lambda)=k^\mu\nabla_\mu\phi(\mathrm{x}(\lambda))$}.

\begin{figure}[t!]
    \centering
    \includegraphics[width=0.4\linewidth]{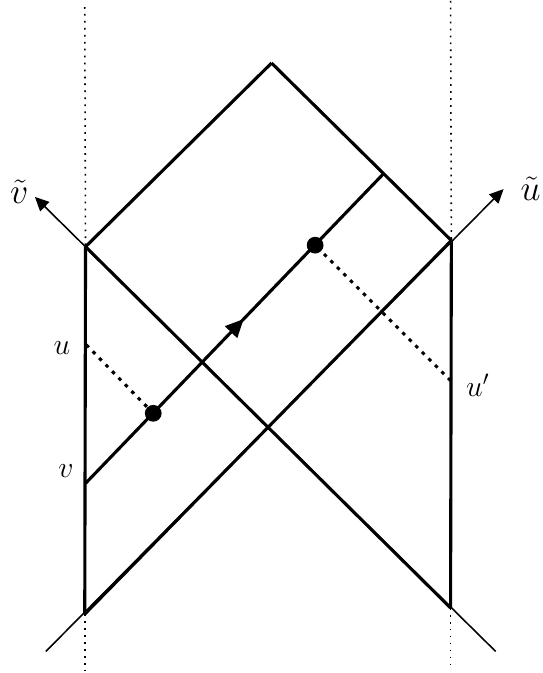}
    \caption{Unruh-DeWitt detector following a null trajectory with fixed $v$. The trajectory can be fully described by two clocks living on the left and right boundaries. The left clock time $v$ fixes the null direction, and boundary times $u$ and $u'$ describe the evolution of the detector in the left exterior and future interior, respectively. In principle, our dressing goes all the way to the top of the diagram, beyond where one would say the classical singularity is present in a higher-dimensional set-up. In practice however, we will turn on our detector along only parts of the trajectory localized around the classical horizon, since our main goal is to deduce horizon physics (which is expected to be universal) and not singularity physics (which is very different in JT gravity and requires UV completion in any case).}
    \label{fig:det}
\end{figure}

In most applications of UDW detectors in the literature, the detector's trajectory $\mathrm{x}(\lambda)$ is timelike, and the parameter $\lambda$ is the detector's proper time. Here, however, we will take $\mathrm{x}(\lambda)$ to be null, and $\lambda$ will then be an affine time for the null geodesic.\footnote{In two spacetime dimensions, every smooth null curve is a geodesic.} Choosing a null trajectory is done partly for convenience, since infalling null geodesics in the black hole patch -- even after coupling directly to the Schwarzian via the gravitational dressing described in Section~\ref{subsec:JTdressing} -- can be easily parametrized by constant values of the past anchoring time $v$ -- see Figure~\ref{fig:det} for an illustration. This will make the technical analysis of the detector's response much more tractable, especially when quantum effects from the Schwarzian dressing are included. From a slightly more physical perspective, one can think of the interaction action~\eqref{eq:UDWaction} as the infinite-boost limit of an infalling timelike observer that couples to the field observable $\frac{d}{d\tau}\phi(\mathrm{x}(\tau))$ where $\tau$ is now the detector's proper time. The motivation to consider an infinitely boosted  timelike trajectory in this case is that we are mainly interested in the physics near the black hole horizon, and the direction of motion of free-falling timelike observers appears to be infinitely boosted relative to static observers in the exterior of the black hole as the horizon is approached. 

The joint state of the detector and the quantum field is generically described by a density matrix $\rho$ acting on the Hilbert space $\mathcal{H}=\mathcal{H}_{\phi}\otimes\mathcal{H}_{\text{d}}$, where $\mathcal{H}_{\phi}$ and $\mathcal{H}_{\text{d}}$ are the Hilbert spaces associated to the field and the detector, respectively. The time evolution of $\rho$ between Cauchy slices labeled by the time $\lambda$ where they intersect the detector's worldline is given in the interaction picture by the von Neumann equation
\begin{equation}
    i\dfrac{d\rho}{d\lambda} = [H_I, \rho],
\end{equation}
with the field operator $\phi(\mathrm{x(\lambda)})$ evolving according to the free dynamics of the QFT, and the monopole operator $\mu(\lambda)$ evolving according to the free internal dynamics of the detector. The evolution of the joint system between times $\lambda_i$ and $\lambda_f$ is then solved by
\begin{equation}
    \rho(\lambda_f) = U(\lambda_f, \lambda_i)\rho(\lambda_i)U^\dagger(\lambda_{f}, \lambda_i),
\end{equation}
where $U$ denotes the time evolution operator in the interaction picture,
\begin{equation}
    U(\lambda_f, \lambda_i) = \mathcal{T}\exp{-i\int_{\lambda_i}^{\lambda_f}H_I(\lambda)d\lambda}.
\end{equation}
In detector calculations, it is customary to take $\lambda_i\to-\infty$ and $\lambda_f\to+\infty$ to schematically model a case where the state of the system is prepared in the far past and then measured again only in the far future. Finite-time effects can be recovered in this setup by suitable choices of the switching function $\chi(\lambda)$ modulating the interaction. 

Now, denoting the initial state of the full system simply by $\rho_0$, the final state by $\rho$, and the time evolution operator by $U$, the reduced state of the detector after coupling to the field is 
\begin{equation}
    \rho_{\text{d}} = \Tr_{\phi}\left(U\rho_0 U^\dagger\right),
\end{equation}
where $\Tr_\phi$ is the partial trace over the Hilbert space associated to the quantum field.

In what follows, we will consider a situation where the field and the detector start in the product state
\begin{equation}
    \rho_0 = \left(\ket{0_P}\bra{0_P}\right)\otimes\left(\ket{g}\bra{g}\right),
\end{equation}
where $\ket{0_P}$ denotes the Poincar\'e vacuum of the massless scalar field in AdS$_2$, and $\ket{g}$ is the ground state of the detector's internal Hamiltonian. Then, the probability that the detector will be found in an excited state $\ket{e}$ after the interaction with the quantum field is simply given by
\begin{equation}\label{eq:generalvacuumexcprob}
    P_{\text{exc}} = \bra{e}\rho_{\text{d}}\ket{e} = \Tr_{\phi}\left(\bra{e}U\rho_0 U^\dagger\ket{e}\right).
\end{equation}
We refer to $P_{\text{exc}}$ as the detector's vacuum excitation probability, as it represents the probability that a detector will transition to an excited state by coupling to the vacuum of the quantum field. 

To compute the vacuum excitation probability in practice, we can use perturbation theory. By taking the Dyson series expansion of the time evolution operator, we can write
\begin{align}
    U &= 1 -i\int d\lambda'H_{\rm I}(\lambda') - \int d\lambda'\int d\lambda''H_{\rm I} (\lambda'')H_{\rm I}(\lambda') + \mathcal{O}(q^3)\nonumber \\
    &\equiv 1 + \underbrace{U^{(1)}}_{\propto \, q} + \underbrace{U^{(2)}}_{\propto \, q^2} + \mathcal{O}(q^3).
\end{align}
The perturbative expansion of the vacuum excitation probability, using the explicit form~\eqref{eq:UDWhamiltonian} for the interaction Hamiltonian, is then given by
\begin{equation}\label{eq:excprobabilitygeneral}
    P_{\text{exc}} = q^2 \int d\lambda \int d\lambda' \chi(\lambda) \chi(\lambda')\bra{g}\mu(\lambda)\ket{e}\bra{e}\mu(\lambda')\ket{g} k^\mu k^\nu \langle \nabla_\mu\phi(\mathrm x(\lambda)) \nabla_\nu\phi(\mathrm{x}(\lambda'))\rangle + \mathcal{O}(q^3),
\end{equation}
where we have denoted the expectation value of field observables in the Poincar\'e vacuum simply as $\langle\cdot\rangle$.

The only missing ingredient here is the time evolution of the matrix element $\bra{g}\mu(\lambda)\ket{e}$, which will generically depend on the detector's internal dynamics. For our purposes, it is natural to take the detector's internal operator $\mu(\lambda)$ to oscillate at a constant frequency $\omega$ in units of affine time. In other words, we will take
\begin{equation}
    \bra{g}\mu(\lambda)\ket{e} = \bra{g}\mu(0)\ket{e}\,e^{-i\omega\lambda}.
\end{equation}
The frequency $\omega$ in this case has the interpretation of the energy gap between the states $\ket{g}$ and $\ket{e}$, sa defined by the free Hamiltonian of the detector. With this, the excitation probability~\eqref{eq:excprobabilitygeneral} becomes 
\begin{equation}\label{eq:excprobgeneral2}
    P_{\text{exc}} = q^2\vert\bra{e}\mu(0)\ket{g}\vert^2 \int d\lambda \int d\lambda' \chi(\lambda) \chi(\lambda')e^{-i\omega(\lambda-\lambda')} k^\mu k^\nu \langle \nabla_\mu\phi(\mathrm x(\lambda)) \nabla_\nu\phi(\mathrm{x}(\lambda'))\rangle + \mathcal{O}(q^3).
\end{equation}
In the limit where $\omega$ is much smaller than the time-scale of the detector-field interaction and when the state of the QFT is Gaussian, it turns out that the leading order term above determines the full nonperturbative excitation probability at all orders in $q$. In this case, we can treat $\mu(\lambda)$ as effectively time-independent throughout the interaction, and the fully nonperturbative expression for the excitation probability becomes $P^{\rm gapless}_{\rm exc} = (1-e^{-2P_{\rm exc}\vert_{\mathcal{O}(q^2)}})/2$. This is often referred to as the ``gapless detector'' in the literature, since it is equivalent to the naive limit where the proper energy gap between what we call the excited and ground states $\ket{g}$ and $\ket{e}$ is taken to $\omega\to 0$. For more details on this particular case of Equation~\eqref{eq:generalvacuumexcprob}, see Appendix~\ref{app:gapless}.

For future reference, it will also be convenient to parametrize the trajectory of the detector by boundary time. Because the time of one single boundary (say, the left one) only covers the trajectory of the infalling detector up to the point where the detector crosses the horizon, we will have to think of the trajectory as a function of the boundary time $u$ when the detector is in the left exterior, and $u'$ when it crosses the horizon and accesses the future interior of the black hole. See Figure \ref{fig:det} once again for reference. The result is that the time evolution operator can be split as
\begin{align}
    U &= \mathcal{T}\exp\left\{-i\int du \, \chi(u')\mu(u')\partial_u\phi_R(u')\right\}\times\mathcal{T}\exp\left\{-i\int du \, \chi(u)\mu(u)\partial_{u}\phi_L(u)\right\} \nonumber \\
    &= (1 + U_R^{(1)} + U_R^{(2)} +\dots)(1 + U_L^{(1)} + U_L^{(2)} +\dots) \nonumber \\
    &= 1 + \underbrace{U_R^{(1)}+U_L^{(1)}}_{\propto \, q}+ \underbrace{U_R^{(1)} U_L^{(1)} + U_R^{(2)} + U_L^{(2)}}_{\propto \, q^2} + \mathcal{O}(q^3).
\end{align}
Then, taking into account the contributions from both left and right boundaries, and changing the integration coordinate from $\lambda$ to $u$ and $u'$, the excitation probability~\eqref{eq:excprobgeneral2} can be written as
\begin{align}\label{eq:exitationprobtwosidedboundary}
    P_{\text{exc}} = q^2\left|\bra{e}\mu(\lambda_i)\ket{g}\right|^2 &\Bigg(\int_{-\infty}^{+\infty}du_1\int_{-\infty}^{+\infty}du_2\,\chi(u_1)\chi(u_2)e^{-i\omega(\lambda(u_1)-\lambda(u_2))}\partial_{u_1}\partial_{u_2}W_{\rm LL}(u_1, u_2) \nonumber \\
    &+ \int_{-\infty}^{+\infty}du\int_{-\infty}^{+\infty}du'\,\chi(u)\tilde{\chi}(u')e^{-i\omega(\lambda(u)-\lambda(u'))}\partial_{u}\partial_{u'}W_{\rm LF}(u, u') \nonumber \\
    &+ \int_{-\infty}^{+\infty}du\int_{-\infty}^{+\infty}du'\,\chi(u')\tilde{\chi}(u)e^{i\omega(\lambda(u)-\lambda(u'))}\partial_{u}\partial_{u'}W_{\rm FL}(u', u) \nonumber \\
    &+ \int_{-\infty}^{+\infty}du_1'\int_{-\infty}^{+\infty}du_2'\,\tilde{\chi}(u_1')\tilde{\chi}(u_2')e^{-i\omega(\lambda(u_1')-\lambda(u_2'))}\partial_{u_1'}\partial_{u_2'}W_{\rm FF}(u_1', u_2')\Bigg).
\end{align}
As denoted, these four contributions come from parts of the excitation probability that are fully in the exterior ($\rm LL$), the future interior $(\rm FF$) or from cross-horizon contributions $(\rm LF$ and $\rm FL$). The functions $\chi(u)$ and $\tilde{\chi}(u')$ correspond to the switching function written in terms of the boundary time on the left and right boundaries respectively, and $\lambda(u)$ denotes the detector's affine time written as a function of boundary time.

In some setups, as we briefly mention in Section~\ref{sec:bdyexp}, it may also be natural to have the matrix element $\bra{g}\mu(\lambda)\ket{e}$ oscillate with a fixed frequency in units of boundary time $u$ (on the left boundary) or $u'$ (on the right boundary). In Appendix~\ref{app:general_monopole}, we introduce a general time dependence of the monopole operator and write the associated generalized expression for $P_{\rm exc}$ which can be relevant in this case.

The vacuum excitation probability will be the figure of merit that we will use in the rest of the paper to evaluate the response of a detector due to its interaction to a quantum field. This is what will allow us to study quantum-gravitational effects from the point of view of a local observer in our model.

\section{Infalling observer: Semiclassical theory}
\label{sec:semiclassical}

In this Section, we compute the excitation probability of an infalling detector coupled to the Poincar\'e vacuum in a fixed AdS$_2$ background. This will be a useful reference point to be compared with the quantum-gravitational effects that we will present later on, and should be seen as the semiclassical limit of the more complete calculation including Schwarzian corrections that we will perform in later sections.

\subsection{Response of an infalling detector in terms of affine time}\label{sub:excprobsemiclassical}

We reviewed the two-point function of a massless scalar field in the Poincar\'e vacuum in Section~\ref{sec:scalarAdS}. To express that two-point function in terms of the affine time along a null geodesic, we note that, if $u$ is an affine parameter for a null geodesic of the metric $d\tilde s^2$ and $\lambda$ is an affine parameter for the same null geodesic\footnote{It is well-known that null geodesics in a given spacetime are also geodesics in any spacetime related to the original one by a Weyl transformation of the metric.} in the Weyl-rescaled metric $ds^2=e^{2\Omega(\mathrm{x})}d\tilde s^2$, where $\Omega(\mathrm x)$ is an arbitrary function, then $\lambda$ and $u$ can be related to each other by
\begin{equation}
    \dfrac{d\lambda}{du}=e^{2\Omega(\mathrm{x})}
\end{equation}
up to a multiplicative constant. Applying this to the AdS$_2$ metric in the black hole patch~\eqref{eq:blackholepatch} written in lightcone coordinates as $ds^2=e^{2\Omega(u, v)}d\tilde s^2$, where $d\tilde{s}^2=-4 du\,dv$, $u=t+\rho$, and $v=t-\rho$, we can pick
\begin{equation}\label{eq:semiclassicalaffinetime}
    \dfrac{d\lambda}{du}=\dfrac{\pi^2}{\beta^2}\dfrac{1}{\sinh^2\left[\frac{\pi}{\beta}(u-v)\right]}.
\end{equation}
We recall that $\kappa=2\pi/\beta$, and that infalling null trajectories are at $v=$ constant. Integrating~\eqref{eq:semiclassicalaffinetime} directly and choosing a reference point where $\lambda=0$ corresponds to the horizon at $u=\infty$, we have
\begin{equation}
    \lambda(u) = \dfrac{\pi}{\beta}\left(1-\coth\left[\frac{\pi}{\beta}(u-v)\right]\right).
\end{equation}
This can also be extended beyond the black hole exterior to also include the interior by instead writing $\lambda$ in terms of the Kruskal coordinates in Equation~\eqref{eq:KruskalGlobal1}. The result gives
\begin{equation}
    \lambda(\tilde{u}) = \kappa\dfrac{ \tilde u \tilde v}{(1+\tilde u \tilde v)},
\end{equation}
or, if we invert the relation above, we find
\begin{equation}\label{eq:kruskaltoaffine}
    \tilde{u}(\lambda) = \dfrac{\lambda}{\tilde v(\kappa-\lambda)}.
\end{equation}
By then replacing this in the vacuum Wightman function as written in Equation~\eqref{eq:WightmanKruskalCoord}, we can directly see that, after pulling back the vacuum two-point function to the null trajectory $\tilde v = $ constant, we have
\begin{equation}\label{eq:derivativetwoptfunctionclassical}
 k^\mu k^\nu \langle \nabla_\mu\phi(\mathrm{x}(\lambda)) \nabla_{\nu}\phi(\mathrm{x}(\lambda'))\rangle \equiv \partial_\lambda \partial_{\lambda '}W(\lambda, \lambda') = -\dfrac{1}{4\pi(\lambda-\lambda')^2}.
\end{equation}
In Equation~\eqref{eq:derivativetwoptfunctionclassical}, we have introduced the more compact notation $\partial_\lambda \partial_{\lambda '}W(\lambda, \lambda')$ for the derivative of the two-point function of the massless scalar field with respect to both of its arguments, evaluated along the detector's trajectory. 
The excitation probability~\eqref{eq:excprobgeneral2} of a probe whose monopole operator oscillates at a constant frequency $\omega$ in units of affine time is then given by
\begin{equation}\label{eq:excitationprobclassicalaffinetime}
    P_{\text{exc}}= -\frac{1}{4\pi}q^2 \abs{\langle e|\mu(0)|g\rangle}^2\int_{-\infty}^{+\infty}d\lambda \int_{-\infty}^{+\infty}d\lambda' \,e^{-i\omega(\lambda-\lambda')}\dfrac{\chi(\lambda)\chi(\lambda')}{(\lambda-\lambda')^2}. 
\end{equation}
The fact that the two-point function~\eqref{eq:derivativetwoptfunctionclassical} depends only on $\lambda-\lambda'$ immediately guarantees that the excitation probability~\eqref{eq:excitationprobclassicalaffinetime} is translationally invariant, in the sense that shifting the switching function from $\chi(\lambda)$ to $\chi(\lambda+c)$ for any constant $c$ does not change $P_{\rm exc}$.  
This is a manifestation of the fact that the horizon is not locally detectable, in accordance with the equivalence principle of classical general relativity. We have reached this conclusion explicitly in this simple model of an AdS$_2$ black hole, but this result should hold quite generally in the near-horizon region of any bifurcate horizon. Since null rays do not feel the conformal factor of the metric, the symmetry transformation is that of null translations $U\to U+c$ of flat 2d Minkowski space. This is symmetry action is not the same as the time translation along a generic worldline, but it does apply close to the black hole horizon where the Rindler approximation is valid and the symmetry group limits to ISO$(1,1)$, and where $\lambda(u) \sim -e^{-\frac{2\pi}{\beta}u}$. As we will see in later sections, however, quantum gravity effects (here manifest in the fluctuations of the Schwarzian wiggles) will break this symmetry, thus breaking the equivalence principle and allowing a local detection of the horizon.

To model the fact that the interaction is most strongly supported in a finite region of spacetime, we the take the switching function to be of Gaussian form,
\begin{equation}\label{eq:gaussianswitching}
\chi(\lambda)=e^{-(\lambda- \lambda_0)^2/2\sigma^2},
\end{equation}
where $\lambda_0$ corresponds to the moment where the coupling is the strongest, and $\sigma$ denotes the characteristic time scale describing the duration of the interaction. The choice of Gaussian switching function is also convenient because it allows for an exact analytical computation of the integral~\eqref{eq:excitationprobclassicalaffinetime}.

As mentioned previously, the translational invariance of the two-point function immediately guarantees that $P_{\rm exc}$ does not depend on $\lambda_0$. Evaluating~\eqref{eq:excitationprobclassicalaffinetime} with the Gaussian switching function~\eqref{eq:gaussianswitching}, we find
\begin{equation}
\label{eq:nearH_SC}
    P_{\text{exc}}= \abs{\langle e|\mu(0)|g\rangle}^2\ \dfrac{q^2}{4} \left(e^{-\sigma^2\omega^2}-\sqrt{\pi}\sigma\omega \erfc(\sigma\omega)\right).
\end{equation}
Equation~\eqref{eq:nearH_SC} provides a closed-form expression for the excitation probability of a detector with energy gap $\omega$ and whose coupling to the field is controlled by a Gaussian switching function of width $\sigma$. 

\paragraph{Large $\sigma$ limit.}
As an interesting limiting case of Equation~\eqref{eq:nearH_SC}, we can consider a situation where the coupling is turned on for arbitrarily long times. This is formally achieved by computing the excitation probability~\eqref{eq:excitationprobclassicalaffinetime} for a family of switching functions of the form $\chi(\lambda)=F(\lambda/\sigma)$, where $F(x)$ is a function that rapidly decays when its argument is larger than $1$, and then taking the limit $\sigma\to\infty$ at the very end of the calculation.\footnote{This strategy for taking the long-time limit of detector calculations is known as \emph{adiabatic switching}. For more applications of adiabatic switching in the context of detector calculations, see e.g.~\cite{Fewster:2015dqb, Fewster:2016ewy}.} The switching~\eqref{eq:gaussianswitching} is already of this form, so we can immediately take the limit of large $\sigma$ (or more precisely, for $\sigma\omega \gg 1$) in Equation~\eqref{eq:nearH_SC} to get
\begin{equation}\label{eq:vacuumproblongtimelimit1}
  P_{\rm exc} = \sigma \abs{\langle e|\mu(0)|g\rangle}^2\dfrac{\sqrt{\pi} q^2}{2}\abs{\omega}\Theta(-\omega).
\end{equation}
The result above is directly proportional to $\sigma$, which encodes the time scale over which the interaction happens. From that, one can naturally define the \emph{transition rate}, given by
\begin{equation}\label{eq:vacuumproblongtimelimit2}
   R_{\rm exc} \coloneqq \lim_{\sigma\to\infty}\dfrac{1}{\sigma} P_{\rm exc} = \abs{\langle e|\mu(0)|g\rangle}^2\dfrac{\sqrt{\pi} q^2}{2}\abs{\omega}\Theta(-\omega).
\end{equation}
The fact that the right-hand side of Equation~\eqref{eq:vacuumproblongtimelimit1} or~\eqref{eq:vacuumproblongtimelimit2} is proportional to $\Theta(-\omega)$ implies that the excitation probability in the $\sigma\to\infty$ limit identically vanishes for $\omega>0$. Therefore, in the limit where the interaction is turned on forever, the probe with the properties above does not get excited from the vacuum. This is very natural: in this limit, the probe can have access to a global Cauchy slice of the spacetime during its lifetime, and should thus be able to conclude that the field is indeed in its ground state. Since any excitation of the detector in the limit of a constant interaction with the field must be accompanied by some energy extraction from the field, we should expect no excitations in the detector if the field starts in the vacuum. This is what general intuition from energy conservation would suggest, and what the vanishing transition rate obtained above confirms.
 
For most of our later purposes, however, it will be more useful to keep in mind the expression for the vacuum excitation probability at finite values of $\sigma$, and refrain from using intuition that is only applicable for the transition rate at very late times. Part of the reason for this is that we are mainly interested in focusing on the response of the detector near the black hole horizon, which is inevitably concentrated on some finite region of spacetime. Another reason is that the limit of very long interaction times will eventually extend the coupling region outside the domain defined in Figure~\ref{fig:penrosediagramAdS2}, because because the detector reaches the region $\tilde u=\infty$ in finite affine time from the horizon. Indeed, Equation~\eqref{eq:kruskaltoaffine} makes it clear that the affine time elapsed between the horizon-crossing moment at $\lambda=0$ and the edge of the domain of Kruskal coordinates is reached at $\lambda_{\text{max}}=\kappa$. We should thus restrict to switching functions whose support is contained in the interval $-\infty<\lambda<\kappa$. Therefore, our detector response will invariably be influenced (at least partially) by finite-time effects.\footnote{The higher-dimensional version of this statement is that it is unphysical to take the infinite-time limit of the response of a particle detector in free fall towards the black hole interior, because the proper time experienced by the detector between crossing the horizon and hitting the singularity is finite~\cite{Shallue:2025zto}.}

\paragraph{Small $\sigma$ limit.}
For later reference, we also consider the alternative limiting case where the interaction is turned-on for a very short amount of affine time, \emph{i.e.} $\sigma\omega\ll 1$. Taking the the leading-order contribution in $\sigma\omega$ in Equation \eqref{eq:nearH_SC}, we find
\begin{equation}
    P_{\rm exc} = \frac{q^2}{4}  + \mathcal{O}(\sigma\omega).
\label{eq:nearH_SC_smallsigma}
\end{equation}
So the transition probability for a detector that is turned on and off quickly around the horizon reaches a constant value as $\sigma\rightarrow 0$, for sufficiently low frequencies ($\sigma\omega \ll 1$).  

\subsection{Excitation probability from the boundary point of view}
\label{sec:bdyexp}

An alternative computation consists of parameterizing the integral with $u$ and $u'$, the left and right boundary times, and then also assume that the detector's monopole operator oscillates at a constant frequency $\omega$ in units of boundary time. In this case, the transition rate becomes
\begin{align}
\label{eq:R(u,v)}
    R_{\rm exc} = q^2\left|\bra{e}\mu(0)\ket{g}\right|^2 &\lim_{T\rightarrow\infty}\frac{1}{T}\Bigg(\int_{-T}^{+T}du_1\int_{-T}^{+T}du_2\,e^{-i\omega(u_1-u_2)}\partial_{u_1}\partial_{u_2}W_{\rm LL}(u_1, u_2) \nonumber \\
    &+ \int_{-T}^{+T}du\int_{-T}^{+T}du'\,2\cos(\omega(u-u'))\partial_{u}\partial_{u'}W_{\rm LF}(u, u') \nonumber  \\
    &+ \int_{-T}^{+T}du_1\int_{-T}^{+T}du_2\,e^{-i\omega(u_1-u_2)}\partial_{u_1}\partial_{u_2}W_{\rm FF}(u_1, u_2)\Bigg),
\end{align}
where we used $W_{\rm LF}(\rm{x},\rm{x}') = W_{\rm FL}(\rm{x}',\rm{x})$ in the semi-classical theory. (See Section~\ref{sec:hermiticity} for a discussion of hermiticity properties in the quantum Schwarzian theory). Here, the switching function is implicitly turned off when crossing the horizon, as we exclude the region $u\in(T,\infty)\cup u'\in(-\infty,-T)$, taking the $T\rightarrow\infty$ limit. This is equivalent to taking a switching function of the form $\chi(u) = \Theta(u+T) - \Theta(u-T)$, where $\Theta$ denotes the Heaviside step function.

\begin{figure}[ht]
    \centering
    \includegraphics[width=0.4\linewidth]{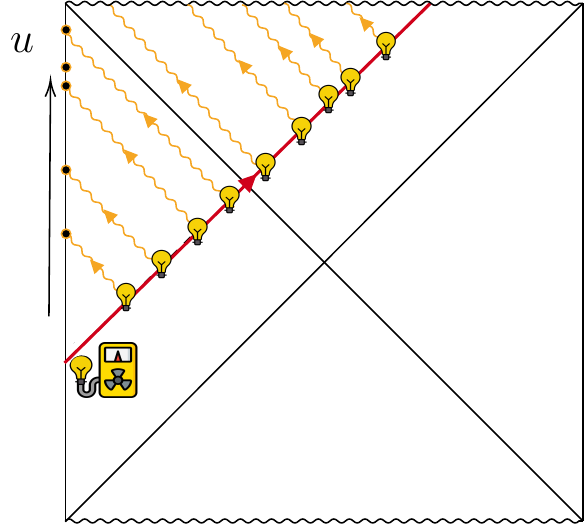}
    \caption{A boundary observer sends a detector into the black hole and designs it such that it emits a signal to the boundary at every detection. After crossing the horizon, no signal reaches the boundary observer. From the light signals received at the boundary, the boundary observer computes the probability to find that the detector has been excited to a state $\omega$.}
    \label{fig:expobs}
\end{figure}

The time used to define the frequency $\omega$ is not the proper time of the trajectory of the detector, making the physical interpretation of the above rate more subtle. The calculation corresponds to the following setup, shown in Figure~\ref{fig:expobs}. A boundary observer constructs a Geiger-type of detector which is designed to emit a light signal every time it detects radiation in the boundary observer's frame. The boundary observer then waits for the light signals sent by his detector. The response rate defined above computes the probability for the boundary observer to conclude that his detector is in a given state.

We compute the response function replacing the correlation function with its semiclassical expression. We find
\begin{align}
    &\lim_{T\rightarrow\infty}\int_{-T}^{+T}du_1\,e^{-i\omega(u_1-u_2)}\partial_{u_1}   \partial_{u_2}W_{\rm LL}(u_1, u_2) \\
    &=\lim_{T\rightarrow\infty}\int_{-T}^{+T}du_1\,e^{-i\omega(u_1-u_2)}\partial_{u_1}   \partial_{u_2}W_{\rm FF}(u_1, u_2) \\
    &= -\frac{\pi}{4\beta^2}\int_{-\infty}^{+\infty} dx \frac{e^{-i\omega x}}{\sinh^2\left(\frac{\pi}{\beta}(x-i\epsilon)\right)} = \frac{\omega}{2}\frac{1}{e^{\omega\beta}-1}, \label{eq:LLresponse}
\end{align}
and
\begin{align}
    &\lim_{T\rightarrow\infty}2\int_{-T}^{+T}du\,\cos(\omega(u-u'))\partial_{u}   \partial_{u'}W_{\rm LF}(u, u') \\
    &= -\frac{\pi}{2\beta^2}\Re\left[\int_{-\infty}^{+\infty} dx \frac{e^{i\omega x}}{\cosh^2\left(\frac{\pi}{\beta}(x+2u')\right)} \right]\\
    &= -\omega\frac{\cos(2\omega u')e^{\beta\omega/2}}{e^{\omega\beta}-1}. \label{eq:LFresponse}
\end{align}
The $LF$-term does not contribute to $R(\omega)$. Taking $\omega\neq 0$, $\lim_{T\rightarrow\infty}\frac{1}{T}\int_{-T}^T du' \cos(2\omega u') = 0$. Then,
\begin{equation}
\label{eq:obsR}
    R_{\rm exc}=q^2\left|\bra{e}\mu(0)\ket{g}\right|^2\frac{2\omega}{e^{\beta\omega}-1}.
\end{equation}
The observer measures a thermal response of the measurement device. This is to be expected, as the setup effectively decouples the exterior and interior modes, as shown in Figure~\ref{fig:expobs}. 

\section{Gravitationally dressed two-point function in JT gravity}
\label{eq:dressedW}

Having introduced the vacuum two-point function of a massless scalar in AdS$_2$, we now dress these operators to the boundary curve $F(t)$ of JT gravity. Following~\cite{Blommaert2019,Mertens:2019bvy, Blommaert:2020yeo, DeVuyst:2022bua, Mertens:2025rpa}, we dress a bulk point $(U,V)$ to the boundary through reparametrization
\begin{equation}
    U=F(u), \quad V=F(v),
\end{equation}
as introduced in Equation~\eqref{eq:dressing} and shown in Figure~\ref{fig:dressout}. Applied to the two-point function \eqref{eq:wightmanleftexterior}, the idea is to express the two-point function of the conformal scalar in AdS$_2$ as a double integral of a Schwarzian bilocal,
\begin{equation}\label{eq:WightmanSchwarzianBilocal}
    -\dfrac{1}{4\pi} \log\left(\dfrac{(F(u) - F(u'))(F(v) - F(v'))}{(F(u) - F(v'))(F(v) - F(u'))}\right) = -\dfrac{1}{4\pi}\int_{v}^{u}dt\,\int_{v'}^{u'} dt' \dfrac{\dot{F}(t)\dot{F}(t')}{(F(t) - F(t'))^2}.
\end{equation}
and then proceed by treating this as an operator insertion in the Schwarzian theory. The quantity $\dot{F}(t)\dot{F}(t')/(F(t) - F(t'))^2$ is precisely the Schwarzian bilocal with weight $\Delta = 1$, whose expectation value in a thermal state is known exactly. In what follows, we will show how this allows us to explicitly calculate both the two-point function in the exterior of the black hole, as well as between points sitting our opposite sides of the black hole horizon.

\subsection{Two-point function in the exterior of the black hole}

 To express the scalar two-point function in the exterior of the black hole in terms of the Schwarzian degree of freedom, we take~\eqref{eq:WightmanSchwarzianBilocal} and write the Schwarzian mode $F(t)$ in the thermal parametrization $F(t) = \tanh(\frac{\pi}{\beta}f_L(t))$ in Lorentzian signature. $f_L(t)$ here parametrizes the left boundary curve in the Schwarzian theory. The result is
\begin{align}\label{eq:doubleintegralschwarzianbilocal}
    W_{\rm LL}(\mathrm{x}, \mathrm x') &= -\dfrac{1}{4\pi}\log\left(\dfrac{\sinh\left(\dfrac{\pi}{\beta}(f_L(u) - f_L(u'))\right)\sinh\left(\dfrac{\pi}{\beta}(f_L(v) - f_L(v'))\right)}{\sinh\left(\dfrac{\pi}{\beta}(f_L(u) - f_L(v'))\right)\sinh\left(\dfrac{\pi}{\beta}(f_L(v) - f_L(u'))\right)}\right) \nonumber \\
    &=-\dfrac{\pi}{4\beta^2}\int_{v}^{u}dt\,\int_{v'}^{u'} dt' \dfrac{\dot{f}_L(t)\dot{f}_L(t')}{\sinh^2\left(\frac{\pi}{\beta}(f_L(t) - f_L(t'))\right)}.
\end{align}
We now evaluate this in the Schwarzian theory in the thermofield double state at inverse temperature $\beta$, obtaining
\begin{equation}\label{eq:dressedWightmanLL}
    W(\mathrm x, \mathrm x') = \dfrac{1}{4\pi}\int_v^u dt \int_{v'}^{u'}dt' \expval{\mathcal{O}_1(it, it')}_\beta,
\end{equation}
where 
\begin{equation}
    \expval{\mathcal{O}_1(z, z')}_{\beta} = \expval{\dfrac{\pi^2}{\beta^2}\dfrac{f'(z)f'(z')}{\sin^2\left[\dfrac{\pi}{\beta}(f(z) - f(z'))\right]}}_\beta
\end{equation}
is the Schwarzian bilocal in the thermal reparametrization in Euclidean signature. The relation between the Schwarzian in Euclidean signature $f(it)$ and that in Lorentzian signature $f_L(t)$ is 
\begin{equation}\label{eq:SchwarzianWickRotation}
    f_L(t) = -i f(it).
\end{equation}
At this point, we simply quote the known result for the thermal expectation value of the Schwarzian bilocal~\cite{Mertens:2017mtv,Bagrets:2016cdf,Yang:2018gdb,Blommaert:2018oro,Iliesiu:2019xuh,Mertens:2022irh}:
\begin{equation}
\label{eq:Sch_bilocal}
    \expval{\mathcal{O}_1(z_1, z_2)}_\beta = \dfrac{e^{S_0}}{Z(\beta)}\int_0^\infty d\mu(k_1)\, d\mu(k_2) e^{-z\frac{k_1^2}{2C}}e^{-(\beta-z)\frac{k_2^2}{2C}} \dfrac{\Gamma(1 \pm ik_1 \pm ik_2)}{8\pi^4 (2C)^2 },
\end{equation}
where the measure in the label $k$ is $d\mu(k)\equiv 2k\,\sinh(2\pi k)\,dk$, and the Schwarzian partition function $Z(\beta)$ is given by
\begin{equation}
    Z(\beta) = \dfrac{1}{4\pi^2}\left(\dfrac{2\pi C}{\beta}\right)^{3/2}e^{S_0 + \frac{2\pi^2 C}{\beta}} \equiv e^{S_0}Z_0(\beta).
\end{equation}
In Equation~\eqref{eq:Sch_bilocal}, $z \equiv z_1 - z_2$, and we assume that $z$ has a positive real part in order for the integrals in $k$ to converge. This is precisely compatible with the regularization of the Wightman function in Lorentzian signature performed by adding a small \emph{negative} imaginary part to the time difference $t-t'$ in real time. 

To obtain the real-time correlation functions from the general expression~\eqref{eq:Sch_bilocal}, we will think of the fundamental Schwarzian $F$ being written as $F(z) = \tan\left(\frac{\pi}{\beta}f(z)\right)$ as a function of a complex parameter $z$, and $z = \tau + it$ is complexified time ($\tau$ here denoting Euclidean time, and $t$ is real time). Plugging this back in Equation~\eqref{eq:doubleintegralschwarzianbilocal} and evaluating the integrals in $t, t'$ gives us
\begin{align}\label{eq:WightmanonesidedSchwarzian}
    W_{\rm LL}(\mathrm x, \mathrm x') =& \dfrac{4}{Z_0(\beta)}\int_0^\infty dM\,  \rho(M)e^{-\beta M}\int_0^{\infty}d E\,\rho(E)\abs{\mathcal{O}^1_{ME}}^2  \nonumber \\
    &\hspace{-1cm}\times\dfrac{e^{-i(E-M)(u+v-u'-v')/2}}{(E-M)^2}\sin\left(\frac{(E-M)(u-v)}{2}\right)\sin\left(\frac{(E-M)(u'-v')}{2}\right).
\end{align}
To write the expression above, we made the change of variables
\begin{equation}
    \dfrac{k_1^2}{2C} \equiv E, \,\,\, \dfrac{k_2^2}{2C} \equiv M,
\end{equation}
and also adopted the shorthand notation\footnote{The insertion of the powers of $e^{S_0}$ above is mostly useful when discussing higher-topology effects on the matter correlators. The effects of higher topology will not play a role in this paper, but we chose to keep the conventions in the definitions above for consistency with related literature.}
\begin{equation}
    \rho(E) = e^{S_0}\dfrac{C}{2\pi^2}\sinh\left(2\pi\sqrt{2CE}\right) \equiv e^{S_0} \rho_0(E), \,\,\,|\mathcal{O}^{1}_{ME}|^2 \equiv 2e^{-2S_0}\dfrac{\Gamma(1\pm i\sqrt{2CM}\pm i\sqrt{2CE})}{(2C)^2}.
\end{equation}
We are denoting the function above with the subscript $W_{\rm LL}$ to emphasize that both points are anchored to the left boundary. This will be an important distinction due to the two-sided dressing that we will discuss in the next section.

\subsection{Two-point function across the horizon}

\begin{figure}[ht]
\centering \includegraphics[width=0.4\textwidth]{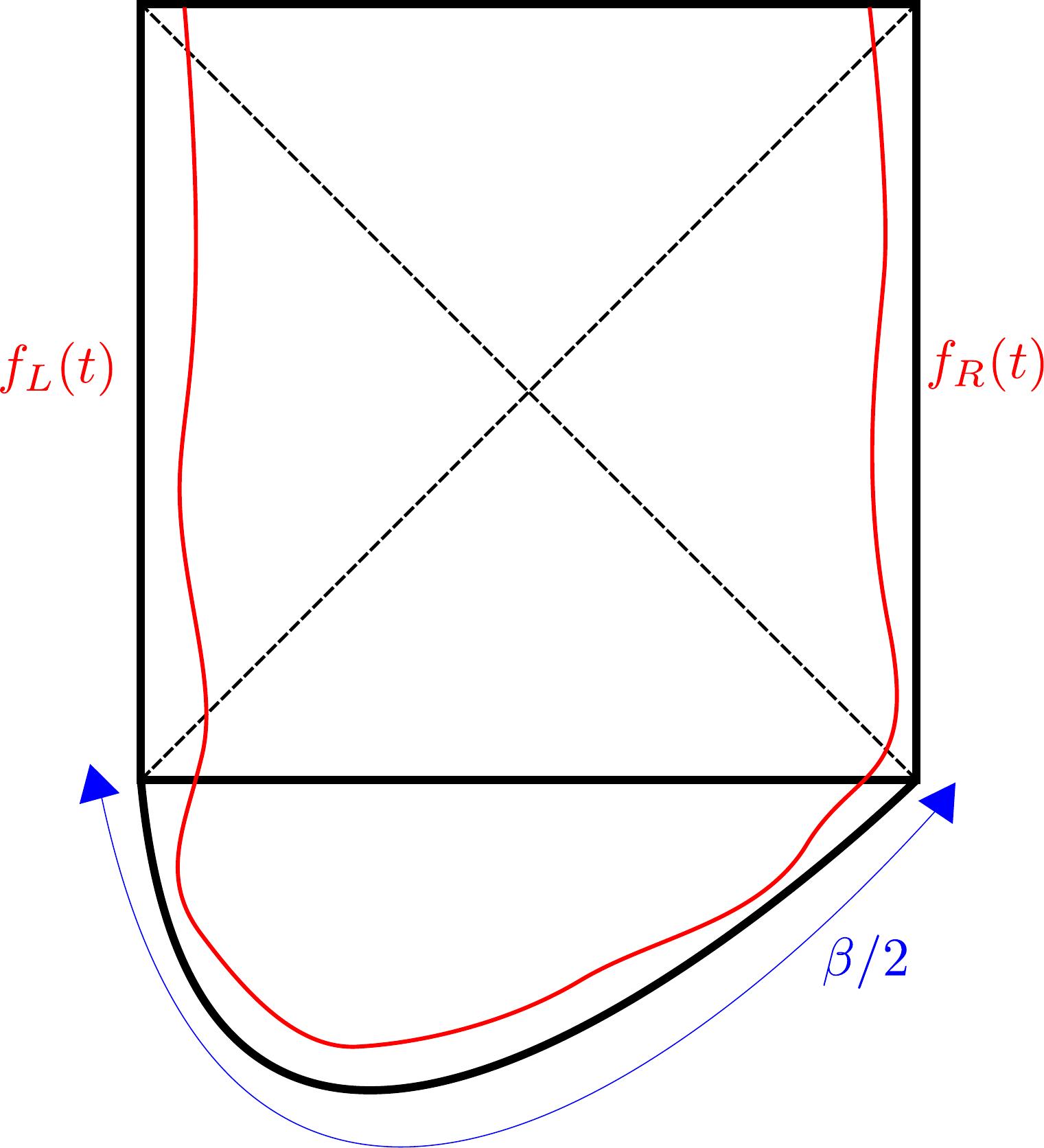}
    \caption{Wiggly boundary curve on both sides of the thermofield double, as analytically continued time reparametrizations of a single $f(t)$. Note that the two-sided extension done in~\eqref{eq:fLR_to_f} and~\eqref{eq:fLR_to_f2} includes both an analytic continuation in the boundary time and a field redefinition to connect the Schwarzians corresponding to each boundary.}
    \label{Fig2bdySchw}
\end{figure}

Consider now the two-point function with one point in the left exterior region and another point in the future interior, Equation~\eqref{eq:wightmanonexterioroneinterior}. We expect to have one Schwarzian for each boundary, with the point $\mathrm{x}'$ depending on a dressing that shoots lightrays to both boundaries. To proceed as before, we first rewrite it in a way that is suggestive of the thermal reparametrization of the Schwarzian mode, with $u\to f_L(u)$, $v\to f_L(v)$, $u' \to f_R(u')$, and $v' \to f_L(v')$. Quoting Equation~\eqref{eq:wightmanonexterioroneinterior}, this gives us
\begin{equation}
    W(\mathrm{x}, \mathrm x') = -\dfrac{1}{4\pi}\log\left(\dfrac{\cosh\left(\dfrac{\pi}{\beta}(f_L(u) + f_R(u'))\right)\sinh\left(\dfrac{\pi}{\beta}(f_L(v) - f_L(v'))\right)}{\sinh\left(\dfrac{\pi}{\beta}(f_L(u) - f_L(v'))\right)\cosh\left(\dfrac{\pi}{\beta}(f_L(v) + f_R(u'))\right)}\right).
\end{equation}
Superficially, this might not look exactly like the kind of expression that would lead to the integral of the Schwarzian bilocal found in~\eqref{eq:doubleintegralschwarzianbilocal}. It is, however, exactly that, if we perform the identification
\begin{align}
\label{eq:fLR_to_f}
    f_L(t) &= -i f(it), \\
    f_R(t') &= i f(-it' 
    -\beta/2) +i\dfrac{\beta}{2}, \label{eq:fLR_to_f2}
\end{align}
where $f(z)$ here is the same as what appeared around Equation~\eqref{eq:SchwarzianWickRotation}. We propose that this should be treated as the key definition of the two-sided expectation values of the bulk observables dressed by the Schwarzian mode in the thermofield double state (Figure~\ref{Fig2bdySchw}). 

On a more pragmatic level, if we now repeat the same steps as in the previous section, we arrive at a result that is simply the analytic continuation of Equation~\eqref{eq:WightmanonesidedSchwarzian} with the replacement of $u'$ by $-u' + i\beta/2$. The result is
\begin{align}\label{eq:WightmanTwosidedSchwarzian}
    W_{\rm LF}(\mathrm x, \mathrm x') =& \dfrac{4}{Z_0(\beta)}\int_0^\infty dM  \rho(M)e^{-\beta M}\int_0^{\infty}d E\,\rho(E) \nonumber \\
    &\times\abs{\mathcal{O}^1_{ME}}^2 e^{-\beta(E-M)/4}\dfrac{e^{-i(E-M)(u+v+u'-v')/2}}{(E-M)^2}\sin\left(\frac{(E-M)(u-v)}{2}\right) \nonumber \\
    &\times\Bigg(-\sin\left(\frac{(E-M)(u'+v')}{2}\right) \cosh\left(\dfrac{\beta(E-M)}{4}\right) \nonumber \\ 
    &+ i \cos\left(\frac{(E-M)(u'+v')}{2}\right)\sinh\left(\dfrac{\beta(E-M)}{4}\right)\Bigg)
\end{align}
The sign in front of $i\beta/2$ in the analytic continuation is chosen to ensure that the $E$ integral is convergent, which requires the real time diference $u-u'$ to have negative real part.

\subsection{Two-point function behind the horizon}

We now consider the two-point function with both points in the future interior (Equation \eqref{eq:FFuv}). Each point is dressed to the left and right boundaries through two lightrays that are shot towards the left and right boundaries. Similarly to the previous computation, we reparameterize the Schwarzian modes with
\begin{align}
    u \rightarrow f_R(u),&~ v\rightarrow f_L(v)\\
    u' \rightarrow f_R(u'),&~ v'\rightarrow f_L(v'),
\end{align}
and perform the identification of Equation \eqref{eq:fLR_to_f}. For a shorter route, we plug $u\rightarrow -u - i\beta/2$ and $u'\rightarrow -u' + i\beta/2$ in Equation \eqref{eq:WightmanonesidedSchwarzian}, and find
\begin{equation}
\begin{aligned}
W_{\rm FF}(\mathrm{x}, \mathrm{x}') &= 
\dfrac{4}{Z_0(\beta)}
\int_{0}^{\infty} \mathrm{d}M \, \rho(M) e^{-\beta M}
\int_{0}^{\infty} \mathrm{d}E \, \rho(E)\abs{\mathcal{O}^1_{ME}}^2 \\
&\quad\times\dfrac{e^{-i(E-M)(-u+v+u'-v')/2}}{(E-M)^2}
\left[
- \sin\!\left( \frac{(E - M)(u + v)}{2} \right)
\cosh\!\left( \frac{\beta(E - M)}{4} \right) \right.\\
&
\left. + i \cos\!\left( \frac{(E - M)(u + v)}{2} \right)
\sinh\!\left( \frac{\beta(E - M)}{4} \right)
\right] \\
&\quad \times 
\left[
- \sin\!\left( \frac{(E - M)(u' + v')}{2} \right)
\cosh\!\left( \frac{\beta(E - M)}{4} \right) \right.\\
&
\left. + i \cos\!\left( \frac{(E - M)(u' + v')}{2} \right)
\sinh\!\left( \frac{\beta(E - M)}{4} \right)
\right].
\end{aligned}
\end{equation}
Again, the sign in the analytic continuation is imposed by demanding that the integral is convergent. Alternatively, one could use $u\rightarrow -u + i\beta/2$ and $u'\rightarrow -u' + i\beta/2$. Blindly making these substitutions at the level of Equation~\eqref{eq:WightmanonesidedSchwarzian} would seemingly lead to a different result; one must remember, however, that the expectation value of the Schwarzian bilocal is only given by Equation~\eqref{eq:Sch_bilocal} in the strip $0<\text{Re}(z)<\beta$, and one then extends this to regions outside the strip by demanding $\beta$-periodicity in Euclidean time. 
 
\subsection{Hermiticity of the extended two-point function}
\label{sec:hermiticity}

One feature of the two-point function of a massless scalar field in the Hartle-Hawking state in AdS$_2$, as reviewed in Section~\ref{sec:scalarAdS}, is that one can obtain the two-point function across the black hole horizon by starting in the exterior region and then adding an imaginary piece to the boundary anchoring time $u$. The expressions obtained in Section~\ref{sec:scalarAdS} made it clear that the  only important piece was the magnitude of the imaginary part being equal to $\beta/2$, but the overall sign did not matter. This turns out not to be the case for the gravitationally dressed two-point function beyond the semiclassical limit of the Schwarzian, however; as we have just seen, the requirement that the integral representation of the gravitationally dressed two-point function converge uniquely selects which sign should be taken in the analytic continuation. In this subsection, we point out that this choice is also consistent with another physically reasonably criterion that one should expect from the dressing -- namely, that the gravitationally dressed two-point function is consistent with the dressed bulk operator being a Hermitian observable in the quantum gravity theory.

Let $A(t)$ be a Hermitian operator, $A^{\dagger}(t)=A(t)$. Then its two-point function on any quantum state must satisfy $\langle{A(t)A(t')\rangle}^*=\langle{A(t')A(t)\rangle}$. The bilocal Schwarzian operator satisfies this property, as can be easily checked from
\begin{equation}
\label{eq:Sch_bilocal2}
    \expval{\mathcal{O}_1(it_1, it_2)}_\beta = \dfrac{1}{Z_0(\beta)}\int_0^\infty d\mu(k_1)\, d\mu(k_2) e^{-it\frac{k_1^2-k_2^2}{2C}}e^{-\beta\frac{k_2^2}{2C}} \dfrac{\Gamma(1 \pm ik_1 \pm ik_2)}{8\pi^4 (2C)^2 },
\end{equation}
where $t=t_1-t_2$. Using this, it is trivial to show that the one-sided correlator \eqref{eq:dressedWightmanLL} satisfies the same property,
\begin{equation}
    W_{\rm LL}(\mathrm{x},\mathrm{x}')^* =W_{\rm LL}(\mathrm{x}',\mathrm{x}).
\end{equation}
This equation can be understood as the statement that the analytic continuation of the Schwarzian bilocal produces two different real-time answers depending on whether we approach the real time axis from above or below (See Figure~\ref{fig:contour}).

However, we want to check that this condition remains valid in our extended definition of the correlator, that includes cases where $\mathrm{x}$ and $\mathrm{x'}$ are on opposite sides of the horizon. Specifically, $ W_{\rm LF}(\mathrm x, \mathrm x')^* = W_{\rm FL}(\mathrm x', \mathrm x) $ and $W_{\rm FF}(\mathrm{x},\mathrm{x}')^* =W_{\rm FF}(\mathrm{x}',\mathrm{x})$. Previously, asking for the convergence of the analytically continued integral led us to the following definitions:
\begin{align}
     W_{\rm LF}(\mathrm x, \mathrm x') &= \dfrac{1}{4\pi}\int_v^u dt \int_{v'}^{-u'+i\beta/2} dt' \expval{\mathcal{O}_1(it, it')}_\beta,\\
     W_{\rm FL}(\mathrm x, \mathrm x') &= \dfrac{1}{4\pi}\int_v^{-u-i\beta/2} dt \int_{v'}^{u'} dt' \expval{\mathcal{O}_1(it, it')}_\beta,\\
     W_{\rm FF}(\mathrm x, \mathrm x') &= \dfrac{1}{4\pi}\int_v^{-u-i\beta/2} dt \int_{v'}^{-u'+i\beta/2} dt' \expval{\mathcal{O}_1(it, it')}_\beta.
\end{align}
Taking the complex conjugate of this expression yields
\begin{equation}
     W_{\rm LF}(\mathrm x, \mathrm x')^* = \dfrac{1}{4\pi}\int_v^u dt \int_{v'}^{-u'-i\beta/2} dt' \expval{\mathcal{O}_1(it', it)}_\beta =  W_{\rm FL}(\mathrm x', \mathrm x),
\end{equation}
and
\begin{equation}
     W_{\rm FF}(\mathrm x, \mathrm x')^* = \dfrac{1}{4\pi}\int_v^{-u+i\beta/2} dt \int_{v'}^{-u'-i\beta/2} dt' \expval{\mathcal{O}_1(it', it)}_\beta =  W_{\rm FF}(\mathrm x', \mathrm x).
\end{equation}
This is ultimately what justifies our different choice of sign for the imaginary part of the boundary anchoring time $u$ depending on whether we are taking the first or second argument of the gravitationally dressed two-point function behind the horizon.

\section{Infalling observer: Schwarzian theory}
\label{sec:Schw}

We now revisit the computations of the excitation probability from Section~\ref{sec:semiclassical}, and compute their value in the Schwarzian theory using the gravitationally dressed two-point function introduced in Section~\ref{eq:dressedW}. The Schwarzian dressing means that the two-point function should be seen as an operator in the Schwarzian theory. Deviations from the semiclassical results are expected to appear close to the horizon, \emph{i.e.} at late boundary times $u \sim C$, where fluctuations in the Schwarzian become strong. Specifically, the bilocal operator~\eqref{eq:Sch_bilocal} goes from its semiclassical value to a power law when the boundary time separation approaches $u-v\sim C$. We therefore expect the detector to effectively see the horizon before crossing it, as the gravitational dressing introduces strong quantum effects near the horizon. 

For conciseness, in the following, we drop the detector matrix element $|\bra{e}\mu(0)\ket{g}|^2$.

\subsection{Response of a detector from the boundary point of view}\label{sec:transitionrateboundaryschwarzian}

As a warm-up, we consider the simple experiment introduced in Section~\ref{sec:bdyexp}, where the frequency of the detector is fixed in boundary time instead of affine time. Practically, we simply replace $W$ by its exact expression in the Schwarzian theory (See Section \ref{eq:dressedW}) in equation \eqref{eq:R(u,v)}. Taking again the same long-time limit applied there and writing $R_{\rm exc} = R_{\rm LL} + R_{\rm LF} + R_{\rm FL} + R_{\rm FF}$, we find:
\begin{align}
\begin{split}
    R_{\rm LL}(\omega) &= R_{\rm FF}(\omega)= q^2\frac{4\pi}{Z_0(\beta)} \int_0^{\infty}dM e^{-\beta M}\rho(M)\rho(M-\omega)|\mathcal{O}^1_{M,M-\omega}|^2 ,\\
    R_{\rm LF}(\omega)&= R_{\rm FL}(\omega) = 0.
\end{split}
\end{align}
Therefore,
\begin{align}
    R(\omega) &= q^2\frac{8\pi}{Z_0(\beta)} \int_0^{\infty}dM e^{-\beta M}\rho(M)\rho(M-\omega)|\mathcal{O}^1_{M,M-\omega}|^2 .
\end{align}
In the semiclassical regime $\beta \ll C$ and taking the probe limit $C/\beta \gg \beta \omega$, we reproduce the previous result~\eqref{eq:obsR}. The ``probe limit'' mentioned here corresponds to the regime the detector's proper frequency $\omega$ is much smaller than the mass of the classical black hole at inverse temperature $\beta$, given by $M=2\pi^2 C/\beta^2$. The result above has similar structure to the excitation rate obtained in~\cite{Blommaert:2020yeo} for a static detector.\footnote{Compared to the expression of~\cite{Blommaert:2020yeo}, there is no greybody factor (or interference factor) in the response function above, as the detector follows an infalling trajectory instead of an accelerated trajectory at fixed distance from the horizon. }

\subsection{Fluctuations of affine time near the horizon}
\label{s:affinetime}
We now present the more intricate calculation where the detector's frequency is fixed in units of affine time. Beyond the semiclassical treatment of the gravitational dressing, the relation between the affine time in the bulk and the boundary time depends on the Schwarzian profile. For each off-shell configuration of the Schwarzian $f(t)$ in the black hole patch, the metric is
\begin{equation}
\label{eq:metricSch}
    ds^2 = -\dfrac{4 f'(u)f'(v) }{\frac{\beta^2}{\pi^2}\sinh^2\left[\frac{\pi}{\beta}(f(u)-f(v))\right]}du\,dv = \mathcal{O}_1(u, v)\,d\tilde s^2,
\end{equation}
where $\mathcal{O}_1(u, v)$ is precisely the thermal Schwarzian bilocal, and $d\tilde s^2 = -4 du\, dv$. From this, we infer that the new affine time $\lambda$ is related to the boundary time $u$ by
\begin{equation}
\label{eq:lambdaSch}
    \dfrac{d\hat{\lambda}}{du}= \mathcal{O}_1(u, v), 
\end{equation}
and therefore the ellapsed affine time between two boundary times $u$ and $u'$ is
\begin{equation}\label{eq:affinetimedifferenceop}
    \hat{\lambda}-\hat{\lambda}' = \int_{u'}^uds\,\mathcal{O}_1(is, iv). 
\end{equation}
The normalization of the affine time chosen above is designed to precisely match the choice made earlier in Equation~\eqref{eq:semiclassicalaffinetime} in the semiclassical limit of the Schwarzian theory. The hat is used to emphasize that $\hat{\lambda}$ is an operator in the quantum gravity theory.

In terms of the boundary time $u$, the horizon is located at $u\to\infty$. In that limit (or more precisely, in the regime $u-v\gg C$), the expectation value of the thermal Schwarzian bilocal behaves like $(u-v)^{-3}$. Therefore,  the (expectation value of) affine time is related to boundary time via
\begin{equation}\label{eq:affinetimeLeft}
    \lambda(u)\approx -\dfrac{A}{(u-v)^2}, \qquad A\equiv \frac{C}{8\pi Z_0(\beta)},
\end{equation}
where we have chosen a reference point such that $\lambda=0$ corresponds to the horizon at $u\to\infty$ once again.\footnote{The normalization of the affine time above is fixed such that, in the semiclassical regime, the relation between affine time and boundary time matches the semiclassical definition~\eqref{eq:semiclassicalaffinetime}. This means that the multiplicative factor of $A$ must be kept in all further calculations for an honest comparison with the semiclassical results.} We will henceforth use $\lambda$ without the hat to denote the expectation value of affine time as a function of boundary time in the Schwarzian theory, in contrast to the affine time operator as defined in Equation~\eqref{eq:affinetimedifferenceop}. Inverting Equation~\eqref{eq:affinetimeLeft}, we find
\begin{equation}
    u(\lambda)\approx v+\sqrt{\dfrac{A}{\abs{\lambda}}}.
\end{equation}

In the black hole interior, the dressing of the bulk point includes a past-directed null ray intersecting the right boundary at boundary time $u'$. Using the proposed analytic continuation $u\to-u+i\beta/2$ to go from the exterior to the interior of the black hole, we find that the relation between affine time and right boundary time in the vicinity of the horizon is
\begin{equation}\label{eq:affinetimeInterior}
    \lambda(u')\approx \dfrac{A}{(u'+v)^2}.
\end{equation}
Inverting this relation and recalling that $u'$ is large and negative in the near-horizon region, we have
\begin{equation}
    u'(\lambda) \approx -v-\sqrt{\dfrac{A}{\lambda}}.
\end{equation}

To check the fundamental limitations in the resolution of affine time defined in Equation~\eqref{eq:lambdaSch}, it is instructive to compute the variance of $\lambda(u)$ as an operator in the Schwarzian theory. From Equation~\eqref{eq:lambdaSch} and setting $\lambda=0$ at the horizon in $u\to\infty$, we have the general expression (valid for points dressed to the left boundary)
\begin{equation}
    \hat{\lambda}(u) = -\int_u^\infty \mathcal{O}_1(is, iv)\,ds.
\end{equation}
Its square is therefore given by
\begin{equation}
    \hat{\lambda}^2 = \int_{u}^\infty ds\int_{u}^\infty ds'\,\mathcal{O}_1(is, iv)\mathcal{O}_1(is', iv),
\end{equation}
and its expectation value then depends on a correlation function involving two Schwarzian bilocals,
\begin{equation}\label{eq:affinetimevariance}
    \langle \hat{\lambda}^2\rangle = \int_{u}^\infty ds\int_{u}^\infty ds'\,\expval{\mathcal{O}_1(is, iv)\mathcal{O}_1(is', iv)}.
\end{equation}
The correlation function of two Schwarzian bilocals is also known exactly~\cite{Mertens:2017mtv}.\footnote{When transferring from semi-classical expressions to full Schwarzian expressions, one always faces ambiguities in how to interpret these as operator insertions. Here we insist that each $\hat{\lambda}$ is an operator in the quantum theory. This implies that the local operators constituting the bilocals are operator-ordered in a way that keeps them together, which in turn means that the Schwarzian path integral that computes them is obtained starting with the Schwarzian diagram in Euclidean signature without crossings \cite{Lam:2018pvp}.} The result, written as a function of Euclidean time coordinates $z_i$ with $i=1, \dots, 4$, is
\begin{align}
    \expval{\mathcal{O}_1(z_1, z_2)\mathcal{O}_1(z_3, z_4)} = \dfrac{1}{2Z_0(\beta)}\int_0^\infty\,\left(\prod_{i=1}^4 d\mu(k_i)\right)&e^{-z_{12}\frac{k_1^2}{2C}-z_{23}\frac{k_2^2}{2C}-z_{34}\frac{k_3^2}{2C}-(\beta-z_{14})\frac{k_4^2}{2C}} \nonumber \\
    &\times \mathcal{A}(k_1, k_2, k_3, k_4),
\end{align}
where $z_{ij}\equiv z_i - z_j$, $d\mu(k)\equiv 2k\sinh(2\pi k)dk$ as before, and the quantity $\mathcal{A}(k_1, k_2, k_3, k_4)$ is given by
\begin{equation}\label{eq:4ptamplitude}
    \mathcal{A}(k_1, k_2, k_3, k_4) \equiv \dfrac{1}{16\pi^6}\dfrac{\Gamma(1\pm ik_1\pm ik_2)}{(2C)^2}\dfrac{\Gamma(1\pm ik_3 \pm ik_2)}{(2C)^2}\dfrac{\delta(k_4-k_2)}{k_4\sinh(2\pi k_4)}.
\end{equation}
Solving the integral over $k_4$ exactly due to the delta function in Equation~\eqref{eq:4ptamplitude}, we can write
\begin{align}
    \expval{\mathcal{O}_1(z_1, z_2)\mathcal{O}_1(z_3, z_4)} = \dfrac{1}{Z_0(\beta)}\int_0^\infty\,&\left(\prod_{i=1}^3 d\mu(k_i)\right)e^{-z_{12}\frac{k_1^2}{2C}-(\beta-z_{14}+z_{23})\frac{k_2^2}{2C}-z_{34}\frac{k_3^2}{2C}} \nonumber \\
    &\times \dfrac{1}{16\pi^6}\dfrac{\Gamma(1\pm ik_1\pm ik_2)}{(2C)^2}\dfrac{\Gamma(1\pm ik_3 \pm ik_2)}{(2C)^2}.
\end{align}
For the purposes of computing the variance in Equation~\eqref{eq:affinetimevariance}, we set $z_1=iu$, $z_3=iu'$, and $z_2=z_4=iv$, which implies $z_{12}=i(u-v)=z_{14}$ and $z_{23}=-i(u'-v)=-z_{34}$. Considering once again that we are in a regime where all $z_{ij}$ are large in magnitude, the variance eventually evaluates to
\begin{equation}\label{eq:finalfluctuationsaffine}
    \langle\hat{\lambda}^2(u)\rangle\sim \dfrac{2\sqrt 2 A}{5\sqrt{\pi C}}\dfrac{(3\sqrt{2}-4)}{(u-v)^{5/2}}.
\end{equation}
The derivation of this result is outlined in Appendix~\ref{app:affinefluctuations}. Since this also decays as a power law with $u-v$, we are safe to assume that there is a regime near the horizon where the spread $\sigma$ of the switching function of the putative detector is larger than the fundamental uncertainty in the value of affine time, and therefore those fluctuations coming from the fact that $\lambda$ is an operator in the quantum gravity theory can be neglected from the point of view of the detector-field coupling.\footnote{Note, however, that the relative error (that is, the ratio $\sqrt{\frac{\langle\Delta\hat{\lambda}\rangle^2}{\langle\hat{\lambda}\rangle^2}}$) grows with $u$ as $(u-v)^{3/4}$.} This will ultimately allow us to replace the switching function $\chi(u)$, which in principle is fixed only in terms of boundary time, by an equivalent reparametrized function in terms of the expectation value of affine time $\lambda$ for a given value of $u$.

\subsection{Response of a detector in affine time near the horizon}

The analogue of Equation~\eqref{eq:excitationprobclassicalaffinetime}, after we also take into account that the derivative Wightman function itself is determined by the Schwarzian bilocal as written in Equation~\eqref{eq:dressedWightmanLL}, is therefore the sum of the four contributions:
\begin{align}
\begin{split}
\label{eq:SchwarzianExcProb}
    P_{\rm LL} &= \dfrac{q^2}{4\pi}\int du\,du'\,\chi(u)\chi(u')e^{-i\omega\int_{u'}^uds\,\mathcal{O}_1(is, iv)} \mathcal{O}_1(iu, iu'),\\
    P_{\rm LF}&= \dfrac{q^2}{4\pi}\int du\,du'\,\chi(u)\chi(u')e^{-i\omega\int_{-u'+i\beta/2}^uds\,\mathcal{O}_1(is, iv)} \mathcal{O}_1(iu, -iu'-\beta/2),\\
    P_{\rm FL}&= \dfrac{q^2}{4\pi}\int du\,du'\,\chi(u)\chi(u')e^{-i\omega\int_{u'}^{-u-i\beta/2}ds\,\mathcal{O}_1(is, iv)}\mathcal{O}_1(-iu+\beta/2, u'),\\
    P_{\rm FF}&= \dfrac{q^2}{4\pi}\int du\,du'\,\chi(u)\chi(u')e^{-i\omega\int_{-u'+i\beta/2}^{-u-i\beta/2}ds\,\mathcal{O}_1(is, iv)} \mathcal{O}_1(-iu+\beta/2, -iu'-\beta/2).\\
\end{split}
\end{align}
When promoting such a classical expression to the quantum gravity model, there are various choices that have to be made. These choices are related to the usual ordering ambiguities when going from a classical to a quantum system. One of these is which of the following corresponds to the actual quantum gravity calculation of interest:
\begin{equation}
\begin{split}
\label{eq:quenched}
    \langle e^{-i\omega \int_b^a ds \,\mathcal{O}_1(is,iv)}\mathcal{O}_1(ia,ib)\rangle_{\beta}, &\qquad
    \langle e^{-i\omega \int_b^a ds \,\mathcal{O}_1(is,iv)}\rangle_\beta\langle\mathcal{O}_1(ia,ib)\rangle_{\beta}\, ,
     \\
     e^{-i\omega \int_b^a ds \,\langle\mathcal{O}_1(is,iv)\rangle_\beta}\langle&\mathcal{O}_1(ia,ib)\rangle_{\beta}, \qquad \hdots
\end{split}
\end{equation}
with many more possible (albeit less natural) options. Ultimately, this boils down to a prescription of precisely how our detector processes the information it gathers from the underlying system within the full quantum gravity model, or alternatively a precise choice of gravitational ensemble. A priori, there is no right or wrong choice here. In Appendix~\ref{app:average}, we briefly sketch two  routes that could be followed to compute the Schwarzian expectation value of Equation~\eqref{eq:SchwarzianExcProb} in different ``ensemble'' choices. 

Below, we instead look at the very-near-horizon regime where the above ambiguities are expected to be less impactful, since one can appeal to the universal late time power law decay of the Schwarzian correlation functions. Indeed, using \eqref{eq:finalfluctuationsaffine}, we can write down quantitatively where the choice of gravitational ensemble becomes irrelevant as one gets close to the black hole horizon as follows. We expand
\begin{equation}
\langle e^{-i\omega \int_b^a ds \,\mathcal{O}_1(is,iv)}\rangle_\beta \approx 1 - i\omega \langle \hat\lambda\rangle +\frac{\omega^2}{2}\langle \hat\lambda^2\rangle + \hdots
\end{equation}
The higher order terms are heavily suppressed when $\omega \langle \hat\lambda^2\rangle \ll \langle \hat\lambda \rangle $ or
\begin{equation}
u \gg \omega^2 / C,
\end{equation}
which is the near-horizon region where we take $u$ (with dimensions of length) to be very large in this way. In this regime, we then get automatically
\begin{equation}
\langle e^{-i\omega \int_b^a ds \,\mathcal{O}_1(is,iv)}\rangle_\beta \approx  e^{-i\omega \int_b^a ds \,\langle\mathcal{O}_1(is,iv)\rangle_\beta}.
\end{equation}
\subsubsection{Two-point function along detector's trajectory}

Using the near-horizon regime, it is straightforward to show that the Schwarzian-corrected derivative two-point function, generalizing \eqref{eq:derivativetwoptfunctionclassical}, can be written in terms of the average affine time of the infalling observer near the horizon as
\begin{align}\label{eq:derivative2ptfunctionnearhorizon}
    \partial_\lambda\partial_{\lambda'}W(\lambda,\lambda') \approx \dfrac{A}{4Z_0(\beta)}&\dfrac{\sgn(\lambda\lambda')}{\abs{\lambda\lambda'}^{3/2}}\int_0^{\infty}dM\,\rho(M)e^{-\beta M}\int_0^{\infty} dE\,\rho(E)\abs{\mathcal{O}^1_{ME}}^2 \\
    &\times \exp{-(E-M)\left[\frac{\beta}{4}\left(1-\sgn(\lambda\lambda')\right)+i\sqrt{A}\left(\frac{1}{\sqrt{\abs{\lambda}}}-\frac{1}{\sqrt{\abs{\lambda'}}}\right)\right]}.\nonumber
\end{align}
It interpolates the gravitationally dressed two-point function both in the exterior and the interior of the black hole, valid in the limit where both $\abs{\lambda}$ and $\abs{\lambda'}$ are close to zero. We are now in a better position to compute the excitation probability of a detector whose coupling to the quantum field is strongly supported in the vicinity of the horizon. 

To mimic what we did explicitly in Section~\ref{sec:semiclassical}, we will again take the switching function to be of the form
\begin{equation}\label{eq:switchingfunctiontranslated}
    \chi(\lambda) = e^{-(\lambda-\lambda_0)^2/2\sigma^2},
\end{equation}
where $\lambda_0$ determines the affine time at which the coupling is centered, and $\sigma$ determines the characteristic duration of the interaction in units of affine time. With this choice of switching, we can use Equation~\eqref{eq:derivative2ptfunctionnearhorizon} as the derivative two-point function in the computation of the excitation probability as long as the coupling between the detector and the field only has strong support in a region where the asymptotic relations~\eqref{eq:affinetimeLeft} and~\eqref{eq:affinetimeInterior} hold. This, in turn, will be true as long as
\begin{equation}
\label{eq:regime}
    \abs{\lambda_0}, \sigma \ll \frac{A}{C^2} =  \frac{1}{8\pi C Z_0(\beta)},
\end{equation}
which we will assume from now on.

\subsubsection{Excitation probability near the horizon}

In the semiclassical case studied in Subsection~\ref{sub:excprobsemiclassical}, the fact that the derivative two-point function depended only on the difference $\lambda-\lambda'$ automatically guaranteed that the excitation probability would depend solely on $\sigma$, and not on $\lambda_0$. Therefore, a simple way to measure the effect of the Schwarzian corrections on the response of the detector near the horizon is to compute the excitation probability with the switching~\eqref{eq:switchingfunctiontranslated} for different values of $\lambda_0$, at fixed $\sigma$. 

We can immediately see that the relation between $\lambda$ and $u$ deviates from its semiclassical value close the horizon, when $u-v\rightarrow \infty$. Specifically, we notice the deviation from the semiclassical computation around $u\sim C$. This suggests we expect $P_{\rm exc}$ to show non-trivial behavior at
\begin{equation}
    \abs{\lambda} \sim  \frac{A}{C^2} \sim \frac{\beta^{3/2}}{C^{5/2}}e^{-2\pi^2C/\beta}.
\end{equation}
Without using the early-time approximation for the Schwarzian bilocal from the two-point function,\footnote{We consider the near-horizon limit for both points $\lambda$ and $\lambda'$, so it could be tempting to take the early time limit of the bilocal operator in the definition of the two-point function. However, as $\lambda,\lambda'\sim 0$, $u,u'\rightarrow \infty$ and points that are close in affine time might be separated by arbitrarily large values of $u$. 
} the excitation probability is given by
\begin{align}\label{eq:excprobLL}
    P_{\rm exc} &= \dfrac{q^2A}{4Z_0(\beta)}\int d\lambda\int d\lambda'\int dM\,\rho(M)\,e^{-\beta M} \int dE\,\rho(E)\abs{\mathcal{O}_{ME}^1}^2 \nonumber \\
    & \sgn(\lambda\lambda')e^{-\frac{\beta}{4}(E-M)(1-\sgn(\lambda\lambda'))}\dfrac{e^{-i\sqrt{A}(E-M)(\frac{1}{\sqrt{\abs{\lambda}}}-\frac{1}{\sqrt{\abs{\lambda '}}})}}{\abs{\lambda\lambda'}^{3/2}}\chi(\lambda)\chi(\lambda')e^{-i\omega(\lambda-\lambda')}.
\end{align}
Recall that
\begin{align}
    \rho(E) &= e^{S_0}\dfrac{C}{2\pi^2}\sinh\left(2\pi\sqrt{2CE}\right), \\
    |\mathcal{O}^{1}_{ME}|^2 &= \frac{\pi^2e^{-2S_0}}{C}\dfrac{E-M}{\sinh(\pi(\sqrt{2CE}+\sqrt{2CM}))\sinh(\pi(\sqrt{2CE}-\sqrt{2CM}))}.
\end{align}
Using $\sgn(\lambda\lambda')=\sgn(\lambda)\sgn(\lambda')$, we can rewrite this integral as
\begin{equation}\label{eq:schwarzianexcprobfinal}
    P_{\rm exc} = \dfrac{q^2A}{4Z_0(\beta)} \int dM\,\rho(M)\,e^{-\beta M} \int dE\,\rho(E)\abs{\mathcal{O}_{ME}^1}^2\mathcal{F}_{\omega}(E-M),
\end{equation}
where we have swapped the order of integration and factorized the integral in $\lambda$ and $\lambda'$ in the function
\begin{equation}
\label{eq:F}
    \mathcal{F}_{\omega}(k) = \abs{\mathcal{I}_\omega(k)}^2 + \abs{\mathcal{J}_\omega(k)}^2 - \mathcal{K}_\omega(k),
\end{equation}
with
\begin{align}
\begin{split}
\label{eq:IJK}
    \mathcal{I}_\omega(k)&\equiv \int_{-\infty}^{0}\dfrac{d\lambda}{\abs{\lambda}^{3/2}}e^{-i\sqrt{A}\frac{k}{\sqrt{\abs{\lambda}}}}\chi(\lambda)e^{-i\omega\lambda}, \\
    \mathcal{J}_\omega(k)&\equiv \int_{0}^{\infty}\dfrac{d\lambda}{\abs{\lambda}^{3/2}}e^{-i\sqrt{A}\frac{k}{\sqrt{\abs{\lambda}}}}\chi(\lambda)e^{-i\omega\lambda},\\
    \mathcal{K_\omega}(k) &\equiv  e^{-\frac{\beta k}{2}}\left(\mathcal{I}_\omega(k)\mathcal{J}^\ast_\omega(k)+ \mathcal{I}^\ast_\omega(k)\mathcal{J}_\omega(k)\right).
\end{split}
\end{align}
The excitation probability~\eqref{eq:schwarzianexcprobfinal} can then be evaluated numerically as a function of the various parameters of the problem.
\subsubsection{Application 1: Detecting the horizon's location}
\label{app1}

Let us then explicitly take the Gaussian switching function~\eqref{eq:switchingfunctiontranslated}. Assuming $\sigma \omega \ll 1$, the integral is most strongly supported near $\lambda=\lambda_0$, and the integrals \eqref{eq:IJK} can be approximated by
\begin{align}
\begin{split}
\label{eq:Iomega_limit}
    \mathcal{I}_\omega(k)&\xrightarrow{\sigma\rightarrow 0} e^{-i\omega\lambda_0}\int_{0}^\infty d\lambda\, \dfrac{e^{-i\sqrt{A}\frac{k}{\sqrt{\lambda}}}}{\lambda^{3/2}} \chi(-\lambda),\\
    \mathcal{J}_\omega(k)&\xrightarrow{\sigma\rightarrow 0} e^{-i\omega\lambda_0} \int_{0}^{\infty}d\lambda\dfrac{e^{-i\sqrt{A}\frac{k}{\sqrt{\lambda}}}}{\lambda^{3/2}}\chi(\lambda).
\end{split}
\end{align}
This gives the leading order term of $P_{\rm exc}$ in the $\sigma\omega\rightarrow 0$ limit.

\begin{figure}
    \centering
    \includegraphics[width=0.5\linewidth]{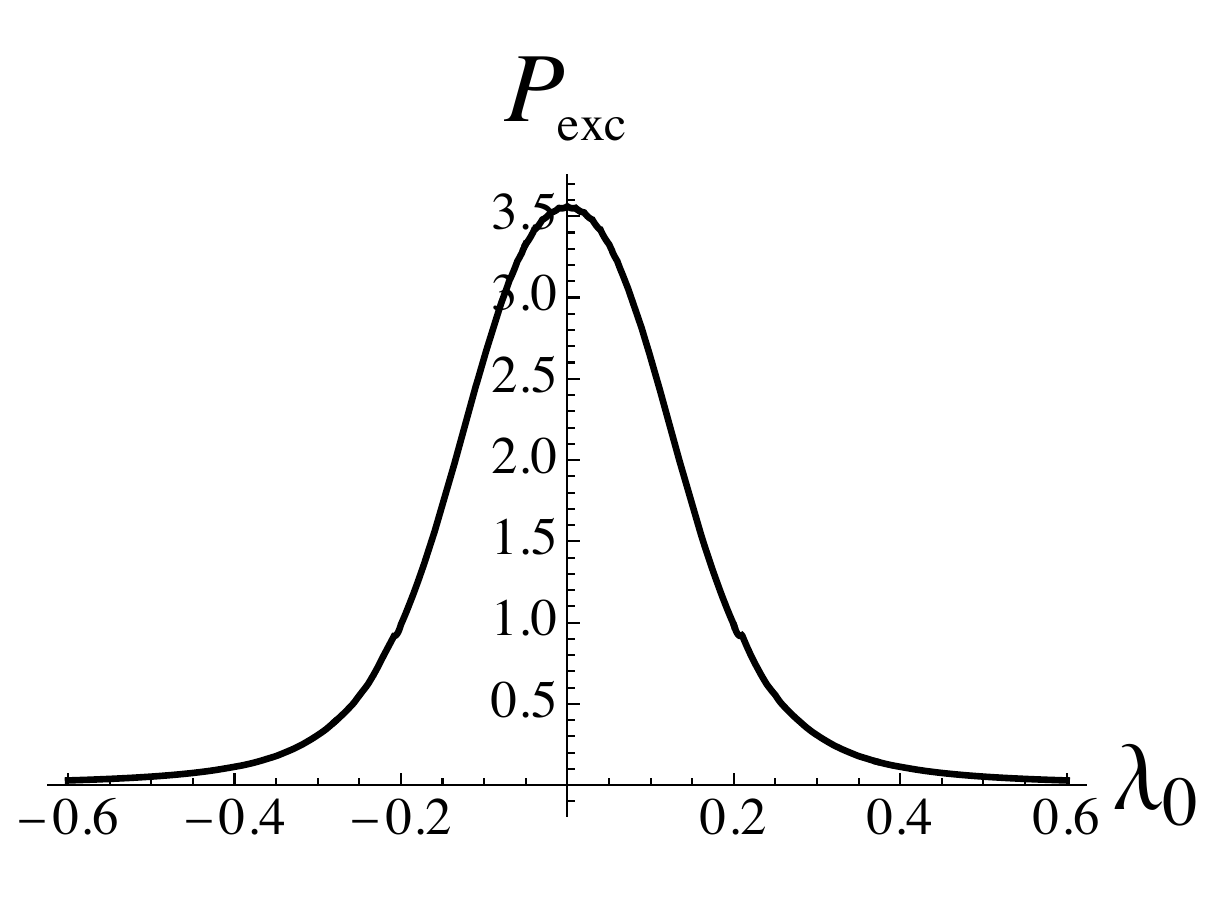}

    \caption{Excitation probability (in units of $q^2$) of a detector in the near-gapless regime ($\sigma \omega \ll 1$) for a gaussian switching function centered around $\lambda_0$, as a function of $\lambda_0$. The numerical computation  is done with $C=1,\beta=15,\sigma=0.1$.}
    \label{fig:Pinlambda0}
\end{figure}

We compute $P_{\rm exc}$ numerically and plot it as a function of $\lambda_0$ in Figure~\ref{fig:Pinlambda0}.\footnote{For all plots in the remaining of the paper, the $\mathcal{I}_{\omega}$ and $\mathcal{J}_{\omega}$ integrals are computed numerically using a regularization $E-M\rightarrow E-M-i\epsilon$. We find that these integrals converge as $\epsilon\rightarrow 0$ with good precision from $\epsilon\sim 0.005$, and take $\epsilon=0.001$ for all numerical computations.}
We find that $P_{\rm exc}$ is peaked around $\lambda_0=0$, and hence an infalling observer detects the horizon as he crosses it. In particular, the width of the peak is larger than $\sigma$. According to the rules of classical general relativity, crossing the horizon of a very large black hole would not be locally detectable. Our results show that, taking into account quantum-gravitational effects, it is possible to detect the horizon before crossing it, and in principle decide to escape.

\subsubsection{Application 2: Measuring the black hole's temperature}
\label{app2}

In the previous subsection, we showed how the detector can detect the black hole horizon itself using quantum gravitational effects near the horizon. This effect is expected as the gravitational dressing breaks the $\lambda \rightarrow \lambda + c$ translation symmetry that we originally found in the semiclassical response function~\eqref{eq:nearH_SC}.

\begin{figure}[ht]
\begin{subfigure}[t] {0.48\linewidth}\centering     \includegraphics[width=1.1\linewidth]{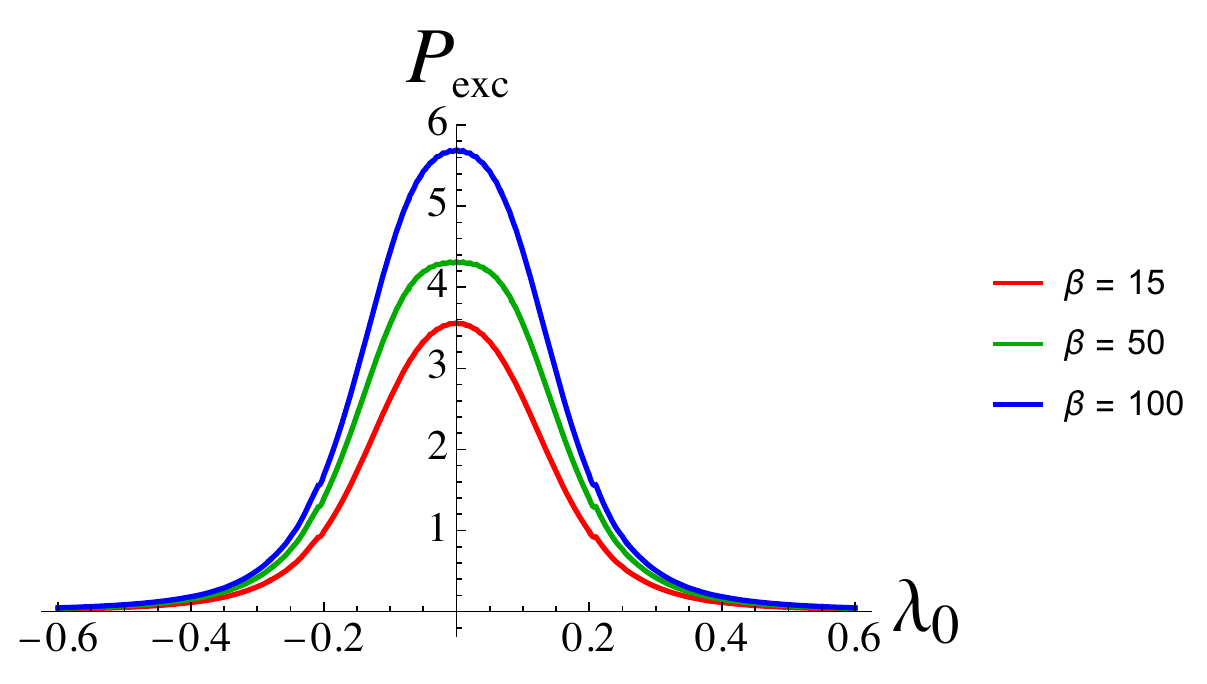}
    \caption{}
    \label{fig:betadep}
    \end{subfigure}
    \begin{subfigure}[t] {0.48\linewidth}\centering 
    \includegraphics[width=0.85\linewidth]{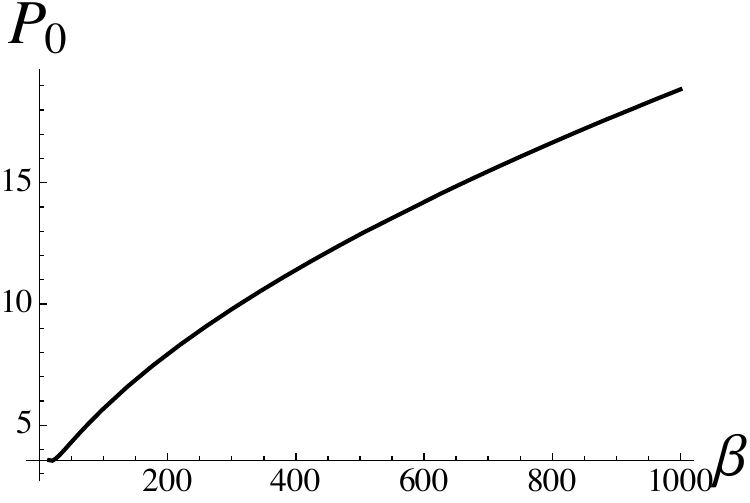}
    \caption{}
    \label{fig:betadepbis}
    \end{subfigure}
    \caption{Excitation probability $P_{\rm exc}$ (in units of $q^2$) of a detector in the near-gapless regime (a) for a gaussian switching function centered around $\lambda_0$, as a function of $\lambda_0$, for different values of the inverse temperature $\beta$. (b) for a gaussian switching function centered around the horizon $\lambda=0$, as function of $\beta$. We used $C=1,\sigma=0.1$.}
\end{figure}

Another feature of the Schwarzian response function is its dependence on the inverse temperature $\beta$. An infalling observer can measure $P_{\rm exc}$, and deduce from it the temperature of the black hole it is about to (or did) fall into, see Figure~\ref{fig:betadep}.

Let us use as a reference the value $P_0\equiv P_{\rm exc}(\lambda_{0}=0)$ of the excitation probability for a Gaussian switching function centered around the horizon, in the low-frequency regime $\sigma\omega \ll 1$. Then the integrals \eqref{eq:IJK} become
\begin{align}
    \mathcal{I}_\omega(k),\, \mathcal{J}_\omega(k) \xrightarrow{\sigma\rightarrow 0}\mathcal{I}_{0}(k)= \int_{0}^\infty d\lambda\, \dfrac{e^{-i\sqrt{A}\frac{k}{\sqrt{\lambda}}}}{\lambda^{3/2}} \chi(\lambda),
\end{align}
and
\begin{equation}
    \mathcal{K}_{\omega}(k) \xrightarrow{\sigma\rightarrow 0} -2e^{-\beta\frac{k}{2}} \abs{\mathcal{I}_0(k)}^2.
\end{equation}
In this regime, the excitation probability does not depend on $\omega$ and reads
\begin{equation}
\label{eq:Pexact}
    P_{\rm exc} = \frac{A e^{S_0}}{2Z(\beta)} \int dM\,\rho(M)e^{-\beta M}\, \int dE\,\rho(E)\,\abs{\mathcal{O}_{ME}^1}^2\left(1-e^{-\beta(E-M)/2}\right)\abs{\mathcal{I}_{0}(E-M)}^2.
\end{equation}
We plot this as a function of $\beta$ in Figure~\ref{fig:betadepbis}. It appears that $P_{\rm exc}(\beta)$ is strictly increasing. This shows that Schwarzian corrections to the response function are larger for quantum black holes, \emph{i.e.} for black holes with increasingly large values of $\beta \gg C$, as one would expect. Furthermore, if we compare it to the leading order term in an expansion in $\sigma\omega$ as we did in Equation~\eqref{eq:nearH_SC_smallsigma}, we also see that the magnitude of the excitation probability is typically much higher than the semiclassical value of $1/4$ (in units of $q^2$, as in Figure~\ref{fig:betadepbis}).

\subsubsection{Application 3: Does the detector meet a firewall?}
\label{sec:firewall}

Lastly, we now consider general values of $\omega$, with no restrictions on the magnitude of $\sigma\omega$. This allows us to study the dependence of the excitation probability on the detector's energy gap. In particular, we want to check whether or not $P_{\rm exc}(\omega)$ satisfies the operational criterion for a firewall that we outlined in Section~\ref{sec:firewallintro}. For convenience, we re-state the criterion here:\\

\textbf{Definition:} \emph{A particle detector is said to encounter a firewall if its excitation probability when coupling smoothly to a quantum field decays at most as a power law in the detector's energy gap $\omega$; i.e., if there is $\alpha\in \mathbb{R}^+$ such that
\begin{equation}
    \lim_{\omega\to\infty} \omega^{\alpha}P_{\text{exc}}(\omega)>0. 
\end{equation}
Alternatively, the detector crosses the horizon ``safely'' (i.e., without meeting a firewall) if its excitation probability goes to $0$ as $\omega\rightarrow \infty$ faster than any polynomial in the detector's energy gap $\omega$.
\\}

The smoothness assumption in the definition above is important because it is well-known that sharp, discontinuous couplings between detector and field can lead to pathological behavior in the detector's response. Even in familiar contexts such as inertial detectors coupled to scalar fields in the Minkowski vacuum, such couplings may lead to excitation probabilities that only decay polynomially with the energy gap of the detector~\cite{Satz:2006kb}. This is of course an artifact of the assumption that the coupling can be turned on or off instantaneously or discontinuously, which is usually deemed unphysical. If a \emph{smooth} coupling still leads to a polynomial decay in the detector's excitation probability, on the other hand, we have strong evidence that the underlying QFT state is indeed singular.\footnote{See~\cite{Louko:2014aba} for an explicit calculation of the UDW detector response across a ``Rindler firewall state'' -- i.e., a state where the correlations between modes on opposite sides of a Rindler horizon are completely severed. The results there are part of the motivation for our general firewall criterion as well.}
 
The firewall criterion we propose can be justified on fairly general grounds by the observation that two-point functions evaluated on finite-energy states (i.e., states with finite expectation values of energy and momentum) are expected to have a universal short-distance structure resembling the vacuum in Minkowski space,\footnote{This statement is most commonly formalized for general QFTs in curved spacetimes by the \emph{Hadamard condition}~\cite{Wald2, Fewster_2013}.} for the vacuum excitation probability decays exponentially with the energy gap. For a slightly more operational point of view on this motivation, one can note that if $P_{\rm exc}(\omega)$ decays as a power law in $\omega$, it is possible to engineer a detector coupling such that the transition probability diverges in the limit of large frequencies. To see this, it is enough to look at the vacuum excitation probability for a detector coupled to a generic field observable, as written in Equation~\eqref{eq:vacuumexcitationprobintro}, and note that if the field observable $O(\lambda)$ is replaced with its $n$-th derivative $O_n(\lambda) \equiv \frac{d^n O}{d\lambda^n}$ along the detector's trajectory, the corresponding excitation probability $P_{\text{exc}}^{(n)}(\omega)$ in the limit of large energy gap $\omega$ will be related to the original $P_{\text{exc}}(\omega)$ by 
\begin{equation}\label{eq:transitionprobscaling}
    P_{\text{exc}}^{(n)}(\omega) \propto \omega^{2n} P_{\text{exc}}(\omega).
\end{equation}
This is easily obtained by replacing $O(\lambda)$ with $\frac{d^n O}{d\lambda^n}$ in Equation~\eqref{eq:vacuumexcitationprobintro} and then repeatedly applying integration by parts.\footnote{When using integration by parts in Equation~\eqref{eq:vacuumexcitationprobintro}, we ignore terms proportional to $\frac{d\chi}{d\lambda}\equiv \chi'(\lambda)$ because we are mainly interested in the limit of large frequencies where $|\chi'(\lambda)| \ll \omega |\chi(\lambda)|$. Here again the smoothness assumption on $\chi(\lambda)$ is important.} 
This would then in principle allow the detector to transition from low-energy states to other states with arbitrarily high energy with high probability by taking $n$ in Equation~\eqref{eq:transitionprobscaling} sufficiently large, even if the coupling to the field itself -- controlled by the switching function $\chi(\lambda)$ -- is smooth. This would ultimately destroy any internal structure of the detector, materializing what one would expect from a firewall.

\begin{figure}[ht]
\centering
   \includegraphics[width=0.6\linewidth]{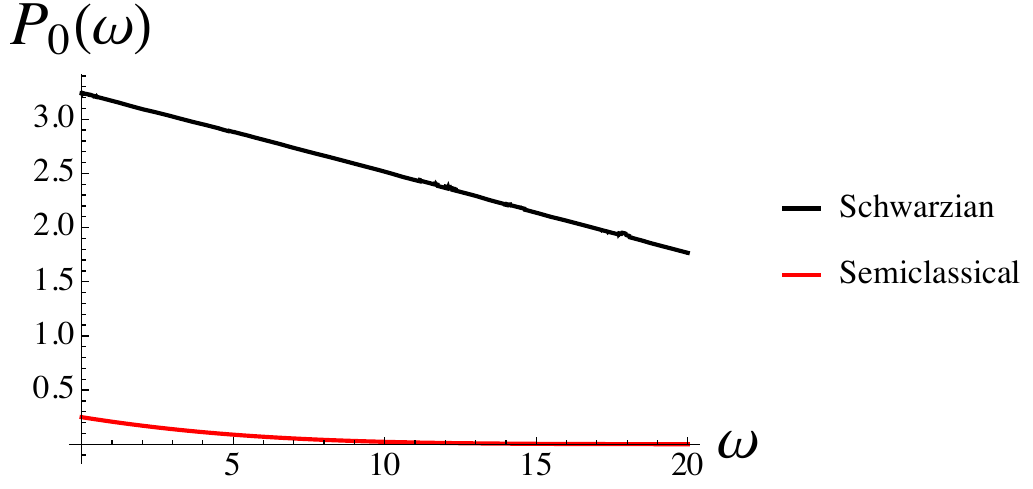}
     \caption{Excitation probability (in units of $q^2$) as a function of the detector's energy gap $\omega$, with $C=1,\beta=15,\sigma=0.1$ and $\lambda_0=0$.}
     \label{fig:firewall}
\end{figure}

The result of numerically evaluating $P_{\rm exc}(\omega)$ over a range of frequencies with our choice of Gaussian switching is displayed in Figure~\ref{fig:firewall}. The plots illustrate once again that the excitation probability is substantially higher than the semiclassical expectation, which is to be expected due to the additional fluctuations caused by the Schwarzian mode. The excitation probability, however, still decays with the detector's energy gap.

The limit of very high frequencies is not immediately visible from Figure~\ref{fig:firewall}, because we lose numerical precision for large values of $\omega$. Fortunately, however, that limit can be tackled analytically. In Appendix~\ref{app:IJK}, we study the asymptotic behavior of the functions $\mathcal{I}_{\omega}(k)$ and $\mathcal{J}_{\omega}(k)$ defined in \eqref{eq:IJK}. We find that, with $\lambda_0=0$,
\begin{align}
    \begin{split}
        P_{\rm exc} \xrightarrow[]{\sigma\omega\rightarrow \infty} q^2\frac{e^{-(\sigma\omega)^2}}{\sigma(\sigma\omega)^{3}} \times \left(\dfrac{\pi A}{4 Z_0(\beta)} \int dM\,\rho(M)\,e^{-\beta M} \int dE\,\rho(E)\abs{\mathcal{O}_{ME}^1}^2(1-e^{-\frac{\beta (E-M)}{2}})\right).
    \end{split}
\end{align}
The exponential decay in $\omega$ therefore ensures that the criterion for a firewall \eqref{def:firewall} is not met, and in that sense, the detector is ``safe'' as it crosses the horizon.

\section{Outlook}
\label{sec:conclusion}
In this paper, we have studied an extension of the gravitational dressing for local bulk operators in JT gravity developed in~\cite{Blommaert2019,Mertens:2019bvy, Blommaert:2020yeo, DeVuyst:2022bua, Mertens:2025rpa} that is able to account for operators in the black hole interior. The main result in this work is that, upon gravitational dressing in JT gravity, an infalling observer can detect the horizon location (thus violating the equivalence principle) and locally measure the black hole temperature. This provides some horizon structure that is not present at the semiclassical level, but is still not dramatic enough to constitute a firewall.

We end our work with some prospects for future work.

\paragraph{Wormhole corrections.}
One of the most important lessons from JT gravity has been the effects that arise from considering higher topologies or wormholes in nonperturbative quantum gravity calculations~\cite{Saad:2019lba,Saad:2019pqd,Almheiri:2019qdq,Penington:2019kki,Iliesiu:2024cnh}. These effects usually lead to features simulating an underlying discreteness of the bulk quantum gravity system, akin to finite-$N$ effects in holographic CFTs in higher dimensions. In the context of JT gravity, such features play a crucial role at extremely long time scales. 

In our work, higher-topology effects are expected to modify the very-near-horizon features observed by an infalling probe when $u-v \sim C e^{S_0}$. From Equation~\eqref{eq:affinetimeLeft}, this corresponds to the time scale where the Schwarzian-corrected affine time away from the horizon is proportional to $e^{-2S_0}$. This is generally much closer to the horizon than where we see our non-trivial detector response $P_{\rm exc}$ studied in this work. For this reason, we expect that the physical conclusions on the horizon location and temperature, as deduced from the infalling observer in subsections \ref{app1} and \ref{app2} respectively, should still remain valid for a wide range of times even if higher topologies are included. There are still some interesting observables where including higher topologies could lead to qualitatively different results, however: for instance, in order to decisively establish the presence or absence of a firewall, it is essential to describe what happens to the observer as they get arbitrarily close to the actual horizon, suggesting that these effects could still play a pronounced role. We will present these modifications in an upcoming work, completing the current story \cite{toap}.

\paragraph{Implications for near-extremal black holes.}
Our choice of gravitational dressing assigns a preferred role to the Schwarzian wiggly boundary curve, which is fixed at constant large value of the dilaton field. Within the context of the higher-dimensional near-extremal black hole, this locus corresponds to the radial location where the long throat transfers into the asymptotic region of the black hole. Whereas this location is not a good reference point for the full higher-dimensional quantum gravity, it is appropriate when restricting to the near-extremal quantum gravity fluctuations. 
As such, we have defined local bulk observables utilizing a semi-classical geometric feature (the end of the throat) of the background quantum state. 
When embedding within this higher-dimensional set-up, depending on the precise ensemble used, an important role is played by additional fluctuations in the charge and non-$s$-wave sectors of the model. Such fluctuations are described in this regime by BF theories based on U(1) and SO(3) respectively \cite{Iliesiu:2020qvm}. We can decorate our gravitational dressing and our calculations with such additional features. E.g. for static observers, the additional U(1) dressing was described in \cite{Mertens:2019bvy}. We leave a more detailed embedding within the near-extremal black hole to future work, and also refer to upcoming work \cite{toappanos}.

\paragraph{Connection to quantum width of the black hole horizon.}
Part of our calculations in subsection \ref{s:affinetime} involved the quantum fluctuations of the affine time $\lambda$ of the infalling null observer. Precisely these fluctuations were very recently discussed in perturbation theory in \cite{Freivogel:2026bsx,Freivogel:2026ofo} in the context of the fluctuation in the horizon's location at much larger scales, as a geometric mean of the Planck length and a black hole length scale \cite{Marolf:2003bb,Verlinde:2019xfb}. In particular, the 2d and 3d case are quite similar, and we can mimic their logic for 3d \cite{Freivogel:2026bsx} in 2d as follows. 
Starting with the 2d black hole geometry, perform an arbitrary diffeomorphism that respects the asymptotic fall-off conditions. This parametrizes arbitrary gravitational fluctuations of the black hole in the 2d and 3d cases where all such fluctuations are diffeomorphisms. For JT gravity, this lands us on
\begin{equation}
    ds^2 = -\dfrac{4 f'(u)f'(v) }{\frac{\beta^2}{\pi^2}\sinh^2\left[\frac{\pi}{\beta}(f(u)-f(v))\right]}du\,dv,
\end{equation}
in terms of the single function $f(.)$, up to small diffeomorphisms. In the 3d case, the authors of \cite{Freivogel:2026bsx} then linearize such expressions around the black hole saddle and compute the two-point function of the affine parameter in this geometry, by relating it to a correlator built from the function $f(.)$ in terms of the boundary CFT stress tensor. 
Following this argument in our 2d case, we would just be performing Schwarzian perturbation theory in $1/C \sim G_N$ of $\mathcal{O}_1(u,v)$ \cite{Maldacena:2016upp}. However, in our set-up, we have available the exact non-perturbative result for the affine time fluctuations. It would be interesting to see how our non-perturbative results connect to the physical conclusions of \cite{Freivogel:2026bsx,Freivogel:2026ofo} for the 2d JT case.

\paragraph{Falling into black hole microstates and other geometries.} 
Our proposal for bulk reconstruction relied heavily on the geometrization of the thermofield double quantum state as the eternal black hole, and the continuation of the boundary clock to both sides of the geometry. 
An interesting extension would be to redo our calculations for a state that looks like a black hole externally, but has a distinct interior, in the form of an end-of-the-world (EOW) brane modeling black hole microstates as defined and studied in this model in \cite{Kourkoulou:2017zaj,Goel:2018ubv,Penington:2019kki}. 
Likewise, another class of semiclassical states with bulk description are those obtained by applying precursor operators and producing a long wormhole in the bulk, as constructed in \cite{Roberts:2014isa}. It would be interesting to understand implications of Schwarzian quantum gravity for horizon-crossing in such bulk geometries, and how to define gravitational dressing explicitly with respect to other features of the state in these more general scenarios.

\paragraph{Connection to von Neumann algebras in quantum gravity.} 
Our proposal for constructing bulk correlators inside the black hole interior has a strong qualitative resemblance to the work of Leutheusser and Liu
\cite{Leutheusser:2021qhd,Leutheusser:2021frk}, in that we have naturally constructed interior operators that require input from \emph{both} holographic boundaries. They explicitly construct a unitary operator (the half-sided modular translation) $U(s)$, generated by a positive Hermitian operator, which maps matter operators in the left exterior to operators in the interior of the black hole. In particular, for an operator $\phi(\mathrm{x})$ in the left exterior, there exists $s_0>0$ which is identified with the Kruskal null distance between $\mathrm{x}$ and the horizon, such that $U^{\dagger}(s)\phi(\mathrm{x})U(s)$ does not commute with operators in the right exterior region for $s>s_0$. This construction yields a holographic definition of infalling time, which is null near the horizon, and generated from time bands of the boundaries. This is reminiscent of the dressing introduced in the present paper, and it would be interesting to strengthen the ties with their formalism. In particular, the construction of the half-sided modular translation is tied to the large $N$ limit and the $1/N$ expansion. It is believed that the black hole interior and horizon are emergent notions that do not exist at finite $N$. We will expand on this idea in the formalism of wormholes in quantum JT gravity in an upcoming paper~\cite{toap}.

\section*{Acknowledgments}
We thank Panagiotis Betzios for general discussions, and Thibaut Verhelst for early discussions in the context of his master thesis work. We acknowledge financial support from the European Research Council (grant BHHQG-101040024). Funded by the European Union. Views and opinions expressed are however those of the author(s) only and do not necessarily reflect those of the European Union or the European Research Council. Neither the European Union nor the granting authority can be held responsible for them.

\appendix 

\section{Comments on gapless detector}
\label{app:gapless}

In the main text, the excitation probability was computed perturbatively in the detector coupling constant $q$. To go beyond the perturbative regime, it is useful to consider a case where the free Hamiltonian of the detector is taken to be zero, $H_F=0$. Physically, this corresponds to the regime where the time scale $\sigma$ for the duration of the interaction between detector and field is much shorter than the detector's Heisenberg time, characterized by the inverse of the detector's proper energy gap $\omega$. When $\omega\lambda_{\text{int}}\ll1$, the detector's monopole operator $\mu(\lambda)$ is effectively time-independent, and the time evolution operator $U$ in the interaction picture can be computed to all orders in the coupling constant $q$. The exact result can be expressed as 
\begin{equation}
    U=e^{i\Lambda}e^{-iq\mu(\lambda_i)\int_{\lambda_i}^\lambda\,d\lambda\,\chi(\lambda)v^\mu\nabla_\mu\phi},
\end{equation}
where $\Lambda$ is a global phase that has no effect on the final state of the system. With this, taking the initial state of the detector to be $\rho_{0}=\ket{0_P}\bra{0_P}$ and the monopole operator $\mu$ to be given by $\mu=\ket{g}\bra{e}+\ket{e}\bra{g}$, where $\{\ket{g}, \ket{e}\}$ form an orthonormal basis for the Hilbert space of the detector, the final density matrix of the detector after tracing out the field can be shown to take the form
\begin{equation}\label{eq:gaplessdetectorfinalstate}
   \rho_d= \begin{pmatrix}
        \expval{\cos^2(g\,\pi[\chi])} & i\expval{\cos(g\,\pi[\chi])\sin(g\,\pi[\chi])} \\
        -i\expval{\cos(g\,\pi[\chi])\sin(g\,\pi[\chi])} &  \expval{\sin^2(g\,\pi[\chi])}
    \end{pmatrix}
\end{equation}
where we have denoted
\begin{align}
    \pi[\chi]&= \int_{\lambda_i}^{\lambda_f}d\lambda\,\chi(\lambda)v^\mu\nabla_\mu\phi, \label{eq:smearedmomentum}\\
    \cos(g\,\pi[\chi])&= \dfrac{e^{ig\pi[\chi]}+e^{-ig\pi[\chi]}}{2}, \\
    \sin(g\,\pi[\chi])&= \dfrac{e^{ig\pi[\chi]}-e^{-ig\pi[\chi]}}{2i}.\label{eq:smearedmomentumsine}
\end{align}
The objects defined in Equations~\eqref{eq:smearedmomentum}-\eqref{eq:smearedmomentumsine} are quantum operators acting on the Hilbert space of the quantum field, and the matrix elements written in Equation~\eqref{eq:gaplessdetectorfinalstate} are the expectation values of functions of these operators computed in the field's initial state. 

The expression~\eqref{eq:gaplessdetectorfinalstate} simplifies even further if we assume that the field starts in a Gaussian state of zero mean, as is the case of the vacuum. In this case, expectation values of exponentials of the field operator are completely defined in terms of the field's two-point function, and we have
\begin{equation}
    \rho_d=\dfrac{1}{2}\begin{pmatrix}1+e^{-2\xi} & 0 \\ 0 & 1-e^{-2\xi}
    \end{pmatrix},
\end{equation}
where
\begin{equation}
    \xi\equiv q^2\int_{\lambda_i}^{\lambda_f}d\lambda \int_{\lambda_i}^{\lambda_f}d\lambda '\chi(\lambda)\chi(\lambda')v^\mu v^\nu\nabla_\mu\nabla_\nu\langle\phi(x)\phi(x')\rangle.
\end{equation}
This tells us, in particular, that the excitation probability $P_{\rm exc}=\bra{e}\rho_d\ket{e}$ of a detector after coupling to the field simply given by
\begin{equation}
    P_{\rm exc}^{\rm gapless}=\dfrac{1-e^{-2\xi}}{2}.
\end{equation}
From a technical point of view, the main advantage of the gapless regime for our purposes is that the computation of the final state of the detector can be done exactly, without relying on perturbation theory. 

\section{General time dependence of the monopole operator and its effect on \texorpdfstring{$P_{\rm exc}$}{Pexc}}
\label{app:general_monopole}

In this appendix, we generalize the evolution of the monopole operator in terms of boundary time. We take the time dependence of $\mu(\lambda)$ parametrized as a function of boundary time $u$ to be of the form
\begin{equation}
    \mu(\lambda(u)) = e^{-i\theta(u)}\ket{g}\bra{e} + e^{i\theta(u)}\ket{e}\bra{g},
\end{equation}
where $\theta(u)$ is a real function, and $\ket{g}$ and $\ket{e}$ are the ground and excited states of the detector, respectively. The definition above makes sure that the operator $\mu(\lambda)$ is Hermitian and equal to its inverse, for any time dependence of the phase $\theta$. The case of a detector that oscillates with constant frequency in units of boundary time corresponds to picking $\theta(u)$ to be linear in $u$, whereas the case of a detector of constant frequency in affine time corresponds to $\theta(u)$ being linear in affine time (which will in general imply a nonlinear functional dependence on $u$). The expression above is valid when the detector is at the left exterior; in the interior of the black hole, a similar-looking expression holds as a function of the boundary time of the right boundary. 

With the parametrization above, making use of our expressions for the gravitationally dressed two-point function and adapting the derivation of Equation~\eqref{eq:exitationprobtwosidedboundary} to the case of a more general time dependence of the detector's monopole operator gives us
\begin{align}
    P_{\text{exc}} = q^2\left|\bra{e}\mu(\lambda_i)\ket{g}\right|^2 &\Bigg(\int_{-\infty}^{+\infty}du_1\int_{-\infty}^{+\infty}du_2\,\chi(u_1)\chi(u_2)e^{-i(\theta(u_1)-\theta(u_2))}\partial_{u_1}\partial_{u_2}W_{\rm LL}(u_1, u_2) \nonumber \\
    &+ \int_{-\infty}^{+\infty}du\int_{-\infty}^{+\infty}du'\,\chi(u)\tilde{\chi}(u')e^{-i(\theta(u)-\theta(u'))}\partial_{u}\partial_{u'}W_{\rm LF}(u, u') \nonumber \\
    &+ \int_{-\infty}^{+\infty}du\int_{-\infty}^{+\infty}du'\,\chi(u')\tilde{\chi}(u)e^{i(\theta(u)-\theta(u'))}\partial_{u}\partial_{u'}W_{\rm FL}(u', u) \nonumber \\
    &+ \int_{-\infty}^{+\infty}du_1'\int_{-\infty}^{+\infty}du_2'\,\tilde{\chi}(u_1')\tilde{\chi}(u_2')e^{-i(\theta(u_1')-\theta(u_2'))}\partial_{u_1'}\partial_{u_2'}W_{\rm FF}(u_1', u_2')\Bigg).
\end{align}
Fixing a given frequency in units of boundary time corresponds to picking $\theta(u)=\omega u$. This is the choice implicitly made in the calculations presented in Section~\ref{sec:bdyexp} in the semiclassical case. Similarly, if we apply this general formula in the case of the gravitationally dressed field and detector, 
the excitation probability becomes
\begin{align}
    P_{\rm exc}= \dfrac{e^{S_0}}{Z(\beta)}&\int_0^\infty dM\,\rho(M)e^{-\beta M}\int_0^\infty dE\,\rho(E)\abs{\mathcal{O}^1_{ME}}^2 \nonumber \\
    \times &\Bigg[\int_{-\infty}^{\infty}du\int_{-\infty}^\infty du' \chi(u)e^{-i\theta(u)}e^{-i(E-M)u}\chi(u')e^{i\theta(u')}e^{i(E-M)u'} \nonumber \\
    +& \int_{-\infty}^{\infty}du\int_{-\infty}^{\infty}du'\tilde{\chi}(u)e^{-i\tilde\theta(u)}e^{i(E-M)u}\tilde{\chi}(u')e^{i\tilde\theta(u')}e^{-i(E-M)u'} \nonumber \\
    -& e^{-\frac{\beta}{2}(E-M)}\int_{-\infty}^{\infty}du\int_{-\infty}^{\infty}du'\chi(u)e^{-i\theta(u)}e^{-i(E-M)u}\tilde{\chi}(u')e^{i\tilde\theta(u')}e^{-i(E-M)u'}\nonumber \\
    -& e^{\frac{\beta}{2}(E-M)}\int_{-\infty}^{\infty}du\int_{-\infty}^{\infty}du'\tilde{\chi}(u)e^{-i\tilde\theta(u)}e^{i(E-M)u}\chi(u')e^{i\theta(u')}e^{i(E-M)u'}\Bigg].
\end{align}
This can be schematically organized as
\begin{align}
\label{eq:Pgen}
    P_{\rm exc}= \dfrac{e^{S_0}}{Z(\beta)}&\int_0^\infty dM\,\rho(M)e^{-\beta M}\int_0^\infty dE\,\rho(E)\abs{\mathcal{O}^1_{ME}}^2 \nonumber \\
    \times &\left[\abs{\mathcal{L}(E-M)}^2 + \abs{\mathcal{M}(E-M)}^2 + \mathcal{N}(E-M)\right], 
\end{align}
where we define
\begin{align}
    \mathcal{L}(k)&=\int_{-\infty}^\infty du\,\chi(u)e^{-i\theta(u)}e^{-iku}, \\
    \mathcal{M}(k)&= \int_{-\infty}^\infty du \,\tilde{\chi}(u)e^{-i\tilde\theta(u)}e^{iku}, \\
    \mathcal{N}(k)&= -e^{-\beta k/2}\left(\mathcal{L}(k)\mathcal{M}^\ast(k) + \mathcal{L}^\ast(k)\mathcal{M}(k)\right).
\end{align}
The late-time limit of this excitation probability, starting with switching functions that are strongly supported in boundary time $u$, will then lead to the transition rate presented in Section~\ref{sec:transitionrateboundaryschwarzian}.
\section{Schwarzian thermal average of the excitation probability}
\label{app:average}

When computing Schwarzian corrections to the detector's response, as in Equation~\eqref{eq:SchwarzianExcProb}, there is an ordering ambiguity in the way we compute the thermal average of $P_{\rm exc}$. Some examples were given in Equation~\eqref{eq:quenched}. Two natural choices of ensemble are the ``annealed'' and ``quenched'' averages. These two kinds of averaging correspond to the evaluation of the gravitational path integral at different stages of the calculation. This has been an important question in the computation of the free-energy $F(\beta)=\log Z_0(\beta)$ in quantum gravity~\cite{Engelhardt:2020qpv,Johnson:2021rsh,Alishahiha:2020jko}. In this context, it has been argued that the correct quantity is the quenched average $\langle\log Z_0(\beta)\rangle_{\beta}$ where the path integral is computed at the very end of the computation. This is contrast with the annealed average $\log \langle Z_0(\beta)\rangle_{\beta}$, where one first compute the quantum gravity partition function, and then compute the free energy from the averaged partition function. \\

\paragraph{Schwarzian corrections in ``quenched'' affine time.}
The ``quenched'' excitation probability is the result of plugging the full expression Equation~\eqref{eq:SchwarzianExcProb} in the Schwarzian path integral. This type of expression appeared in \cite{Maldacena:2017axo}. We do not expect that the thermal expectation value of the form~\eqref{eq:quenched} has an exact analytic form. To find an analytical expressions, two approximations could be used. The first consists of keeping the quenched average but taking small frequencies $\omega$ (with respect to the time scale $|u-u'|$) to expand in perturbation theory:
\begin{align}
\begin{split}
  &\langle e^{-i\omega\int_{u'}^uds\,\mathcal{O}_1(is, iv)} \mathcal{O}_1(iu, iu')\rangle_{\beta}\\
  &\simeq \langle \left(1-i\omega\int_{u'}^uds\,\mathcal{O}_1(is, iv)+\dots \right) \mathcal{O}_1(iu, iu')\rangle_{\beta}\\
 &\simeq \langle \mathcal{O}_1(iu, iu')\rangle_{\beta} - i\omega\int_{u'}^uds\,\langle \mathcal{O}_1(is, iv) \mathcal{O}_1(iu, iu')\rangle_{\beta} + \dots
\end{split}
\end{align}
The first term has an exact expression, given in Equation \eqref{eq:Sch_bilocal}. The integrant in the second term is the four-point function of the Schwarzian theory and also has a closed form expression (See Equation~\eqref{eq:4pts}) that depends on the location of the four points on the thermal circle~\cite{Mertens:2017mtv}.\\ 

\paragraph{Schwarzian corrections in ``annealed'' affine time.}

Another possible direction, which we call the ``annealed'' averaging, is to neglect interactions between the two Schwarzian bilocals:
\begin{equation}
  \langle e^{-i\omega\int_{u'}^uds\,\mathcal{O}_1(is, iv)} \mathcal{O}_1(iu, iu')\rangle_{\beta} \simeq e^{-i\omega\int_{u'}^uds\,\langle \mathcal{O}_1(is, iv)\rangle_{\beta}} \langle \mathcal{O}_1(iu, iu')\rangle_{\beta},
\end{equation}
thus neglecting correlation between fluctuations of the gravitational dressing of the null trajectory and fluctuations of the two point function. Plugging in the closed form expression of the Schwarzian bilocal thermal average, we find
\begin{align}
\begin{split}
    P_{\rm LL}(\omega) &= \dfrac{1}{4\pi Z}q^2|\bra{\omega}\mu(0)\ket{0}|^2\int_{-\infty}^{+\infty}du\int_{-\infty}^{+\infty}du'\,\chi(u)\chi(u')\\
    &\times \exp\left(-\omega\frac{2C}{Z}\int_0^{\infty}d\mu(k_1)d\mu(k_2)e^{-\frac{\beta k_2^2 + iv(k_1^2-k_2^2)}{2C}}\frac{\Gamma(1\pm ik_1\pm ik_2)}{k_1^2-k_2^2}\left(e^{-iu'\frac{k_1^2-k_2^2}{2C}}-e^{-iu\frac{k_1^2-k_2^2}{2C}}\right)\right)\\
    &\times\int_0^{\infty}d\mu(k_1)d\mu(k_2)e^{-i\frac{(u-u')(k_1^2-k_2^2)}{2C}}e^{-\beta\frac{k_2^2}{2C}}\Gamma(1\pm ik_1\pm ik_2).
\end{split}
\end{align}
As before, we find the $LF$ and $FF$ contribution through the euclidean continuations $u'\rightarrow - u'\pm i\beta/2$, $u$ fixed and $u'\rightarrow - u'\pm i\beta/2$, $u\rightarrow - u\pm i\beta/2$.\footnote{Note that this euclidean continuation $u\rightarrow -u\pm i\beta/2$ also gives a mapping between the metric in the left exterior and future interior: $ds^2 = -\frac{4\pi^2}{\beta^2}dudv/\sinh^2(\frac{2\pi}{\beta}\frac{u-v}{2})\rightarrow ds^2-\frac{4\pi^2}{\beta^2}dudv/\cosh^2(\frac{2\pi}{\beta}\frac{u+v}{2})$.} We find
\begin{align}
\begin{split}
    P_{\rm LF}(\omega) &= \dfrac{1}{4\pi Z}q^2|\bra{\omega}\mu(0)\ket{0}|^2\int_{-\infty}^{+\infty}du\int_{-\infty}^{+\infty}du'\,\chi(u)\chi(u')\\
    &\times \exp\left(-\omega\frac{2C}{Z}\int_0^{\infty}d\mu(k_1)d\mu(k_2)e^{-\frac{\beta k_2^2 + iv(k_1^2-k_2^2)}{2C}}\frac{\Gamma(1\pm ik_1\pm ik_2)}{k_1^2-k_2^2}\left(e^{\beta\frac{k_1^2-k_2^2}{4C}}e^{iu'\frac{k_1^2-k_2^2}{2C}}-e^{-iu\frac{k_1^2-k_2^2}{2C}}\right)\right)\\
    &\times\int_0^{\infty}d\mu(k_1)d\mu(k_2)e^{-i\frac{(u+u')(k_1^2-k_2^2)}{2C}}e^{-\beta\frac{(k_1^2+k_2^2)}{4C}}\Gamma(1\pm ik_1\pm ik_2),
\end{split}
\end{align}
and
\begin{align}
\begin{split}
    P_{\rm FF}(\omega) &= \dfrac{1}{4\pi Z}q^2|\bra{\omega}\mu(0)\ket{0}|^2\int_{-\infty}^{+\infty}du\int_{-\infty}^{+\infty}du'\,\chi(u)\chi(u')\\
    &\times \exp\left(-\omega\frac{2C}{Z}\int_0^{\infty}d\mu(k_1)d\mu(k_2)e^{\beta\frac{k_1^2-k_2^2}{4C}}e^{-\frac{\beta k_2^2 + iv(k_1^2-k_2^2)}{2C}}\frac{\Gamma(1\pm ik_1\pm ik_2)}{k_1^2-k_2^2}\left(e^{iu'\frac{k_1^2-k_2^2}{2C}}-e^{iu\frac{k_1^2-k_2^2}{2C}}\right)\right)\\
    &\times\int_0^{\infty}d\mu(k_1)d\mu(k_2)e^{i\frac{(u-u')(k_1^2-k_2^2)}{2C}}e^{-\beta\frac{k_2^2}{2C}}\Gamma(1\pm ik_1\pm ik_2).
\end{split}
\end{align}
One should keep in mind the prescription chosen in \cite{Blommaert2019} to define the off-shell metric as a \textit{hermitian} operator. In particular, the Schwarzian bilocal is not hermitian, as
\begin{equation}
    \mathcal{O}_1(u, v)^* = \mathcal{O}_1(v, u).
\end{equation}
In \cite{Blommaert2019}, a suitable hermitian operator definition of the metric was obtained as an average of the time-orderings:
\begin{equation}
  \hat{ ds^2 } = \frac{ \mathcal{O}_1(u,v) + \mathcal{O}_1(v,u)}{2}.
\end{equation}
Following this prescription would lead us to the alternative definition of annealed affine time:
\begin{equation}
    \lambda-\lambda' = \int_{u'}^u ds \frac{\mathcal{O}_1(is,iv)+\mathcal{O}_1(iv,is)}{2}.
\end{equation}
We leave further investigations of these quenched and annealed averages to future work.

\section{Details on fluctuation of affine time}\label{app:affinefluctuations}
The relevant object in the computation of fluctuations of affine time is the correlation function of two Schwarzian bilocals, 
 \begin{align}
 \label{eq:4pts}
    \expval{\mathcal{O}_1(z_1, z_2)\mathcal{O}_1(z_3, z_4)} = \dfrac{1}{Z_0(\beta)}\int_0^\infty\,&\left(\prod_{i=1}^3 d\mu(k_i)\right)e^{-z_{12}\frac{k_1^2}{2C}-(\beta-z_{14}+z_{23})\frac{k_2^2}{2C}-z_{34}\frac{k_3^2}{2C}} \nonumber \\
    &\times \dfrac{1}{16\pi^6}\dfrac{\Gamma(1\pm ik_1\pm ik_2)}{(2C)^2}\dfrac{\Gamma(1\pm ik_3 \pm ik_2)}{(2C)^2},
\end{align}
where $z_1=iu$, $z_3=iu'$, and $z_2=z_4=iv$, which implies $z_{12}=i(u-v)=z_{14}$ and $z_{23}=-i(u'-v)=-z_{34}$. Note this also means that $z_{23}-z_{14}=-i(u'-v)-i(u-v)=-i(u+u'-2v)$. This is the quantity that appears in the expression for the Schwarzian expectation value of the square of the affine time of our gravitationally dressed infalling observer. 

We are mainly interested in the case where $u, u'\gg v$, or more precisely, in the case where both $u-v$ and $u'-v$ are much greater than $C$.
The idea to solve the integral is to note that, in this regime, there are wildly oscillating factors in the integrand, and there is no point in the interior of the domain where the phase is stationary. This implies that the integral in the limit of $u, u'\gg v$ will be dominated by the edge of the domain of integration, where $k_{1, 2, 3}\ll 1$. We can therefore approximate the measure as
\begin{equation}
    \prod_{i=1}^3 d\mu(k_i) \simeq (4\pi)^3 \prod_{i=1}^3 k_i^2\,dk_i,
\end{equation}
and also take
\begin{equation}
    \dfrac{1}{16\pi^6}\dfrac{\Gamma(1\pm ik_1\pm ik_2)}{(2C)^2}\dfrac{\Gamma(1\pm ik_3 \pm ik_2)}{(2C)^2} \simeq \dfrac{1}{16\pi^6(2C)^4}.
\end{equation}
Our target integral is therefore a product of three integrals of the form
\begin{equation}
    \int_0^\infty k_i^2\, dk_i\,e^{-\alpha_i k_i^2}.
\end{equation}
These can all be evaluated using the trick
\begin{equation}
     \int_0^\infty k_i^2\, dk_i\,e^{-\alpha_i k_i^2}= -\dfrac{d}{d\alpha_i} \int_0^\infty dk_i\,e^{-\alpha_i k_i^2} = -\dfrac{1}{2}\dfrac{d}{d\alpha_i}\left(\sqrt{\dfrac{\pi}{\alpha_i}}\right)=\dfrac{1}{4}\sqrt{\dfrac{\pi}{\alpha_i^3}}.
\end{equation}
Putting all of this together and restoring the various values of the constants $\alpha_i$, we get
\begin{equation}
    \expval{\mathcal{O}_1(z_1, z_2)\mathcal{O}_1(z_3, z_4)} = \dfrac{\sqrt{2C}}{Z_0(\beta)} \dfrac{1}{16 \pi^{3/2}} \dfrac{1}{(u-v)^{3/2}}\dfrac{1}{(u'-v)^{3/2}}\dfrac{1}{(u+u'-2v)^{3/2}}.
\end{equation}
The expectation value of the square of affine time then becomes
\begin{align}
     \langle\lambda^2\rangle &= \int_{u}^\infty ds\int_{u}^\infty ds'\,\langle\mathcal{O}_1(is, iv)\mathcal{O}_1(is', iv)\rangle \nonumber \\
     &=\dfrac{\sqrt{2C}}{Z_0(\beta)} \dfrac{1}{16 \pi^{3/2}}\int_{u}^\infty ds\int_{u}^\infty ds'\dfrac{1}{(s-v)^{3/2}}\dfrac{1}{(s'-v)^{3/2}}\dfrac{1}{(s+s'-2v)^{3/2}} \nonumber \\
     &=\dfrac{\sqrt{2C}}{Z_0(\beta)} \dfrac{1}{16 \pi^{3/2}}\int_{u-v}^\infty ds\int_{u-v}^\infty ds'\dfrac{1}{s^{3/2}}\dfrac{1}{s'^{3/2}}\dfrac{1}{(s+s')^{3/2}} \nonumber \\
     &= \dfrac{\sqrt{2C}}{Z_0(\beta)}\dfrac{1}{16 \pi^{3/2}} \dfrac{1}{(u-v)^{5/2}}\int_1^\infty dx\int_1^\infty dy\dfrac{1}{x^{3/2}y^{3/2}(x+y)^{3/2}}\nonumber \\
     &= \dfrac{\sqrt{2C}}{Z_0(\beta)}\dfrac{1}{16 \pi^{3/2}} \dfrac{1}{(u-v)^{5/2}}\dfrac{4(3\sqrt{2}-4)}{5},
\end{align}
where in the manipulations above we shifted the integration variable from $s\to s-v$, and subsequently rescaled $s-v=(u-v)x, s'-v=(u-v)y$, where $x, y$ are dimensionless. Re-expressing $1/Z_{0}(\beta)=8\pi A/C$, we finally arrive at
\begin{equation}
    \langle \lambda^2(u)\rangle \simeq \dfrac{2\sqrt 2 A}{5\sqrt{\pi C}}\dfrac{3\sqrt{2}-4}{(u-v)^{5/2}},
\end{equation}
which is Equation~\eqref{eq:finalfluctuationsaffine}.

\section{Asymptotic behavior of \texorpdfstring{$\mathcal{F}_{\omega}(k)$}{Fomega(k)}}
\label{app:IJK}

Let us study the asymptotic behavior of the integrals $\mathcal{I}_{\omega}(k)$ and $\mathcal{J}_{\omega}(k)$ (Equation~\eqref{eq:IJK}) that define the function \eqref{eq:F}, when the switching function is a Gaussian. Recall that
\begin{equation}
    \mathcal{F}_{\omega}(k) = \abs{\mathcal{I}_\omega(k)}^2 + \abs{\mathcal{J}_\omega(k)}^2 - \mathcal{K}_\omega(k),
\end{equation}
where
\begin{align}
\begin{split}
    \mathcal{I}_\omega(k)&\equiv \int_{-\infty}^{0}\dfrac{d\lambda}{\abs{\lambda}^{3/2}}e^{-i\sqrt{A}\frac{k}{\sqrt{\abs{\lambda}}}}e^{-\frac{(\lambda-\lambda_0)^2}{2\sigma^2}}e^{-i\omega\lambda}, \\
    \mathcal{J}_\omega(k)&\equiv \int_{0}^{\infty}\dfrac{d\lambda}{\abs{\lambda}^{3/2}}e^{-i\sqrt{A}\frac{k}{\sqrt{\abs{\lambda}}}}e^{-\frac{(\lambda-\lambda_0)^2}{2\sigma^2}}e^{-i\omega\lambda},\\
    \mathcal{K_\omega}(k) &\equiv  e^{-\frac{\beta k}{2}}\left(\mathcal{I}_\omega(k)\mathcal{J}^\ast_\omega(k)+ \mathcal{I}^\ast_\omega(k)\mathcal{J}_\omega(k)\right).
\end{split}
\end{align}
Integrals $\mathcal{I}_{\omega}(k)$ and $\mathcal{J}_{\omega}(k)$ have a singularity at $\lambda=0$. It is useful to rewrite them as
\begin{align}
    \begin{split}
        \mathcal{I}_{\omega}(k) &= 2\int_0^{\infty} dt \,e^{\sigma\omega\,S_+(t)},\\
        \mathcal{J}_{\omega}(k) &= 2\int_0^{\infty} dt\, e^{\sigma\omega\,S_-(t)},
    \end{split}
\end{align}
where
\begin{equation}
\label{eq:Spm}
    S_{\pm}(t) = -i\frac{\sqrt{A}}{\sigma\omega}kt - \frac{1}{2\sigma^3\omega}\left(\frac{1}{t^2}\pm \lambda_0 \right)^2+\frac{i}{\sigma t^2}.
\end{equation}
We use the saddle-point approximation in the limit $\sigma\omega \rightarrow \infty$. The location of the saddles is not very sensitive to the first term in $S_{\pm}(t)$ as long as $\sqrt{A}$ is small compared to $1/\sigma^2$. Then, the saddles are located at
\begin{equation}
    t_*^2 \sim \frac{1}{i\sigma^2 \omega},
\end{equation}
and
\begin{equation}
    \mathcal{I}_{\omega}(k), \,\mathcal{J}_{\omega}(k) \xrightarrow[]{\sigma\omega \rightarrow \infty} e^{i\pi/4}\sqrt{\frac{\pi}{2\sigma}} (\sigma\omega)^{-3/2}e^{-\frac{(\sigma\omega)^2}{2}}.
\end{equation}
We thus find the asymptotic behavior of $\mathcal{F}_{\omega}$:
\begin{equation}
    \mathcal{F}_{\omega}(k) \xrightarrow[]{\sigma\omega \rightarrow \infty}  \frac{\pi}{\sigma}(1-e^{-\frac{\beta k}{2}})\frac{e^{-(\sigma\omega)^2}}{(\sigma\omega)^{3}}.
\end{equation}

\bibliographystyle{JHEP} 
\bibliography{references.bib}

\providecommand{\href}[2]{#2}\begingroup\raggedright\begin{thebibliography}{10}

\bibitem{Rovelli:1990ph}
C.~Rovelli, \emph{{What Is Observable in Classical and Quantum Gravity?}}, \href{https://doi.org/10.1088/0264-9381/8/2/011}{\emph{Class. Quant. Grav.} {\bfseries 8} (1991) 297}.

\bibitem{Giddings:2005id}
S.B.~Giddings, D.~Marolf and J.B.~Hartle, \emph{{Observables in effective gravity}}, \href{https://doi.org/10.1103/PhysRevD.74.064018}{\emph{Phys. Rev. D} {\bfseries 74} (2006) 064018} [\href{https://arxiv.org/abs/hep-th/0512200}{{\ttfamily hep-th/0512200}}].

\bibitem{Donnelly:2015hta}
W.~Donnelly and S.B.~Giddings, \emph{{Diffeomorphism-invariant observables and their nonlocal algebra}}, \href{https://doi.org/10.1103/PhysRevD.93.024030}{\emph{Phys. Rev. D} {\bfseries 93} (2016) 024030} [\href{https://arxiv.org/abs/1507.07921}{{\ttfamily 1507.07921}}].

\bibitem{Donnelly:2016rvo}
W.~Donnelly and S.B.~Giddings, \emph{{Observables, gravitational dressing, and obstructions to locality and subsystems}}, \href{https://doi.org/10.1103/PhysRevD.94.104038}{\emph{Phys. Rev. D} {\bfseries 94} (2016) 104038} [\href{https://arxiv.org/abs/1607.01025}{{\ttfamily 1607.01025}}].

\bibitem{Goeller:2022rsx}
C.~Goeller, P.A.~Hoehn and J.~Kirklin, \emph{{Diffeomorphism-invariant observables and dynamical frames in gravity: reconciling bulk locality with general covariance}},  \href{https://arxiv.org/abs/2206.01193}{{\ttfamily 2206.01193}}.

\bibitem{Hamilton:2005ju}
A.~Hamilton, D.N.~Kabat, G.~Lifschytz and D.A.~Lowe, \emph{{Local bulk operators in AdS/CFT: A Boundary view of horizons and locality}}, \href{https://doi.org/10.1103/PhysRevD.73.086003}{\emph{Phys. Rev. D} {\bfseries 73} (2006) 086003} [\href{https://arxiv.org/abs/hep-th/0506118}{{\ttfamily hep-th/0506118}}].

\bibitem{Hamilton:2006az}
A.~Hamilton, D.N.~Kabat, G.~Lifschytz and D.A.~Lowe, \emph{{Holographic representation of local bulk operators}}, \href{https://doi.org/10.1103/PhysRevD.74.066009}{\emph{Phys. Rev. D} {\bfseries 74} (2006) 066009} [\href{https://arxiv.org/abs/hep-th/0606141}{{\ttfamily hep-th/0606141}}].

\bibitem{Kraus:2002iv}
P.~Kraus, H.~Ooguri and S.~Shenker, \emph{{Inside the horizon with AdS / CFT}}, \href{https://doi.org/10.1103/PhysRevD.67.124022}{\emph{Phys. Rev. D} {\bfseries 67} (2003) 124022} [\href{https://arxiv.org/abs/hep-th/0212277}{{\ttfamily hep-th/0212277}}].

\bibitem{Almheiri:2013hfa}
A.~Almheiri, D.~Marolf, J.~Polchinski, D.~Stanford and J.~Sully, \emph{{An Apologia for Firewalls}}, \href{https://doi.org/10.1007/JHEP09(2013)018}{\emph{JHEP} {\bfseries 09} (2013) 018} [\href{https://arxiv.org/abs/1304.6483}{{\ttfamily 1304.6483}}].

\bibitem{Marolf:2013dba}
D.~Marolf and J.~Polchinski, \emph{{Gauge/Gravity Duality and the Black Hole Interior}}, \href{https://doi.org/10.1103/PhysRevLett.111.171301}{\emph{Phys. Rev. Lett.} {\bfseries 111} (2013) 171301} [\href{https://arxiv.org/abs/1307.4706}{{\ttfamily 1307.4706}}].

\bibitem{Papadodimas:2012aq}
K.~Papadodimas and S.~Raju, \emph{{An Infalling Observer in AdS/CFT}}, \href{https://doi.org/10.1007/JHEP10(2013)212}{\emph{JHEP} {\bfseries 10} (2013) 212} [\href{https://arxiv.org/abs/1211.6767}{{\ttfamily 1211.6767}}].

\bibitem{Papadodimas:2013wnh}
K.~Papadodimas and S.~Raju, \emph{{Black Hole Interior in the Holographic Correspondence and the Information Paradox}}, \href{https://doi.org/10.1103/PhysRevLett.112.051301}{\emph{Phys. Rev. Lett.} {\bfseries 112} (2014) 051301} [\href{https://arxiv.org/abs/1310.6334}{{\ttfamily 1310.6334}}].

\bibitem{Papadodimas:2013jku}
K.~Papadodimas and S.~Raju, \emph{{State-Dependent Bulk-Boundary Maps and Black Hole Complementarity}}, \href{https://doi.org/10.1103/PhysRevD.89.086010}{\emph{Phys. Rev. D} {\bfseries 89} (2014) 086010} [\href{https://arxiv.org/abs/1310.6335}{{\ttfamily 1310.6335}}].

\bibitem{Grinberg:2020fdj}
M.~Grinberg and J.~Maldacena, \emph{{Proper time to the black hole singularity from thermal one-point functions}}, \href{https://doi.org/10.1007/JHEP03(2021)131}{\emph{JHEP} {\bfseries 03} (2021) 131} [\href{https://arxiv.org/abs/2011.01004}{{\ttfamily 2011.01004}}].

\bibitem{Leutheusser:2021frk}
S.A.W.~Leutheusser and H.~Liu, \emph{{Emergent Times in Holographic Duality}}, \href{https://doi.org/10.1103/PhysRevD.108.086020}{\emph{Phys. Rev. D} {\bfseries 108} (2023) 086020} [\href{https://arxiv.org/abs/2112.12156}{{\ttfamily 2112.12156}}].

\bibitem{Maldacena:2001kr}
J.M.~Maldacena, \emph{{Eternal black holes in anti-de Sitter}}, \href{https://doi.org/10.1088/1126-6708/2003/04/021}{\emph{JHEP} {\bfseries 04} (2003) 021} [\href{https://arxiv.org/abs/hep-th/0106112}{{\ttfamily hep-th/0106112}}].

\bibitem{Hamilton:2006fh}
A.~Hamilton, D.N.~Kabat, G.~Lifschytz and D.A.~Lowe, \emph{{Local bulk operators in AdS/CFT: A Holographic description of the black hole interior}}, \href{https://doi.org/10.1103/PhysRevD.75.106001}{\emph{Phys. Rev. D} {\bfseries 75} (2007) 106001} [\href{https://arxiv.org/abs/hep-th/0612053}{{\ttfamily hep-th/0612053}}].

\bibitem{Lewkowycz:2016ukf}
A.~Lewkowycz, G.J.~Turiaci and H.~Verlinde, \emph{{A CFT Perspective on Gravitational Dressing and Bulk Locality}}, \href{https://doi.org/10.1007/JHEP01(2017)004}{\emph{JHEP} {\bfseries 01} (2017) 004} [\href{https://arxiv.org/abs/1608.08977}{{\ttfamily 1608.08977}}].

\bibitem{Almheiri:2017fbd}
A.~Almheiri, T.~Anous and A.~Lewkowycz, \emph{{Inside out: meet the operators inside the horizon. On bulk reconstruction behind causal horizons}}, \href{https://doi.org/10.1007/JHEP01(2018)028}{\emph{JHEP} {\bfseries 01} (2018) 028} [\href{https://arxiv.org/abs/1707.06622}{{\ttfamily 1707.06622}}].

\bibitem{Jafferis:2020ora}
D.L.~Jafferis and L.~Lamprou, \emph{{Inside the hologram: reconstructing the bulk observer{\textquoteright}s experience}}, \href{https://doi.org/10.1007/JHEP03(2022)084}{\emph{JHEP} {\bfseries 03} (2022) 084} [\href{https://arxiv.org/abs/2009.04476}{{\ttfamily 2009.04476}}].

\bibitem{Gao:2021tzr}
P.~Gao and L.~Lamprou, \emph{{Seeing behind black hole horizons in SYK}}, \href{https://doi.org/10.1007/JHEP06(2022)143}{\emph{JHEP} {\bfseries 06} (2022) 143} [\href{https://arxiv.org/abs/2111.14010}{{\ttfamily 2111.14010}}].

\bibitem{deBoer:2022zps}
J.~de~Boer, D.L.~Jafferis and L.~Lamprou, \emph{{On black hole interior reconstruction, singularities and the emergence of time}},  \href{https://arxiv.org/abs/2211.16512}{{\ttfamily 2211.16512}}.

\bibitem{Leutheusser:2022bgi}
S.~Leutheusser and H.~Liu, \emph{{Subregion-subalgebra duality: Emergence of space and time in holography}}, \href{https://doi.org/10.1103/PhysRevD.111.066021}{\emph{Phys. Rev. D} {\bfseries 111} (2025) 066021} [\href{https://arxiv.org/abs/2212.13266}{{\ttfamily 2212.13266}}].

\bibitem{Leutheusser:2021qhd}
S.~Leutheusser and H.~Liu, \emph{{Causal connectability between quantum systems and the black hole interior in holographic duality}}, \href{https://doi.org/10.1103/PhysRevD.108.086019}{\emph{Phys. Rev. D} {\bfseries 108} (2023) 086019} [\href{https://arxiv.org/abs/2110.05497}{{\ttfamily 2110.05497}}].

\bibitem{VanRaamsdonk:2010pw}
M.~Van~Raamsdonk, \emph{{Building up spacetime with quantum entanglement}}, \href{https://doi.org/10.1142/S0218271810018529}{\emph{Gen. Rel. Grav.} {\bfseries 42} (2010) 2323} [\href{https://arxiv.org/abs/1005.3035}{{\ttfamily 1005.3035}}].

\bibitem{Jackiw:1984je}
R.~Jackiw, \emph{{Lower Dimensional Gravity}}, \href{https://doi.org/10.1016/0550-3213(85)90448-1}{\emph{Nucl. Phys.} {\bfseries B252} (1985) 343}.

\bibitem{Teitelboim:1983ux}
C.~Teitelboim, \emph{{Gravitation and Hamiltonian Structure in Two Space-Time Dimensions}}, \href{https://doi.org/10.1016/0370-2693(83)90012-6}{\emph{Phys. Lett.} {\bfseries 126B} (1983) 41}.

\bibitem{Almheiri:2014cka}
A.~Almheiri and J.~Polchinski, \emph{{Models of AdS$_{2}$ backreaction and holography}}, \href{https://doi.org/10.1007/JHEP11(2015)014}{\emph{JHEP} {\bfseries 11} (2015) 014} [\href{https://arxiv.org/abs/1402.6334}{{\ttfamily 1402.6334}}].

\bibitem{Mertens:2022irh}
T.G.~Mertens and G.J.~Turiaci, \emph{{Solvable models of quantum black holes: a review on Jackiw{\textendash}Teitelboim gravity}}, \href{https://doi.org/10.1007/s41114-023-00046-1}{\emph{Living Rev. Rel.} {\bfseries 26} (2023) 4} [\href{https://arxiv.org/abs/2210.10846}{{\ttfamily 2210.10846}}].

\bibitem{Fabbri:2005mw}
A.~Fabbri and J.~Navarro-Salas, \emph{{Modeling black hole evaporation}}, World Scientific, Singapore (2005), \href{https://doi.org/10.1142/p378}{10.1142/p378}.

\bibitem{Nayak:2018qej}
P.~Nayak, A.~Shukla, R.M.~Soni, S.P.~Trivedi and V.~Vishal, \emph{{On the Dynamics of Near-Extremal Black Holes}}, \href{https://doi.org/10.1007/JHEP09(2018)048}{\emph{JHEP} {\bfseries 09} (2018) 048} [\href{https://arxiv.org/abs/1802.09547}{{\ttfamily 1802.09547}}].

\bibitem{Iliesiu:2020qvm}
L.V.~Iliesiu and G.J.~Turiaci, \emph{{The statistical mechanics of near-extremal black holes}}, \href{https://doi.org/10.1007/JHEP05(2021)145}{\emph{JHEP} {\bfseries 05} (2021) 145} [\href{https://arxiv.org/abs/2003.02860}{{\ttfamily 2003.02860}}].

\bibitem{Turiaci:2024cad}
G.J.~Turiaci, \emph{{Les Houches lectures on two-dimensional gravity and holography}}, \href{https://doi.org/10.21468/SciPostPhysLectNotes.113}{\emph{SciPost Phys. Lect. Notes} {\bfseries 113} (2026) 1} [\href{https://arxiv.org/abs/2412.09537}{{\ttfamily 2412.09537}}].

\bibitem{Jensen:2016pah}
K.~Jensen, \emph{{Chaos in AdS$_2$ Holography}}, \href{https://doi.org/10.1103/PhysRevLett.117.111601}{\emph{Phys. Rev. Lett.} {\bfseries 117} (2016) 111601} [\href{https://arxiv.org/abs/1605.06098}{{\ttfamily 1605.06098}}].

\bibitem{Maldacena:2016upp}
J.~Maldacena, D.~Stanford and Z.~Yang, \emph{{Conformal symmetry and its breaking in two dimensional Nearly Anti-de-Sitter space}}, \href{https://doi.org/10.1093/ptep/ptw124}{\emph{PTEP} {\bfseries 2016} (2016) 12C104} [\href{https://arxiv.org/abs/1606.01857}{{\ttfamily 1606.01857}}].

\bibitem{Engelsoy:2016xyb}
J.~Engels{\"o}y, T.G.~Mertens and H.~Verlinde, \emph{{An investigation of AdS$_{2}$ backreaction and holography}}, \href{https://doi.org/10.1007/JHEP07(2016)139}{\emph{JHEP} {\bfseries 07} (2016) 139} [\href{https://arxiv.org/abs/1606.03438}{{\ttfamily 1606.03438}}].

\bibitem{toap}
V.~Franken, T.G.~Mertens,  and B.~de~S.~L.~Torres, \emph{To appear}, .

\bibitem{Blommaert2019}
A.~Blommaert, T.G.~Mertens and H.~Verschelde, \emph{{Clocks and Rods in Jackiw-Teitelboim Quantum Gravity}}, \href{https://doi.org/10.1007/JHEP09(2019)060}{\emph{JHEP} {\bfseries 09} (2019) 060} [\href{https://arxiv.org/abs/1902.11194}{{\ttfamily 1902.11194}}].

\bibitem{Mertens:2019bvy}
T.G.~Mertens, \emph{{Towards Black Hole Evaporation in Jackiw-Teitelboim Gravity}}, \href{https://doi.org/10.1007/JHEP07(2019)097}{\emph{JHEP} {\bfseries 07} (2019) 097} [\href{https://arxiv.org/abs/1903.10485}{{\ttfamily 1903.10485}}].

\bibitem{Blommaert:2020yeo}
A.~Blommaert, T.G.~Mertens and H.~Verschelde, \emph{{Unruh detectors and quantum chaos in JT gravity}}, \href{https://doi.org/10.1007/JHEP03(2021)086}{\emph{JHEP} {\bfseries 03} (2021) 086} [\href{https://arxiv.org/abs/2005.13058}{{\ttfamily 2005.13058}}].

\bibitem{DeVuyst:2022bua}
J.~De~Vuyst and T.G.~Mertens, \emph{{Operational islands and black hole dissipation in JT gravity}}, \href{https://doi.org/10.1007/JHEP01(2023)027}{\emph{JHEP} {\bfseries 01} (2023) 027} [\href{https://arxiv.org/abs/2207.03351}{{\ttfamily 2207.03351}}].

\bibitem{Mertens:2025rpa}
T.G.~Mertens, T.~Tappeiner and B.~de~S.~L.~Torres, \emph{{Fiducial observers and the thermal atmosphere in the black hole quantum throat}}, \href{https://doi.org/10.1007/JHEP04(2026)145}{\emph{JHEP} {\bfseries 04} (2026) 145} [\href{https://arxiv.org/abs/2507.20983}{{\ttfamily 2507.20983}}].

\bibitem{Witten:2021unn}
E.~Witten, \emph{{Gravity and the crossed product}}, \href{https://doi.org/10.1007/JHEP10(2022)008}{\emph{JHEP} {\bfseries 10} (2022) 008} [\href{https://arxiv.org/abs/2112.12828}{{\ttfamily 2112.12828}}].

\bibitem{Witten:2018zxz}
E.~Witten, \emph{{APS Medal for Exceptional Achievement in Research: Invited article on entanglement properties of quantum field theory}}, \href{https://doi.org/10.1103/RevModPhys.90.045003}{\emph{Rev. Mod. Phys.} {\bfseries 90} (2018) 045003} [\href{https://arxiv.org/abs/1803.04993}{{\ttfamily 1803.04993}}].

\bibitem{Araki:1964lyc}
H.~Araki, \emph{{Type of von Neumann Algebra Associated with Free Field}}, \href{https://doi.org/10.1143/ptp.32.956}{\emph{Prog. Theor. Phys.} {\bfseries 32} (1964) 956}.

\bibitem{Driessler:1976ky}
W.~Driessler, \emph{{On the Type of Local Algebras in Quantum Field Theory}}, \href{https://doi.org/10.1007/BF01609853}{\emph{Commun. Math. Phys.} {\bfseries 53} (1977) 295}.

\bibitem{Chandrasekaran:2022cip}
V.~Chandrasekaran, R.~Longo, G.~Penington and E.~Witten, \emph{{An algebra of observables for de Sitter space}}, \href{https://doi.org/10.1007/JHEP02(2023)082}{\emph{JHEP} {\bfseries 02} (2023) 082} [\href{https://arxiv.org/abs/2206.10780}{{\ttfamily 2206.10780}}].

\bibitem{Chandrasekaran:2022eqq}
V.~Chandrasekaran, G.~Penington and E.~Witten, \emph{{Large N algebras and generalized entropy}}, \href{https://doi.org/10.1007/JHEP04(2023)009}{\emph{JHEP} {\bfseries 04} (2023) 009} [\href{https://arxiv.org/abs/2209.10454}{{\ttfamily 2209.10454}}].

\bibitem{ShadiCrossedProduct}
S.~Ali~Ahmad and R.~Jefferson, \emph{{Crossed product algebras and generalized entropy for subregions}}, \href{https://doi.org/10.21468/SciPostPhysCore.7.2.020}{\emph{SciPost Phys. Core} {\bfseries 7} (2024) 020} [\href{https://arxiv.org/abs/2306.07323}{{\ttfamily 2306.07323}}].

\bibitem{Jensen:2023yxy}
K.~Jensen, J.~Sorce and A.J.~Speranza, \emph{{Generalized entropy for general subregions in quantum gravity}}, \href{https://doi.org/10.1007/JHEP12(2023)020}{\emph{JHEP} {\bfseries 12} (2023) 020} [\href{https://arxiv.org/abs/2306.01837}{{\ttfamily 2306.01837}}].

\bibitem{Fewster:2024pur}
J.C.~Fewster, D.W.~Janssen, L.D.~Loveridge, K.~Rejzner and J.~Waldron, \emph{{Quantum Reference Frames, Measurement Schemes and the Type of Local Algebras in Quantum Field Theory}}, \href{https://doi.org/10.1007/s00220-024-05180-7}{\emph{Commun. Math. Phys.} {\bfseries 406} (2025) 19} [\href{https://arxiv.org/abs/2403.11973}{{\ttfamily 2403.11973}}].

\bibitem{Unruh1976}
W.G.~Unruh, \emph{Notes on black-hole evaporation}, \href{https://doi.org/10.1103/PhysRevD.14.870}{\emph{Phys. Rev. D} {\bfseries 14} (1976) 870}.

\bibitem{DeWitt:1980hx}
B.S.~DeWitt, \emph{{Quantum gravity: the new synthesis}},  in \emph{{General Relativity}: {An Einstein Centenary Survey}}, pp.~680--745, {Cambridge University Press}, (1980).

\bibitem{Sciama1977}
P.~Candelas and D.W.~Sciama, \emph{Irreversible thermodynamics of black holes}, \href{https://doi.org/10.1103/PhysRevLett.38.1372}{\emph{Phys. Rev. Lett.} {\bfseries 38} (1977) 1372}.

\bibitem{Reznik2003}
B.~Reznik, \emph{Entanglement from the vacuum}, \href{https://doi.org/10.1023/A:1022875910744}{\emph{Found. Phys.} {\bfseries 33} (2003) 167}.

\bibitem{Reznik:2003mnx}
B.~Reznik, A.~Retzker and J.~Silman, \emph{{Violating Bell's inequalities in the vacuum}}, \href{https://doi.org/10.1103/PhysRevA.71.042104}{\emph{Phys. Rev. A} {\bfseries 71} (2005) 042104} [\href{https://arxiv.org/abs/quant-ph/0310058}{{\ttfamily quant-ph/0310058}}].

\bibitem{jose}
J.~Polo-G\'omez, L.J.~Garay and E.~Mart\'{\i}n-Mart\'{\i}nez, \emph{A detector-based measurement theory for quantum field theory}, \href{https://doi.org/10.1103/PhysRevD.105.065003}{\emph{Phys. Rev. D} {\bfseries 105} (2022) 065003}.

\bibitem{Pozas-Kerstjens:2015}
A.~Pozas-Kerstjens and E.~Mart\'{i}n-Mart\'{i}nez, \emph{Harvesting correlations from the quantum vacuum}, \href{https://doi.org/10.1103/PhysRevD.92.064042}{\emph{Phys. Rev. D} {\bfseries 92} (2015) 064042}.

\bibitem{kelly}
B.~de~S.~L.~Torres, K.~Wurtz, J.~Polo-G{\'o}mez and E.~Mart{\'\i}n-Mart{\'\i}nez, \emph{{Entanglement structure of quantum fields through local probes}}, \href{https://doi.org/10.1007/JHEP05(2023)058}{\emph{JHEP} {\bfseries 05} (2023) 058} [\href{https://arxiv.org/abs/2301.08775}{{\ttfamily 2301.08775}}].

\bibitem{Mertens:2017mtv}
T.G.~Mertens, G.J.~Turiaci and H.L.~Verlinde, \emph{{Solving the Schwarzian via the Conformal Bootstrap}}, \href{https://doi.org/10.1007/JHEP08(2017)136}{\emph{JHEP} {\bfseries 08} (2017) 136} [\href{https://arxiv.org/abs/1705.08408}{{\ttfamily 1705.08408}}].

\bibitem{Almheiri:2012rt}
A.~Almheiri, D.~Marolf, J.~Polchinski and J.~Sully, \emph{{Black Holes: Complementarity or Firewalls?}}, \href{https://doi.org/10.1007/JHEP02(2013)062}{\emph{JHEP} {\bfseries 02} (2013) 062} [\href{https://arxiv.org/abs/1207.3123}{{\ttfamily 1207.3123}}].

\bibitem{Bousso:2012as}
R.~Bousso, \emph{{Complementarity Is Not Enough}}, \href{https://doi.org/10.1103/PhysRevD.87.124023}{\emph{Phys. Rev. D} {\bfseries 87} (2013) 124023} [\href{https://arxiv.org/abs/1207.5192}{{\ttfamily 1207.5192}}].

\bibitem{Nomura:2012sw}
Y.~Nomura, J.~Varela and S.J.~Weinberg, \emph{{Complementarity Endures: No Firewall for an Infalling Observer}}, \href{https://doi.org/10.1007/JHEP03(2013)059}{\emph{JHEP} {\bfseries 03} (2013) 059} [\href{https://arxiv.org/abs/1207.6626}{{\ttfamily 1207.6626}}].

\bibitem{Harlow:2013tf}
D.~Harlow and P.~Hayden, \emph{{Quantum Computation vs. Firewalls}}, \href{https://doi.org/10.1007/JHEP06(2013)085}{\emph{JHEP} {\bfseries 06} (2013) 085} [\href{https://arxiv.org/abs/1301.4504}{{\ttfamily 1301.4504}}].

\bibitem{Stanford:2022fdt}
D.~Stanford and Z.~Yang, \emph{{Firewalls from wormholes}},  \href{https://arxiv.org/abs/2208.01625}{{\ttfamily 2208.01625}}.

\bibitem{Blommaert:2024ftn}
A.~Blommaert, C.-H.~Chen and Y.~Nomura, \emph{{Firewalls at exponentially late times}}, \href{https://doi.org/10.1007/JHEP10(2024)131}{\emph{JHEP} {\bfseries 10} (2024) 131} [\href{https://arxiv.org/abs/2403.07049}{{\ttfamily 2403.07049}}].

\bibitem{Chandrasekaran:2026gvk}
V.~Chandrasekaran, \emph{{Smooth horizons from topology change in canonical quantum gravity}},  \href{https://arxiv.org/abs/2606.06404}{{\ttfamily 2606.06404}}.

\bibitem{Harlow:2018tqv}
D.~Harlow and D.~Jafferis, \emph{{The Factorization Problem in Jackiw-Teitelboim Gravity}}, \href{https://doi.org/10.1007/JHEP02(2020)177}{\emph{JHEP} {\bfseries 02} (2020) 177} [\href{https://arxiv.org/abs/1804.01081}{{\ttfamily 1804.01081}}].

\bibitem{Danielsson:1998wt}
U.H.~Danielsson, E.~Keski-Vakkuri and M.~Kruczenski, \emph{{Vacua, propagators, and holographic probes in AdS / CFT}}, \href{https://doi.org/10.1088/1126-6708/1999/01/002}{\emph{JHEP} {\bfseries 01} (1999) 002} [\href{https://arxiv.org/abs/hep-th/9812007}{{\ttfamily hep-th/9812007}}].

\bibitem{Spradlin:1999bn}
M.~Spradlin and A.~Strominger, \emph{{Vacuum states for AdS(2) black holes}}, \href{https://doi.org/10.1088/1126-6708/1999/11/021}{\emph{JHEP} {\bfseries 11} (1999) 021} [\href{https://arxiv.org/abs/hep-th/9904143}{{\ttfamily hep-th/9904143}}].

\bibitem{Fewster:2015dqb}
C.J.~Fewster, B.A.~Ju{\'a}rez-Aubry and J.~Louko, \emph{{Asymptotically thermal responses for smoothly switched detectors}},  in \emph{{14th Marcel Grossmann Meeting on Recent Developments in Theoretical and Experimental General Relativity, Astrophysics, and Relativistic Field Theories}}, vol.~4, pp.~3801--3806, 2017, \href{https://doi.org/10.1142/9789813226609_0501}{DOI} [\href{https://arxiv.org/abs/1511.00701}{{\ttfamily 1511.00701}}].

\bibitem{Fewster:2016ewy}
C.J.~Fewster, B.A.~Ju{\'a}rez-Aubry and J.~Louko, \emph{{Waiting for Unruh}}, \href{https://doi.org/10.1088/0264-9381/33/16/165003}{\emph{Class. Quant. Grav.} {\bfseries 33} (2016) 165003} [\href{https://arxiv.org/abs/1605.01316}{{\ttfamily 1605.01316}}].

\bibitem{Shallue:2025zto}
C.J.~Shallue and S.M.~Carroll, \emph{{What Hawking radiation looks like as you fall into a black hole}}, \href{https://doi.org/10.1103/y7kj-4zjw}{\emph{Phys. Rev. D} {\bfseries 112} (2025) 085013} [\href{https://arxiv.org/abs/2501.06609}{{\ttfamily 2501.06609}}].

\bibitem{Bagrets:2016cdf}
D.~Bagrets, A.~Altland and A.~Kamenev, \emph{{Sachdev{\textendash}Ye{\textendash}Kitaev model as Liouville quantum mechanics}}, \href{https://doi.org/10.1016/j.nuclphysb.2016.08.002}{\emph{Nucl. Phys. B} {\bfseries 911} (2016) 191} [\href{https://arxiv.org/abs/1607.00694}{{\ttfamily 1607.00694}}].

\bibitem{Yang:2018gdb}
Z.~Yang, \emph{{The Quantum Gravity Dynamics of Near Extremal Black Holes}}, \href{https://doi.org/10.1007/JHEP05(2019)205}{\emph{JHEP} {\bfseries 05} (2019) 205} [\href{https://arxiv.org/abs/1809.08647}{{\ttfamily 1809.08647}}].

\bibitem{Blommaert:2018oro}
A.~Blommaert, T.G.~Mertens and H.~Verschelde, \emph{{The Schwarzian Theory - A Wilson Line Perspective}}, \href{https://doi.org/10.1007/JHEP12(2018)022}{\emph{JHEP} {\bfseries 12} (2018) 022} [\href{https://arxiv.org/abs/1806.07765}{{\ttfamily 1806.07765}}].

\bibitem{Iliesiu:2019xuh}
L.V.~Iliesiu, S.S.~Pufu, H.~Verlinde and Y.~Wang, \emph{{An exact quantization of Jackiw-Teitelboim gravity}}, \href{https://doi.org/10.1007/JHEP11(2019)091}{\emph{JHEP} {\bfseries 11} (2019) 091} [\href{https://arxiv.org/abs/1905.02726}{{\ttfamily 1905.02726}}].

\bibitem{Lam:2018pvp}
H.T.~Lam, T.G.~Mertens, G.J.~Turiaci and H.~Verlinde, \emph{{Shockwave S-matrix from Schwarzian Quantum Mechanics}}, \href{https://doi.org/10.1007/JHEP11(2018)182}{\emph{JHEP} {\bfseries 11} (2018) 182} [\href{https://arxiv.org/abs/1804.09834}{{\ttfamily 1804.09834}}].

\bibitem{Satz:2006kb}
A.~Satz, \emph{{Then again, how often does the Unruh-DeWitt detector click if we switch it carefully?}}, \href{https://doi.org/10.1088/0264-9381/24/7/003}{\emph{Class. Quant. Grav.} {\bfseries 24} (2007) 1719} [\href{https://arxiv.org/abs/gr-qc/0611067}{{\ttfamily gr-qc/0611067}}].

\bibitem{Louko:2014aba}
J.~Louko, \emph{{Unruh-DeWitt detector response across a Rindler firewall is finite}}, \href{https://doi.org/10.1007/JHEP09(2014)142}{\emph{JHEP} {\bfseries 09} (2014) 142} [\href{https://arxiv.org/abs/1407.6299}{{\ttfamily 1407.6299}}].

\bibitem{Wald2}
R.M.~Wald, \emph{Quantum Field Theory in Curved Spacetime and Black Hole Thermodynamics}, The University of Chicago Press (1994).

\bibitem{Fewster_2013}
C.J.~Fewster and R.~Verch, \emph{The necessity of the {H}adamard condition}, \href{https://doi.org/10.1088/0264-9381/30/23/235027}{\emph{Class. Quant. Grav.} {\bfseries 30} (2013) 235027}.

\bibitem{Saad:2019lba}
P.~Saad, S.H.~Shenker and D.~Stanford, \emph{{JT gravity as a matrix integral}},  \href{https://arxiv.org/abs/1903.11115}{{\ttfamily 1903.11115}}.

\bibitem{Saad:2019pqd}
P.~Saad, \emph{{Late Time Correlation Functions, Baby Universes, and ETH in JT Gravity}},  \href{https://arxiv.org/abs/1910.10311}{{\ttfamily 1910.10311}}.

\bibitem{Almheiri:2019qdq}
A.~Almheiri, T.~Hartman, J.~Maldacena, E.~Shaghoulian and A.~Tajdini, \emph{{Replica Wormholes and the Entropy of Hawking Radiation}}, \href{https://doi.org/10.1007/JHEP05(2020)013}{\emph{JHEP} {\bfseries 05} (2020) 013} [\href{https://arxiv.org/abs/1911.12333}{{\ttfamily 1911.12333}}].

\bibitem{Penington:2019kki}
G.~Penington, S.H.~Shenker, D.~Stanford and Z.~Yang, \emph{{Replica wormholes and the black hole interior}}, \href{https://doi.org/10.1007/JHEP03(2022)205}{\emph{JHEP} {\bfseries 03} (2022) 205} [\href{https://arxiv.org/abs/1911.11977}{{\ttfamily 1911.11977}}].

\bibitem{Iliesiu:2024cnh}
L.V.~Iliesiu, A.~Levine, H.W.~Lin, H.~Maxfield and M.~Mezei, \emph{{On the non-perturbative bulk Hilbert space of JT gravity}}, \href{https://doi.org/10.1007/JHEP10(2024)220}{\emph{JHEP} {\bfseries 10} (2024) 220} [\href{https://arxiv.org/abs/2403.08696}{{\ttfamily 2403.08696}}].

\bibitem{toappanos}
P.~Betzios, P.~Ghiringhelli and O.~Papadoulaki, \emph{Work in progress}, .

\bibitem{Freivogel:2026bsx}
B.~Freivogel and U.~Moitra, \emph{{Large Quantum Gravity Fluctuations of BTZ Black Holes}},  \href{https://arxiv.org/abs/2606.28160}{{\ttfamily 2606.28160}}.

\bibitem{Freivogel:2026ofo}
B.~Freivogel, A.~Speranza and E.~Verlinde, \emph{{Quantum Fluctuations of the Black Hole Horizon}},  \href{https://arxiv.org/abs/2606.28243}{{\ttfamily 2606.28243}}.

\bibitem{Marolf:2003bb}
D.~Marolf, \emph{{On the quantum width of a black hole horizon}}, \href{https://doi.org/10.1007/3-540-26798-0_9}{\emph{Springer Proc. Phys.} {\bfseries 98} (2005) 99} [\href{https://arxiv.org/abs/hep-th/0312059}{{\ttfamily hep-th/0312059}}].

\bibitem{Verlinde:2019xfb}
E.P.~Verlinde and K.M.~Zurek, \emph{{Observational signatures of quantum gravity in interferometers}}, \href{https://doi.org/10.1016/j.physletb.2021.136663}{\emph{Phys. Lett. B} {\bfseries 822} (2021) 136663} [\href{https://arxiv.org/abs/1902.08207}{{\ttfamily 1902.08207}}].

\bibitem{Kourkoulou:2017zaj}
I.~Kourkoulou and J.~Maldacena, \emph{{Pure states in the SYK model and nearly-$AdS_2$ gravity}},  \href{https://arxiv.org/abs/1707.02325}{{\ttfamily 1707.02325}}.

\bibitem{Goel:2018ubv}
A.~Goel, H.T.~Lam, G.J.~Turiaci and H.~Verlinde, \emph{{Expanding the Black Hole Interior: Partially Entangled Thermal States in SYK}}, \href{https://doi.org/10.1007/JHEP02(2019)156}{\emph{JHEP} {\bfseries 02} (2019) 156} [\href{https://arxiv.org/abs/1807.03916}{{\ttfamily 1807.03916}}].

\bibitem{Roberts:2014isa}
D.A.~Roberts, D.~Stanford and L.~Susskind, \emph{{Localized shocks}}, \href{https://doi.org/10.1007/JHEP03(2015)051}{\emph{JHEP} {\bfseries 03} (2015) 051} [\href{https://arxiv.org/abs/1409.8180}{{\ttfamily 1409.8180}}].

\bibitem{Engelhardt:2020qpv}
N.~Engelhardt, S.~Fischetti and A.~Maloney, \emph{{Free energy from replica wormholes}}, \href{https://doi.org/10.1103/PhysRevD.103.046021}{\emph{Phys. Rev. D} {\bfseries 103} (2021) 046021} [\href{https://arxiv.org/abs/2007.07444}{{\ttfamily 2007.07444}}].

\bibitem{Johnson:2021rsh}
C.V.~Johnson, \emph{{On the Quenched Free Energy of JT Gravity and Supergravity}},  \href{https://arxiv.org/abs/2104.02733}{{\ttfamily 2104.02733}}.

\bibitem{Alishahiha:2020jko}
M.~Alishahiha, A.~Faraji~Astaneh, G.~Jafari, A.~Naseh and B.~Taghavi, \emph{{Free energy for deformed Jackiw-Teitelboim gravity}}, \href{https://doi.org/10.1103/PhysRevD.103.046005}{\emph{Phys. Rev. D} {\bfseries 103} (2021) 046005} [\href{https://arxiv.org/abs/2010.02016}{{\ttfamily 2010.02016}}].

\bibitem{Maldacena:2017axo}
J.~Maldacena, D.~Stanford and Z.~Yang, \emph{{Diving into traversable wormholes}}, \href{https://doi.org/10.1002/prop.201700034}{\emph{Fortsch. Phys.} {\bfseries 65} (2017) 1700034} [\href{https://arxiv.org/abs/1704.05333}{{\ttfamily 1704.05333}}].

\end{thebibliography}\endgroup
\end{document}